\documentclass{aa}
 
\usepackage{graphics}
\usepackage{latexsym}

\begin{document}
\thesaurus{08.01.1;  
08.06.3; 
08.12.1; 
10.01.1; 
10.05.1 
}
\title{Abundances in metal-rich stars \thanks{Based on observations 
at the McDonald Observatory.} \thanks{Table 11 is only available in electronic form at the CDS via anonymous ftp to cdsarc.u-strasbg.fr (130.79.128.5) or via http://cdsweb.u-strasbg.fr/Abstract.html}
}
 
\subtitle{Detailed  abundance analysis of 47 G and K dwarf stars \\
 with [Me/H] $>$ 0.10 dex}
 
\author{Sofia Feltzing \inst{1,} \inst{2}
   \and Bengt Gustafsson \inst{2}
}

\offprints{S. Feltzing \\ sofia@ast.cam.ac.uk}
 
\institute{
Royal Greenwich Observatory, Madingley Road, Cambridge CB3 0EZ, United Kingdom
   \and Astronomiska observatoriet, Box 515, S-751\,20 Uppsala, Sweden.
            }

\date{Received 25 June 1997;  accepted 29 August 1997}
 
\maketitle
 
\begin{abstract}

We have derived elemental abundances of O, Na, Mg, Al, Si, Ca, Ti, Cr,
Mn, Fe, Co, Ni as well as for a number of s-elements for 47 G and K
dwarf, with [Me/H]$>0.1$ dex. The selection of stars was based on their
kinematics as well as on their $uvby-\beta$-photometry.  One sample of
stars on rather eccentric orbits traces the chemical evolution
interior to the solar orbit and another, on circular orbits, the
evolution around the solar orbit. A few Extreme Population I stars
were included in the latter sample.

The stars have $-0.1$ dex $<  {\rm [Fe/H]} < 0.42$ dex. The
spectroscopic [Fe/H] correlate well with the [Me/H] derived from
$uvby-\beta$ photometry.  We find that the elemental abundances of
Mg, Al, Si, Ca, Ti, Cr and  Ni all follow [Fe/H]. Our data put further
constraints on models of galactic chemical evolution, in particular of
Cr,  Mn and Co which has not previously been studied for  dwarf stars
with ${\rm [Me/H]} >0.1$ dex.  The increase in [Na/Fe] and [Al/Fe] as
a function of [Fe/H] found previously by Edvardsson et al. (1993a) has
been confirmed for [Na/Fe].  This upturning relation, and the scatter
around it, are shown not to be due to a mixture of populations with
different mean distances to the galactic centre. We do not confirm the
same trend for aluminium, which is somewhat  surprising since both
these elements are thought to be produced in the  same environments in
the pre-supernova stars. Nor have  we been able to trace any tendency
for relative abundances of O, Si, and Ti relative to Fe to vary with
the stellar velocities, i.e. the stars present mean distance to the
galactic centre. These results imply that there is no significant
difference in the chemical evolution of the different stellar
populations for stars with [Me/H]$>0.1$ dex. We find that [O/Fe]
continue to decline with increasing [Fe/H] and that oxygen and
europium  correlate well. However  [Si/Fe] and [Ca/Fe] seem to stay
constant. A real ("cosmic") scatter in [Ti/Fe] at given [Fe/H] is
suggested as well as a decreasing abundance of the s-elements relative
to iron for the most metal-rich  dwarf stars.  We discuss our results
in the context of recent models of galactic chemical evolution.

In our sample we have included a few very metal rich stars, sometimes
called SMR (super metal rich) stars. We find these stars to be among
the most iron-rich in our sample  but far from as metal-rich as
indicated by their photometric metallicities. SMR stars on highly
eccentric  orbits, alleged to trace the evolution of the chemical
evolution in the galactic Bulge, have  previously been found
overabundant in O, Mg and Si. We have included three such stars from
the study by Barbuy \& Grenon (1990). We find them to be less metal
rich and  the other elemental abundances remain puzzling.  

Detailed spectroscopic abundance analyses of  K dwarf stars are
rare. Our study includes 5 K dwarf stars and has revealed what appears
to be a striking example of overionization. The overionization is
especially prominent for Ca, Cr and Fe. The origin of this apparent
overionization is not clear and we discuss different explanations in
some detail.

      \keywords{Galaxy: abundances --
                 evolution --              
	        Stars: abundances --
                 late-type  --
                fundamental parameters
               }
   \end{abstract}

\section{Introduction}

The time that F and G dwarf stars spend on the main-sequence span a range
from $10^9$ to several times $10^{10}$ years. This means that such
stars may be used as tracers of the chemical and dynamical evolution
of the Galaxy; in fact, a combination of chemical and kinematical data is
a very powerful tool for studying the galactic chemical evolution,
cf. e.g. Edvardsson et al. (1993a) and Wyse \& Gilmore (1995).

Spectroscopic abundance analysis of stars have now become accurate
enough to admit determinations of rather small ($\la 0.10$ dex) relative
abundance differences in differential studies of stellar samples of
large size. These developments make it possible to explore the
galactic chemical evolution in considerable detail. 

Edvardsson et al. (1993a) analysed 189 F and G dwarf stars, with $-1.1$ dex $
< {\rm{[Fe/H]}} <0.25$ dex. Accurate abundances were determined for a number of
key elements, O, Na, Mg, Al, Si, Ca, Ti, Fe and Ni, as well as  a number
of s-process elements. The abundance results combined with accurate
velocity data enabled a detailed study of the chemical evolution of
kinematically distinct populations.  Extensions of this study were
made by Tomkin et al. (1995) and Woolf et al. (1995) who measured carbon and
europium abundances, respectively, for about half of the stars in the
Edvardsson et al. (1993a) sample. 

The study  by Edvardsson et al. (1993a)  raised a number of new
questions concerning the most metal-rich stars in the galactic disk,
not the least concerning the build-up of sodium, magnesium and
aluminium. Sodium, aluminium and, possibly, magnesium relative iron vs
[Fe/H] showed an increase for ${\rm{[Fe/H]}} > 0.0 $ dex (cf. Edvardsson et
al. 1993a Fig. 15a-l). Are these ``upturns'' real? Large star-to-star
scatter was also encountered for the abundance of certain elements:
magnesium, aluminium and titanium, relative to iron at a given
[Fe/H]. Could this scatter be reduced by using more/better abundance
criteria or is the scatter intrinsic to the stellar population? One
suggestion was that ``upturns'' and scatter could be due to a mixing
of populations with different ages and with different mean distances
from the galactic centre, e.g. a mixture of old metal-rich stars, more
concentrated to the centre and young Extreme Population I stars on
solar like orbits. We have therefore studied a sample of 47 metal-rich
dwarf stars, with photometric metallicities ${\rm{[Me/H]}}
>0.10$ dex, chosen to represent different mean perigalactic distances and
presumably different ages.

The paper is organized as follows: in Sect. 2 and 3, we describe the
selection criteria of the stellar sample, the observations and
reductions, Sect. 4 contains a description of the analysis while the
errors are discussed in detail in Sect. 5. Abundance results and their
interpretation in terms of models of galactic chemical evolution are
presented in Sect. 6 and, finally, Sect. 7 contains a summary and discussion.

\section{Selection of stars}

The stars have been selected to include both stars from the old
metal-rich disk population, as well as more recently formed Extreme
Population I dwarf stars. The sample of stars was confined
photometrically by  $0\fm38  <b-y< 0\fm63$ and with  $\delta c_1$ and
$ \Delta m_1 $ such that  with the calibrations of Edvardsson et
al. (1993a) and Olsen (1984), solar-type dwarfs and subgiants are
singled out with $4500 {\rm K} \leq T_{\rm eff} \leq 6000 $ K, $4.0
\leq \log g \leq 4.6$ and [Me/H] $ > 0.10$ dex.  Magnitude and
metallicity data were taken from the catalogues by Olsen (1983, 1993, 
1994 and priv. comm.). These
were combined with accurate velocity data (UVW) calculated by Olsen
from available proper motions and unpublished CORAVEL  radial
velocities by Andersen, Mayor \& Nordstr\"om,  in the selection of
programme stars.

From the catalogue we seleted 42 G and K dwarf stars with [Me/H] $>
0.10$ for our study.  13 of these stars have $V_{\rm LSR}  < - 50$
km/s and/or $Q_{\rm LSR}>$ 60 km/s, where $V_{\rm LSR}$ is the stellar
space velocity component in the direction of the rotation of the
Galaxy  relative to the local standard of rest (LSR) and $Q_{\rm LSR}$
is the total space velocity relative to LSR. These stars were selected
to represent the chemical evolution galactic regions different from
those represented by the solar orbit, in particular located more
closely to the galactic centre.  As shown in Edvardsson et
al. (1993b), Fig. 1, $V_{\rm LSR}  < - 50$ km/s singles out stars with
mean perigalactica less than 7 kpc. The 17 stars with $Q_{\rm LSR} <
30$ km/s represent the chemical evolution in the gas close to the
solar orbit.

Since we also wished to explore the oxygen abundances in metal-rich
disk dwarf stars on highly eccentric orbits we added 3 additional
stars from  the sample of stars for which Barbuy \& Grenon (1990)
found abnormally high [O/Fe] ratios  (HD37986, HD77338,
HD87007). Moreover, we included  2 stars from Schuster \& Nissen
(1988) (HD175518 and HD182572). 

The total final sample contains 47 stars and has 3 in common with
Edvardsson et al.(1993a): HD30562, HD67228 and  HD144585.

\section{Observations and reductions}
\subsection{Observations}
\label{sect:observations}

The observations were carried out with the 2.7~m telescope of the
McDonald Observatory, University of Texas,  during two observing
runs, March $1-9$, 1994 and  April 28 -- May 2, 1994, using the
2d-coud\'e cross dispersion echelle  spectrometer (Tull et al. 1995). 

 At the time of the observations the new 2k$\times$2k chip, which
covers most of the visual spectrum of the observed star in one
exposure, was not yet available at the  spectrometer. The CCD-chip
used has 800$\times$864 pixels and covers approximately 18 echelle
orders of length 25$-$35 \AA.  Settings of the detector for the
observations were selected to enable reliable analysis of  the
spectral lines from the elements oxygen, sodium, magnesium  and
aluminium, silicon, calcium, titanium, as well as chromium, iron,
nickel, yttrium and europium.  Many sources were utilized to make the
selection of lines optimized for the current project, the most
important being Edvardsson et al. (1993a), Morell (1994), M\"ackle et
al. (1975), Ruland et al. (1981). The  Arcturus Atlas by Griffin
(1968) and the Solar Spectrum Atlas by Delbouille et al. (1973),
were carefully inspected, in order to avoid  blends and define
suitable continuum regions. Altogether 3 different settings of the CCD
in the focal plane of the spectrometer camera were used.

The nominal spectral resolution (slit width 0.25 $\mu$m) of  $\sim$
100\,000,  was verified by observations of a thorium-\\argon lamp and
by measuring telluric lines. This relatively high resolution was
judged to be important in view of the relatively crowded spectra, in
particular for the cooler stars.

\subsection{Reductions of stellar spectra}

The reductions were carried out with standard IRAF\footnote{IRAF is
 distributed by National Optical Astronomy Observatories, operated by
 the Association of Universities for Research in Astronomy, Inc.,
 under contract with the National Science Foundation, USA.} packages
 for reduction and extraction of echelle spectra. The procedure
 adopted was: all raw frames were first corrected for read-out noise
 and bias, the flat fields added together,  all object frames were
 trimmed, the new flat field frames were normalized, the stellar
 frames were divided by the normalized flat field, corrected for
 scattered light and the one-dimensional stellar spectra
 extracted. Finally, the spectrum of the comparison lamp, a
 thorium-argon lamp, was reduced, extracted and the lines in it
 identified. The comparison spectra were used to carry out the
 wavelength calibration of the stellar spectra. A Legendre polynomial
 was fitted to each stellar spectrum to define the continuum, using
 the CONTINUUM task in IRAF. 

The results of this procedure for defining the continuum were
 inspected visually and judged to give  good continua for all the
 lines used in the abundance analysis; however, for the strong
 Ca\,{\sc i} line at 6162 \AA, used for checking surface gravities,
 this definition of the continuum was not satisfactory for all stars.
 Instead, a blaze function was constructed for each night by fitting a
 Legendre polynomial to the spectrum of a B star observed during the
 night. The stellar spectra were then divided by this blaze
 function. Then a low order Legendre polynomial was fitted to the
 stellar spectrum to rectify it. The change in continuum was negligible
 for half of the stars, as compared with the result of using
 CONTINUUM; this also includes half of the  K dwarf stars. 

Our spectra are generally of  high quality, reaching S/N of $\sim$ 200
in most cases. Some problems with fringes in the red were
encountered. The measurements of spectral line equivalent widths were
done  using the SPLOT task in IRAF. Karin Eriksson is gratefully
thanked for carrying out almost half of the measurements. Care was
exercised to guarantee that the measurements agreed well between the
two operators. 

\subsection{Solar observations and reductions}
\label{sol}

The solar observations were performed by illuminating the slit of the
spectrograph by sky light each afternoon. The solar spectra have
$\sim$ 2 times higher S/N ratios than the stellar spectra.  The
reductions and measurements of the solar spectra were made in the same
way as for the stellar spectra.  

The solar observations were used to determine ``astrophysical'' $\log
gf$-values. For lines, denoted by K in Table \ref{linelist},  shifted
by the difference in radial velocities between the stars and the Sun
outside our solar spectral recordings, but judged to be of interest
for the analyses, solar equivalent widths were instead obtained from
the solar spectrum atlas by Kurucz et al. (1984).

\section{Analysis}

We have performed a standard Local Thermodynamic Equilibrium (LTE)
analysis, strictly differential with respect to the Sun, to derive  chemical
abundances from the measured equivalent widths. 

\subsection{Model atmospheres}

We used the MARCS program,  first described by Gustafsson et
al. (1975), to generate the model atmospheres.  Since then, the
program has been further developed in various ways and updated in
order to handle the line blanketing of millions of absorption lines
more accurately,  Asplund et al. (1997). The
following assumptions enter into the calculation of the models: the
atmosphere is assumed to be plane-parallel and in hydrostatic
equilibrium,  the total flux (including mixing-length convection) is
constant, the source function is described by the Planck function at
the local temperature with a scattering term, the populations of
different excitation levels and ionization stages are governed by
LTE. Since the analysis is differential relative to the Sun we have
also used a solar model atmosphere calculated with the same program as
the stellar models, in spite of the fact that the empirically derived 
Holweger-M\"uller
model  better reproduces the solar observed limb darkening; see
Blackwell et al. (1995) for a discussion of this.

\subsection{Fundamental parameters of model atmospheres} 

\begin{table*}
\caption[]{Stellar parameters of the observed stars. The fifth column
gives the spectral classification according to Olsen(1983, 1993, 
1994 and priv. comm.), except
where indicated 1) from  Eggen (1960),
and 2) spectral class from Hoffleit \& Jascheck (1982); the sixth and seventh
columns give effective temperature and surface gravity as derived from
photometry; the eight and ninth give surface gravities as derived from
spectra, wings of Ca\,{\sc i}, and parallaxes, respectively; the tenth
and eleventh give iron abundances as determined from our spectral
analysis and metallicities as determined from photometry; the
following four columns contain the space velocity data for the stars,
in km/s relative to the LSR, from Olsen (priv. comm.). In the last column
the stellar space velocity is given.}  {\setlength{\tabcolsep}{0.8mm}
\begin{tabular}{llllllrrrrrrrrrrc}
\hline\noalign{\smallskip}
HD  & HR & Name &   \multicolumn{1}{c}{V} & Spec.class &   T$_{\rm eff}$ & log{\it g}$_{_{\rm phot}}$ & log{\it g}$_{_{\rm spec}}$ &log{\it g}$_{_{\pi}}$ & \multicolumn{1}{c}{[Fe/H]} & \multicolumn{1}{c}{[M/H]}& $\xi_t$& \multicolumn{1}{c}{U} &  \multicolumn{1}{c}{V} &  \multicolumn{1}{c}{W} &  \multicolumn{1}{c}{Q}   \\ 
\noalign{\smallskip}
\hline\noalign{\smallskip}
30562   & 1536 &  & 5.77 & F8V  & 5876 & 4.00 & 4.00 & 3.9 &  0.19 & 0.14&1.4 &  40.8 &--70.9 &--10.7 & 82.5 & \\
36130   &      &  & 7.76 & G0M  & 5986 & 4.34 & 4.40 & 4.0 &  0.15 & 0.15&1.1 &--14.3 &--52.5 &--43.7 & 69.8 &  \\
37088   &      &  & 8.51 & G0   & 5856 & 4.26 & 4.35 &     &  0.10 & 0.31&1.8 &--17.4 &--12.4 &--43.0 & 48.0 &  \\
37216   &      &  & 7.84 & G5M  & 5527 & 4.47 & 4.90 &     &--0.02 & 0.17&1.1 &  9.8  &   1.0 & --4.3 & 10.7 &  \\
49178   &      &  & 8.07 & G0E   & 5683 & 4.37 & 4.45 &     &  0.01 & 0.06&1.2 &  32.3 &  10.3 &--10.5 & 35.5 &  \\
54322   &      &  & 8.40 & G5    & 5894 & 4.50 & 4.55 &     &  0.15 & 0.33&0.9 &--20.2 & --0.7 &   7.1 & 21.4 &  \\
55693   &      &  & 7.18 & G5M   & 5845 & 4.15 & 4.40 &     &  0.26 & 0.24&1.3 &  33.9 &   3.3 & --5.0 & 34.4 &  \\
67228   & 3176 &  & 5.30 & GIVB  & 5831 & 4.14 & 3.90 & 3.6 &  0.16 & 0.22&1.4 &--41.9 &  16.6 & --9.2 & 46.0 & \\                                        
68988   &      &  & 8.20 & G0    & 5956 & 4.03 & 4.25 &     &  0.37 &     &1.4 &--94.1 & --8.8 & 11.8 &  95.2 &  \\
69582   &      &  & 7.56 & G5    & 5652 & 4.34 & 4.74 &     &  0.08 & 0.06&1.2 &--20.0 &  11.5 & 10.5 &  25.3 &  \\
69830   & 3259 &  & 5.96 & G7.5V & 5484 & 4.30 & 4.95 & 4.5 &--0.03 & 0.10&1.1 &--41.3 &--51.7 &--4.4 &  66.3 &  \\
71479   &      &  & 7.18 & G0    & 6036 & 4.18 & 4.48 &     &  0.25 & 0.32&1.5 &  34.8 &--40.4 &--9.1 &  54.1 &  \\
72946B  & 3396 &  & 7.20 & G5V   & 5911 & 4.40 & 5.00 &     &  0.24 & 0.40&1.3 &  15.0 &--18.0 &--3.4 &  23.7 &  \\
75782   &      &  & 7.08 & G0    & 5930 & 3.87 & 3.77 &     &  0.18 & 0.11&1.5 &  12.3 &--23.0 &--9.1 &  27.7 &  \\
76780   &      &  & 7.64 & G5M   & 5869 & 4.30 & 4.80 &     &  0.21 & 0.30&1.3 &  22.3 &--6.7  & 10.1 &  25.3 &  \\
80607A  &      &  & 9.15 & G5R   & 5457 & 4.28 &      &     &  0.27 & 0.30&1.1 &--16.9 &  15.0 & 19.3 &  29.7 &  \\
87646   &      &  & 8.07 & G0    & 5961 & 4.06 & 4.41 &     &  0.30 & 0.39&1.4 &  21.9 &--11.4 &  2.6 &  24.8 &  \\
91204   &      &  & 7.82 & G0    & 5864 & 4.05 & 4.00 &     &  0.17 & 0.24&1.4 &--19.6 &   9.7 &  3.6 &  22.2 &  \\
94835   & \multicolumn{2}{l}{G\,058-030} & 9.11 & K0/G0 & 5896 & 4.06 &      &     &   0.13 &   0.04&1.4 &         &--65.1 &        &      &  \\
101242  &      &  & 7.61 & G5    & 5790 & 4.28 & 4.68 &     &  0.07 & 0.19&1.3 &  45.2 &--50.0 &--2.6 &  67.4  &  \\
106156  &      &  & 7.92 & K0    & 5437 & 4.27 & 4.77 &     &  0.13 & 0.20&1.1 &--61.9 &--13.0 &-11.3 &  64.3  & \\
110010  &      &  & 7.01 & G0    & 5965 & 4.08 & 4.58 & 4.7 &  0.35 & 0.34&1.4 &   9.0 &--14.1 &--6.6 &  17.9  &  \\
117243  &      &  & 8.35 & G0/G5III$^{1}$&  5902 & 4.01 & 4.36 &     &   0.24 & 0.33&1.0 & 13.2  &--59.9 & 4.6  &  61.5  &\\
125968  &      &   & 7.77 &  G0/G5IV-V$^{1}$ & 5868 & 4.12 & 4.32 & 4.7 &   0.15 & 0.25&1.4 &  22.6 &--89.6 &--27.7  & 96.5 &  \\
126511  &      &   & 8.37 & G5   & 5472 & 4.28 & 4.70 &     &  0.06 & 0.10&1.3 &--14.5 &--47.6 & --9.9  &  50.7 &  \\
128987  &      &   & 7.24 & G5   & 5588 & 4.35 & 5.00 &     &  0.05 & 0.10&0.9 & 14.7  &   2.9 &  --7.1 &  16.6 & \\
130087  &      &   & 7.52 & F5   & 6023 & 4.01 & 4.41 &     &  0.25 & 0.25&1.5 & 13.0  &--16.6  &  4.9  & 21.7 & \\
134474  &      &   & 8.88 & G5   & 5375 & 4.46 & 5.06 &     &   0.16 & 0.33&1.0 &   7.7 &  22.8 & --27.1 &   36.2 & \\
134987  & 5657 & 23\,Lib & 6.47 & G4V & 5833 & 4.11 & 4.31 & 4.0 &   0.36 & 0.58&1.3 &  10.3 &--24.9 &   26.9 &   38.1 & \\
137510  & 5740 &   & 6.27 & G0IV-V    & 5929 & 3.91 & 3.91 &     &   0.25 & 0.28&1.5 & --0.2 &  2.0  &    7.6 &    7.9 &\\
144585  & 5996 &   & 6.31 & G4IV-V    & 5831 & 4.03 & 4.38 &     &   0.27  & 0.26&1.4  & 23.6  &--17.5 & 33.6   & 44.6 &  \\
171999A &      &   & 8.33 & G5       & 5249 & 4.32 & 4.65 & 4.0 &   0.40 & 0.06&0.9 &   7.6 &--61.3 &  --3.3 &   61.9 &   \\
175518  &      &   &      & K0IV-V $^{1}$ & 5713 & 3.93 &4.73  & 4.3 &   0.32 & 0.67&1.5 &  14.6 &--99.8 &   12.6 &  101.7 &  \\
178911A & 7272 &   & 6.73 & G5R   & 5910 & 4.24 & 4.44 & 4.0 &   0.06 & 0.35& &  35.1 & --8.8 &   6.0  &   36.7 &  \\
180890  &      &   & 8.35 & G5    & 5530 & 4.23 & 4.53 &     &0.14 &   0.09&1.5 &  16.9 &--50.0 & --9.4  &   53.6 & \\
182572  & 7373 & 31\,Aql & 6.36 & G8IV$^{2}$ & 5739 & 3.83 & 4.43 & 4.1 &   0.42 & 0.50&1.4 & --1.0 &         &          &  123.2 &  \\
183263  &      &   & 7.87 & G5/G2IV  & 5837 & 4.05 & 4.40 &     &   0.15 & 0.22&1.5 &  18.6 &--32.2 &   8.6  &   38.2 &  \\
186427  & 7504 & 16\,Cyg\,B & 6.23 & G3V & 5773 & 4.17 & 4.42 &     &   0.06 &   0.21&1.3 &--27.9 &--17.9 &   5.8  &   33.6 & \\
187055  &      &   & 9.00 & G5   & 5298 & 4.56 & 4.96 & 3.9 &  0.16 & 0.25&0.9 & 83.0 &   5.3 & --12.7 &   84.1 &  \\   
\hline\noalign{\smallskip}
\multicolumn{14}{c}{K dwarf stars} \\
\hline\noalign{\smallskip}
32147   & 1614 &  & 6.21 & K3V  & 4625 & 4.57 & 4.55 & 4.4 &  0.28 & 0.17&1.0 &--11.5 &--36.1 &--3.7  & 38.0 &  \\
61606A  &      &  & 7.17 & K2V   & 4833 & 4.55 & 4.85 & 4.6 &--0.08 & 0.11&1.0 &--34.9 &   9.4 & --1.0 & 36.1 &  \\
103932  &      &  & 6.95 & K5V   & 4510 & 4.58 & 4.85 & 4.6 &  0.16 & 0.21&1.0 &   9.9 &--59.9 &  0.4 &  60.7  &  \\
131977A & 5568 &   & 5.72 & K4V  & 4585 & 4.58 & 4.70 & 4.6 &   0.04  & 0.18&1.0 &--57.2 & --9.5 & --26.1 &   63.6 &\\
136834  &      &   & 8.26 & K0        & 4765 & 4.56 & 4.47 & 4.5 &   0.19 & 0.23 &1.0&   9.6 &--44.3 &  --8.7 &   46.2 &\\
\hline\noalign{\smallskip}
\multicolumn{14}{c}{Stars in common with Barbuy \& Grenon (1990)}\\
\hline\noalign{\smallskip}
37986$^3$   &      &  & 7.37 & G5/K0IV & 5455 & 4.50 & 4.40 & 4.3 &  0.27  &0.47 &1.0  & & & &  \\
77338$^3$   &      &  & 8.63 & K0IV&  5290 & 4.50 & 4.90 &     &  0.22 & 0.45 &1.0 & & & &  \\
87007$^3$   &      &  & 8.81 & K2    & 5300 & 4.50 & 4.70 &     &  0.27 & 0.43 &1.0 &40.3$^4$  & --42.5$^4$ & --14.2$^4$ & &   \\
\noalign{\smallskip}
\hline\noalign{\smallskip}
\multicolumn{15}{l}{$^3$ All stellar parameters from Barbuy \& Grenon (1990)} \\
\multicolumn{15}{l}{$^4$ Barbuy, private communication}\\
\label{parameters}
\end{tabular}}
\end{table*}

The effective temperature and surface gravity for each star were
derived from $uvby-\beta$ photometry from Olsen (1983, 1993, 1994 and
priv. comm.).  We have used an extension of the calibration in
Edvardsson et al. (1993a) and the calibration in Olsen (1984). The
calibration by Edvardsson et al. (1993a) is valid for the parameter
space $2.58<\beta<2.72$ and $0.04<(\delta c_1 + 0.5\cdot \delta m_1)<
0.16$. This translates roughly  to $5600~ {\rm K} < T_{\rm eff} <
7000~ {\rm K} $, $3.8 < \log  g < 4.5$ and  metallicity $-1 ~ {\rm
dex}< {\rm [Me/H]} <  +0.3~ {\rm dex}$. The Edvardsson et al. (1993a)
calibration was preferred for all stars to which it is applicable.
For the three stars previously studied by Barbuy \& Grenon (1990)
$uvby-\beta$ photometry is not available, and we used the stellar
parameters cited by them.

\subsection{Surface gravity}

\begin{figure}
\resizebox{\hsize}{!}{\includegraphics{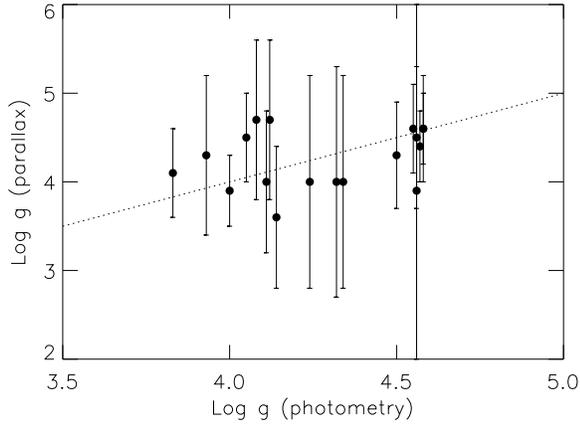}}
\caption[]{Surface gravities derived from parallaxes vs surface
gravities derived from photometry. The one-to-one relation is
indicated by the dotted line. The error bars reflect the errors
in the parallaxes. Note the different scales on x- and y-axis. }
\label{figgravity}
\end{figure}

Surface gravities were derived from the $c_1$ index, which primarily
measures the Balmer jump and is sensitive to the surface gravity in
solar type dwarf stars. In addition to surface gravities estimated
from photometry we have also used the wings of the strong calcium line
at 6162 {\AA} (see Blackwell \& Willis (1977) and Edvardsson (1988)),
to derive surface gravities. The results are given in Table
\ref{parameters}.   For these results  we are indebted to Matthias
Palmer and Mikael Nilsson, who carried out these laborious
determinations.  Edvardsson et al. (1993a) estimated the error in
surface gravity determined from their calibration of the $c_1$ index
to 0.2 dex. Olsen (1984) quotes a similar accuracy for his
calibration.  For the stars with trigonometric parallaxes in van
Altena et al. (1991) we have also estimated surface gravities from
parallaxes.  This exercise was meant to be a consistency check on the
photometric and spectroscopic determinations of the surface
gravities. Assuming that the stars have masses of 0.8 M$_{\odot}$ and
interpolating  the bolometric corrections  in the table given in Allen
(1973)  we derived, from the parallaxes, the surface gravities (see
e.g. Gustafsson et al. 1974) shown in Fig. \ref{figgravity} and given
in Table \ref{parameters} \footnote{ With the release of the
Hipparcos parallaxes this can now be addressed in greater detail and
for most  of our stars. The results will be published elsewhere.}.

The agreement between surface gravities determined with the different
methods is in general good. We note, however, that there is in the
mean an offset in the surface gravities determined from the wings of
the strong Ca\,{\sc i} line as compared to the surface gravities
determined from photometry, see Table \ref{parameters}, of about
$+0.27$ dex.  The reason for this offset is not clear but,
a corresponding uncertainty in surface gravity is of minor
significance for the abundance results, see Table
\ref{tabellpartest}. A tendency for the trigonometric $\log g$
determiantions to agree closer with the photometric than with the
spectroscopic ones may also be traced.  It is interesting to note that
the K dwarf stars in our study have  $\log g$ derived from
parallaxes that are in good agreement with those determined from
photometry.

\begin{table*}
\caption[]{Column 1: Wavelength as quoted in Moore et al. (1966). Column 2
: Excitation energy of the lower level involved in the
transition. Column 3: Astrophysically derived oscillator strengths
(based on the solar equivalent widths). Column 4: Correction factor to
Uns\"old's damping constant. Column 5: Radiative damping
constant. Column 6: Equivalent width as measured in the Sun. If
nothing else is indicated in the 7th column observations of the
daylight sky with the same spectrograph were used. 
Column 7: Lines used to check the excitation
equilibrium are denoted by t and lines for which the solar equivalent
width was measured from the Kurucz et al.\,(1984) Solar Flux Atlas by
a K, see Sect. \ref{sol}.  Cr\,{\sc ii} and Fe\,{\sc ii}  
lines  excluded from the analysis
of the K dwarfs stars due to possible blends are denoted by $+$, 
see Sect. \ref{kdwarfs}. For
each element the solar logarithmic abundance is given according to in
Anders \& Grevesse (1989), except for iron for which the value 7.51 was used,
see  Holweger et al. (1991) and Bi\'emont et al. (1991).  } 
\begin{tabular}{lrrrrrrlclrrrrrrl}
\hline\noalign{\smallskip}
{ $\lambda$} & { $\chi_l$} & {$ \log gf$} & {$\delta \Gamma_6$} & {  $\Gamma _{\rm rad}$} & { W$_{\lambda \odot}$} & Note &&{ $\lambda$} & { $\chi_l$} &{$ \log gf$} & {$\delta \Gamma_6$} & {  $\Gamma _{\rm rad}$} & { W$_{\lambda \odot}$} & Note \\ 
{[\AA] }      &[eV]       &            &                     &[s$^{-1}$]                & [m\AA]                 &    && {[\AA] }      &[eV]       &            &                     &[s$^{-1}$]                & [m\AA]                 &  \\ 
\noalign{\smallskip}
\hline\noalign{\smallskip}
\multicolumn{7}{l} {{\bf O\,{\sc i}}; $\log \epsilon_\odot = 8.93$}&& \multicolumn{7}{l} {{\bf V\,{\sc i}}; $\log \epsilon_\odot= 4.00$}\\ 
 6300.31 & 0.0  & --9.84 & 2.5 & 1.0+8 &  4.6 &   & &              6452.31 & 1.19 & --0.81 & 2.5 & 4.0+7 &  8.9 &    \\  
 6156.80 &10.74 & --0.43 & 2.5 & 1.0+8 &  4.1 &   & &              6504.18 & 1.18 & --0.47 & 2.5 & 3.9+7 & 27.0 &   \\ 
 6158.17 &10.74 & --0.65 & 2.5 & 1.0+8 &  2.6 &   & &              \multicolumn{7}{l} {{\bf V\,{\sc ii}}; $\log \epsilon_\odot= 4.00$}\\
 7771.95 & 9.14 &  0.29  & 2.5 & 1.0+8 & 72.2 &   & &              5303.22 & 2.28 & --2.02 & 2.5 & 2.7+8 &  4.1 &   \\ 
 7774.18 & 9.14 &  0.14  & 2.5 & 1.0+8 & 61.5 &   & &              5384.87 & 2.28 & --2.50 & 2.5 & 2.7+8 &  1.4 &   \\    
 7775.39 & 9.14 & --0.05 & 2.5 & 1.0+8 & 50.9 &   & &              \multicolumn{7}{l} {{\bf Cr\,{\sc i}}; $\log \epsilon_\odot= 5.67$}\\
\multicolumn{7}{l}{{\bf Na\,{\sc i}}; $\log \epsilon_\odot=6.33$}&&    5220.91 & 3.38 & --1.01 & 2.5 & 8.1+7 & 10.9 &   \\    
 6160.75 & 2.10 & --1.30 & 2.1 & 1.0+8 & 58.8 &   & &              5238.96 & 2.71 & --1.43 & 2.5 & 2.5+8 & 16.7 &   \\ 
 6154.23 & 2.10 & --1.62 & 2.1 & 1.0+8 & 37.2 &   & &               5304.18 & 3.46 & --0.79 & 2.5 & 2.5+8 & 14.9 &   \\    
\multicolumn{7}{l}{{\bf Mg\,{\sc i}}; $\log \epsilon_\odot= 7.58$}&&    5312.86 & 3.45 & --0.69 & 2.5 & 2.5+8 & 18.5 &   \\   
6319.24 & 5.11 & --2.20 & 2.5 & 1.0+8 & 31.7 &   & &                5318.77 & 3.44 & --0.78 & 2.5 & 2.5+8 & 15.7 &   \\  
 7759.37 & 5.93 & --1.76 & 2.5 & 1.0+8 & 17.8 &   & &               5480.51 & 3.50 & --0.90 & 2.5 & 7.4+7 & 11.1 &    \\   
 7930.82 & 5.94 & --1.04 & 2.5 & 1.0+8 & 56.6 &   & &               5574.39 & 4.45 & --0.70 & 2.5 & 6.7+7 &  2.5 &    \\    
\multicolumn{7}{l}{{\bf Al\,{\sc i}}; $\log \epsilon_\odot =6.47$}&&   6630.03 & 1.03 & --3.56 & 2.5 & 2.4+7 &  7.1 &   \\   
6696.03 & 3.14 & --1.58 & 2.5 & 1.0+8 & 38.1 &   & &                6636.33 & 4.14 & --1.20 & 2.5 & 4.4+7 &  1.6 &   \\     
 6698.67 & 3.14 & --1.89 & 2.5 & 1.0+8 & 22.4 &   & &               6643.00 & 3.84 & --1.20 & 2.5 & 7.4+7 &  3.1 &    \\   
\multicolumn{7}{l} {{\bf Si\,{\sc i}}; $\log \epsilon_\odot=7.55$}&&   6796.49 & 4.40 & --0.29 & 2.5 & 1.6+8 &  7.3 &   \\   
6518.74 & 5.95 & --1.40 & 2.5 & 1.0+8 & 26.6 &   & &                \multicolumn{7}{l}{{\bf Cr\,{\sc ii}}; $\log \epsilon_\odot=5.67$}\\
 7415.96 & 5.61 & --0.80 & 2.5 & 1.0+8 & 93.5 &   & &              5305.86 & 3.83 & --2.09 & 2.5 & 2.6+8 & 24.7 &   \\                 
 7760.64 & 6.20 & --1.35 & 2.5 & 1.0+8 & 19.4 &   & &              5308.42 & 4.07 & --1.82 & 2.5 & 2.6+8 & 26.9 & +  \\               
\multicolumn{7}{l} {{\bf Ca\,{\sc i}}; $\log \epsilon_\odot=6.36$}&&  5310.69 & 4.07 & --2.26 & 2.5 & 2.6+8 & 12.9 &   \\                 
6166.44 & 2.52 & --1.36 & 5.2 & 1.9+7 & 70.4 & K & &	           \multicolumn{7}{l} {{\bf Mn\,{\sc i}}; $\log \epsilon_\odot = 5.39$ }\\
 6455.60 & 2.52 & --1.41 & 2.3 & 4.7+7 & 58.6 &   & &              5388.50 & 3.37 & --1.69 & 2.5 & 7.1+7 &  5.3 &    \\                
 6508.85 & 2.53 & --2.35 & 2.0 & 4.4+7 & 13.7 &   & &              5399.47 & 3.85 & --0.18 & 2.5 & 9.0+7 & 38.1 &    \\                
 6709.87 & 2.93 & --2.76 & 4.5 & 3.8+8 &  2.5 &   & &              5470.64 & 2.16 & --1.38 & 2.5 & 4.0+8 & 58.4 &    \\                
 6798.47 & 2.71 & --2.42 & 3.7 & 1.9+7 &  8.7 &   & &              6440.93 & 3.77 & --1.27 & 2.5 & 7.1+7 &  6.3 &   \\                 
\multicolumn{7}{l}{{\bf Sc\,{\sc i}}; $\log \epsilon_\odot= 3.10$}&&  7764.66 & 5.37 &   0.17 & 2.5 & 9.8+7 &  5.9 &    \\                
5484.64 & 1.85 &   0.04 & 1.5 & 1.5+8 & 2.2 &   & &	           \multicolumn{7}{l} {{\bf Fe\,{\sc i}}; $\log \epsilon_\odot = 7.51$}\\ 
\multicolumn{7}{l}{{\bf Sc\,{\sc ii}}; $\log \epsilon_\odot=3.10$}&&  5308.69 & 4.26 & --2.43 & 2.0 & 2.1+8 &  7.6 &   t\\                
5239.82 & 1.45 & --0.76  & 1.5 & 1.0+8 & 48.5 &   & &              5315.07 & 4.37 & --1.54 & 2.0 & 1.8+8 & 32.5 &   t  \\              
5318.36 & 1.36 & --1.77  & 1.5 & 1.5+8 & 12.4 &   & &              5223.18 & 3.63 & --2.31 & 2.0 & 7.9+7 & 28.5 &   t\\                
 6300.69 & 1.51 & --2.00  & 1.5 & 2.3+8 &  6.2 &   & &             5308.69 & 4.26 & --2.43 & 2.0 & 2.1+8 &  7.6 &   t\\                
 6320.84 & 1.50 & --1.88  & 1.5 & 2.3+8 &  8.2 &   & &             5315.07 & 4.37 & --1.54 & 2.0 & 1.8+8 & 32.5 &   t  \\              
\multicolumn{7}{l}{{\bf Ti\,{\sc i}}; $\log \epsilon_\odot= 4.99$}&&  5320.03 & 3.64 & --2.54 & 2.0 & 3.1+8 & 19.3 &   \\                 
5219.70 & 0.02 & --2.25 & 2.5 & 6.0+6 & 26.7 & K & &               5321.11 & 4.43 & --1.30 & 2.0 & 1.7+8 & 42.0 & K \\                 
 5299.98 & 1.05 & --1.44 & 2.5 & 3.4+6 & 20.4 & K & &              5322.04 & 2.28 & --2.89 & 2.0 & 1.0+8 & 62.2 & K \\                 
 5389.16 & 0.81 & --2.24 & 2.5 & 8.3+7 &  6.7 &    & &             5386.34 & 4.15 & --1.76 & 2.0 & 2.3+8 & 31.6 &   t\\                
 5471.20 & 1.44 & --1.61 & 2.5 & 1.1+8 &  7.0 &   & &              5395.22 & 4.44 & --1.81 & 2.0 & 1.8+8 & 18.7 &   t\\                
 5473.55 & 2.33 & --0.83 & 2.5 & 1.1+8 &  5.8 &    & &             5398.28 & 4.44 & --0.81 & 2.0 & 1.9+8 & 70.5 &   t  \\              
 5474.23 & 1.46 & --1.31 & 2.5 & 8.4+7 & 12.7 &   & &              5473.16 & 4.19 & --2.02 & 2.0 & 2.2+8 & 19.5 &   t\\                
 5474.46 & 2.34 & --0.96 & 2.5 & 1.1+8 &  4.3 &   & &              5483.10 & 4.15 & --1.49 & 2.0 & 2.6+8 & 45.5 &   t   \\             
 5490.15 & 1.46 & --0.98 & 2.5 & 1.5+8 & 21.6 & K & &              5560.22 & 4.43 & --1.16 & 2.0 & 1.6+8 & 49.5 &   t \\               
 6303.76 & 1.44 & --1.60 & 2.5 & 1.7+8 &  7.8 &    & &             5577.02 & 5.03 & --1.51 & 2.0 & 6.9+8 & 11.7 &   t \\               
 6312.24 & 1.46 & --1.60 & 2.5 & 1.7+8 &  7.5 &    &&              6165.36 & 4.14 & --1.55 & 2.0 & 8.8+7 & 43.8 &   t               \\ 
 7440.58 & 2.26 & --0.76 & 2.5 & 1.4+8 &  9.5 &    &&              6303.46 & 4.32 & --2.59 & 2.0 & 1.9+8 & 5.1  & \\
 7949.15 & 1.50 & --1.45 & 2.5 & 2.0+6 & 11.3 &    &&              6380.75 & 4.19 & --1.27 & 2.0 & 7.4+7 & 56.9 & \\
\noalign{\smallskip}
\hline
\end{tabular}
\label{linelist}
\end{table*}

\begin{table*}
{\footnotesize
\begin{tabular}{lrrrrrrlclrrrrrrl}
\hline\noalign{\smallskip}
{ $\lambda$} & { $\chi_l$} & {$ \log gf$} & {$\delta \Gamma_6$} & {  $\Gamma _{\rm rad}$} & { W$_{\lambda \odot}$} &Note &&{ $\lambda$} & { $\chi_l$} & {$ \log gf$} & {$\delta \Gamma_6$} & {  $\Gamma _{\rm rad}$} & { W$_{\lambda \odot}$} & Note \\ 
{[\AA] }      &[eV]       &            &                     &[s$^{-1}$]                & [m\AA]                 &    && {[\AA] }      &[eV]       &            &                     &[s$^{-1}$]                & [m\AA]                 &   \\
\noalign{\smallskip}
\hline\noalign{\smallskip}
 6436.41 & 4.19 & --2.41 & 2.0 & 3.0+7 & 10.6 &    t &&          \multicolumn{7}{l} {{\bf Ni\,{\sc i}}; $\log\epsilon_\odot = 6.25$}\\ 
 6501.67 & 4.83 & --1.25 & 2.0 & 1.0+8 & 27.7 &      &&		      5220.29 & 3.74 & --1.29 & 2.5 & 8.7+7 & 28.3 &   \\              
 6696.32 & 4.83 & --1.50 & 2.0 & 2.4+8 & 18.0 &    t && 	      5388.35 & 1.94 & --3.46 & 2.5 & 1.1+8 & 13.0 &   \\              
 6699.13 & 4.59 & --2.12 & 2.0 & 1.4+8 &  8.3 &    t &&	           5392.33 & 4.15 & --1.30 & 2.5 & 2.0+8 & 14.5 &   \\                 
 6703.57 & 2.76 & --3.01 & 2.0 & 1.0+8 & 38.1 &    t &&	          5468.11 & 3.85 & --1.68 & 2.5 & 1.6+8 & 11.8 & K \\  
 6704.50 & 4.22 & --2.58 & 2.0 & 1.0+8 &  6.6 &    t &&	          6175.37 & 4.09 & --0.59 & 2.5 & 2.3+8 & 49.6 & K \\  
 6710.32 & 1.48 & --4.81 & 2.0 & 1.7+7 & 16.6 &    t &&	          6176.81 & 4.09 & --0.34 & 2.5 & 1.4+8 & 64.2 & K \\  
 6713.04 & 4.61 & --1.29 & 2.0 & 2.4+8 & 36.0 & K    &&	      6316.58 & 4.15 & --1.89 & 2.5 & 2.0+8 &  4.4 &    \\      
 6713.74 & 4.79 & --1.43 & 2.0 & 2.4+8 & 22.3 & K    &&  	      6502.22 & 3.40 & --2.85 & 2.5 & 1.0+8 &  2.6 &   \\   
 6796.12 & 4.14 & --2.27 & 2.0 & 1.4+8 & 14.9 &    t &&	      6635.13 & 4.42 & --0.74 & 2.5 & 1.5+8 & 26.5 &   \\
 6806.85 & 2.73 & --3.10 & 2.0 & 1.0+8 & 35.8 &      &&              6813.61 & 5.34 & --0.41 & 2.5 & 8.2+8 &  9.9 &   \\  
 6810.26 & 4.61 & --1.01 & 2.0 & 2.3+8 & 51.1 &    t &&              7110.90 & 1.94 & --2.89 & 2.5 & 5.2+7 & 38.3 &   \\    
 7107.46 & 4.19 & --1.96 & 2.0 & 4.8+7 & 24.2 &    t &&	          7126.71 & 3.54 & --2.34 & 2.5 & 4.8+7 &  6.4 &   \\  
 7114.57 & 2.69 & --4.02 & 2.0 & 8.0+7 &  7.7 &    t &&	          7414.51 & 1.99 & --2.04 & 2.5 & 1.0+8 & 81.9 &   \\ 
 7120.58 & 4.14 & --3.40 & 2.0 & 8.0+8 &  1.3 &      &&           \multicolumn{7}{l} {{\bf Cu\,{\sc i}}; $\log \epsilon_\odot$ = 4.21 }\\
 7127.57 & 4.99 & --1.07 & 2.0 & 4.9+8 & 30.2 &    t &&           5220.08 & 3.82 & --0.61  & 2.50 & 1.0+8 & 16.4 &   \\  
 7130.92 & 4.22 & --0.66 & 2.0 & 2.1+8 & 97.7 &    t &&	         7933.12  & 3.78 & --0.27  & 2.50 & 1.0+8 & 35.9 &   \\  
 7418.33 & 4.14 & --2.84 & 2.0 & 5.5+7 &  4.7 &    t &&            \multicolumn{7}{l} {{\bf Y\,{\sc i}}; $\log \epsilon_\odot=2.24$}\\ 
 7418.67 & 4.14 & --1.47 & 2.0 & 1.1+8 & 50.7 &    t &&	          6435.04 & 0.07 & --0.98  & 2.50 & 1.0+8 &  1.8 &   \\   
 7421.55 & 4.64 & --1.68 & 2.0 & 2.5+8 & 19.0 &    t &&	          6687.50 & 0.00 & --0.67  & 2.50 & 1.0+8 &  4.3 &   \\  
 7440.91 & 4.91 & --0.62 & 2.0 & 5.0+8 & 60.6 &    t &&	         \multicolumn{7}{l} {{\bf Y\,{\sc ii}}; $\log \epsilon_\odot=2.24$}\\ 
 7751.11 & 4.99 & --0.76 & 2.0 & 6.4+8 & 48.9 &    t &&           5402.78 & 1.84 & --0.64 &  2.5 & 1.0+8 & 11.2 \\   
 7941.09 & 3.27 & --2.47 & 2.0 & 1.4+8 & 43.7 &    t &&	         6795.42 & 1.70 & --1.14 &  2.5 & 1.0+8  &  5.9 &   \\ 
 7955.71 & 5.03 & --1.17 & 2.0 & 6.4+8 & 25.5 &    t &&           5473.39 & 1.74 & --0.83 &  2.5 & 1.0+8 &  9.4 &   \\  
 7959.14 & 5.03 & --1.13 & 2.0 & 6.4+8 & 27.5 &    t &&          \multicolumn{7}{l} {{\bf Zr\,{\sc i}}; $\log \epsilon_\odot=2.60$}\\
 \multicolumn{7}{l}{{\bf Fe\,{\sc ii}}; $\log \epsilon_\odot=7.51$}&&	         5385.12 & 0.52 & --0.97  & 2.5 & 1.0+8 &  1.3 &   \\  
 6416.93 & 3.89 & --2.69  & 2.0 & 3.4+8 & 41.5 & +  &&	         6506.35 & 0.63 & --0.64  & 2.5 & 1.0+8 &  2.5 & \\
 6432.68 & 2.89 & --3.62 & 2.0 & 2.9+8 & 42.3 &  &&          7439.87 & 0.54 & --1.00  & 2.5 & 1.0+8 &  1.5 &    \\                 
 6456.39 & 3.90 & --2.21 & 2.0 & 3.4+8 & 65.0 &  && \multicolumn{7}{l} {{\bf Mo\,{\sc i}}; $\log \epsilon_\odot=1.92$}\\ 
 6516.08 & 2.89 & --3.36 & 2.0 & 2.9+8 & 55.6 &  &&          5570.39 & 1.33 &  0.43  & 2.5 & 1.0+8 &  9.6 &   \\ 
 \multicolumn{7}{l}{{\bf Co\,{\sc i}}; $\log \epsilon_\odot=4.92$}  && \multicolumn{7}{l}{{\bf La\,{\sc ii}}; $\log \epsilon_\odot=1.22$}\\
 5301.04 & 1.71 & --1.89 & 2.5 & 1.2+8 & 21.7 &   &&             6320.42 & 0.17 & --1.39 & 2.5 & 1.0+8 & 5.2 &   \\  
 5312.65 & 4.21 & --0.02 & 2.5 & 2.5+8 &  7.7 &   &&             6390.49 & 0.32 & --1.47 & 2.5 & 1.0+8 & 3.2 &    \\  
 5483.36 & 1.71 & --1.25 & 2.5 & 1.9+7 & 50.6 &   &&             \multicolumn{7}{l}{{\bf Nd\,{\sc ii}}; $\log \epsilon_\odot=1.50$}\\
 6455.00 & 3.63 & --0.24 & 2.5 & 7.4+7 & 16.2 &    &&	            5319.82 & 0.55 &  0.02  & 2.5 & 1.0+8 & 18.8 &   \\ 
 6632.47 & 2.28 & --1.73 & 2.5 & 6.5+6 & 11.8 &   &&            \multicolumn{7}{l}{{\bf Eu\,{\sc ii}}; $\log \epsilon_\odot=0.51$}\\
 6814.96 & 1.96 & --1.76 & 2.5 & 2.1+7 & 20.6 &   &&  6645.12 & 1.38 &  0.28  & 2.5 & 1.0+8 &  5.7 &   \\         
 7417.39 & 2.04 & --2.00 & 2.5 & 2.2+7 & 11.9 &    &&	          \multicolumn{7}{l}{{\bf Hf\,{\sc ii}}; $\log \epsilon_\odot=0.88$}\\
 7437.07 & 5.98 &   1.13 & 2.5 & 7.0+7 &  2.6 &   &&              5311.63 & 1.78 &  0.13  & 2.5 & 1.0+8 &  4.4 &       \\
\noalign{\smallskip}\hline \\
\end{tabular}}
\end{table*}

On the basis of these comparisons we found no reason to change the
surface gravities to be used in the abundance analysis but kept those
determined from photometry.

\subsection{Microturbulence parameters} 
\label{sec:mic}

We have determined microturbulence parameters, $\xi_t$, from both the
   Ca\,{\sc i} lines and the Fe\,{\sc i}  lines for 12 of the stars
   with enough calcium and iron lines measured. For the determination from the
   Ca\,{\sc i} lines, abundances from the individual lines were
   derived with ${\xi}_t$  ranging from 0.10 to 1.90 km/s. The
   microturbulence parameter for each star was then determined as the
   $\xi_t$ value which gave the smallest abundance scatter  (the
   inflexion point, see e.g. Smith 1981). The microturbulence
   parameters  were also determined by plotting  the iron abundance
   versus the reduced equivalent widths, $\log  (W_{\lambda}/
   \lambda)$, derived for each line. If the correct microturbulence
   parameter is used the slope of a fit to the data points should be
   zero. Both methods agreed well for these 12 stars.

Using the values obtained we derived a relation between effective
temperature,  surface gravity and the microturbulence parameter, which
was then used to determine microturbulence parameters for the rest of
the stars: $ \xi _t = 4.5\cdot 10^{ -4} T_{\rm eff} - 0.31\cdot \log
g$ and predicts, with $T_{\rm eff}$ and $\log g$ adopted, $\xi_t$
values to an accuracy of $\pm 0.1$ km/s for the 12 stars.  This
relation is valid for $ 3.93 < \log g < 4.50$ and $ 5530~{\rm K} < T_{\rm eff}
< 6036~{\rm K}$. For a few stars just outside the validity range we
extrapolated the relation to calculate approximative microturbulence
parameters. This procedure was checked to yield consistent abundances
for individual iron lines of different strengths. For five stars a
somewhat lower microturbulence parameter was preferred (HD36130,
HD134474, HD171999, HD182572, HD187055). Our relation for
microturbulence parameters yield slightly higher values than the
relation presented by Edvardsson et al. (1993a).

For the K dwarf stars none of the described methods seemed to yield
definite values for the microturbulence parameter. A microturbulence
parameter of 1.0 km/s was adopted for these stars.

\subsection{Atomic line data}

The oscillator strengths were determined  by requiring that the
abundances calculated for the solar model ($T_{\rm eff}$=5780K, $\log
g = 4.44$, [Me/H]=0.00, $\xi _t$=1.00) should reproduce the observed
equivalent widths of the solar spectrum. The resulting $\log gf$
values are given in Table \ref{linelist}.

Different line broadening mechanisms, van der Waals damping, radiation
damping, thermal Doppler broadening and microturbulence were
considered in the calculations of equivalent widths and abundances.
Enhancement factors for the van der Waals damping were compiled from
the literature. For iron lines values from Hannaford et al. (1992) and
Holweger et al. (1991)  were used, for calcium values from Smith (1981)
and references therein, for sodium values from Holweger (1971) and
for scandium values from Neuforge (1992).  For the remaining lines a
correction factor of 2.5 was adopted to the classical Uns\"old value,
according to M\"ackle et al. (1975).  For the (unimportant)
radiation damping parameter
values  from Kurucz (1989) were adopted for lines from calcium
through nickel.

\section{Errors in resulting abundances}

\begin{table*}
\caption[]{{Effects on  abundance estimates for a number of elements
from changes in fundamental parameters  of the model atmosphere. The
resulting abundances are shown for 3 stars: HD72946 (T$_{\rm
eff}$/$\log{\it g}$/[Fe/H])=(5911/4.4/0.24), HD103932
(4510/4.58/0.16), and HD110010 (5965/4.08/0.35).  The first line for
each star gives [X/H], X being the element indicated in the header, derived using a model
atmosphere constructed with the stellar parameters adopted in this
study. The following lines contain the differences when effective
temperature or surface gravity has been changed as indicated in the
first column. Note that the changes in parameters are different, and
larger, for HD103932.}}
\begin{tabular}{lrrrrrrrrrrrrrrrr}
\hline\noalign{\smallskip}
 \multicolumn{2}{l}{ID / $\Delta$ } & [O\,{\sc i}] &Na\,{\sc i} & Mg\,{\sc i} & Al\,{\sc i} & S\,{\sc i} &  Ca\,{\sc i} & Ti\,{\sc i} & Cr \,{\sc i} & Cr\,{\sc ii}&  Fe\,{\sc i} & Fe\,{\sc ii} &   Ni\,{\sc i} &   Eu\,{\sc ii}\\
\noalign{\smallskip}
\hline\noalign{\smallskip}
 \multicolumn{2}{l}{{\bf HD72946}}&  0.16  &  0.22  &  0.24  &  0.23  &  0.13  &  0.22  & 0.38   &  0.24 &  0.09  &  0.24 & 0.00  &  0.18    &  0.09\\
 \multicolumn{2}{l}{+100 K}       & +0.01  & +0.04  & +0.04  & +0.06  & +0.01  & +0.06  & +0.08  & +0.06 &--0.03  & +0.05 & +0.01 & +0.05    & 0.00  \\
  \multicolumn{2}{l}{+0.2 dex}    & +0.09  & --0.03 & --0.01 & --0.01 & --0.01 & --0.02 & 0.00   & 0.00  & +0.08  &--0.01 & +0.12 & +0.01    & +0.08  \\
\noalign{\smallskip}
\hline\noalign{\smallskip}
 \multicolumn{2}{l}{{\bf HD103932}}& 0.28 & &0.30   &0.14  &0.27   &--0.12 &  +0.03 &   0.01 & 0.44     & 0.14  &  0.57 &0.41 &  0.32  \\
  \multicolumn{2}{l}{+200 K}      & +0.05 & &--0.08 &+0.12 &--0.18 &+0.18  &  +0.24 &  +0.12 & --0.18   & --0.04& --0.32&--0.05   &--0.01 \\
  \multicolumn{2}{l}{+0.4 dex}    & +0.16 & &+0.02  &--0.04 &+0.10 &--0.04 & --0.02 &   +0.02 &  +0.19 &  +0.07 & +0.21&+0.09 & +0.15  \\
\noalign{\smallskip}
\hline\noalign{\smallskip}
 \multicolumn{2}{l}{{\bf HD110010}}& --0.04 & 0.53   &  0.48  &  0.40  &  0.51  & 0.35   & 0.29   &   0.30 &  0.22  &  0.35 & 0.33  &  0.31   &  0.16  \\
  \multicolumn{2}{l}{+100 K}       &  +0.01 & +0.05  &  +0.04 &  +0.04 &  +0.01 &  +0.06 & +0.08  &  +0.06 & +0.03  & +0.06 &--0.05 & +0.05   &  0.00   \\
  \multicolumn{2}{l}{+0.2 dex}     &  +0.09 & --0.02 & --0.01 & --0.02 & --0.01 & --0.01 & 0.00   &  0.00  & +0.08  &--0.01 &--0.03 & 0.00    & +0.09  \\
\hline\noalign{\smallskip}
\end{tabular}
\label{tabellpartest}
\end{table*}

\subsection{Measurement of equivalent widths and fitting of
continua}
\label{sec:5.1}

If we assume that the true continuum  level is not wildly different
from the fitted continuum the error in measured equivalent width can
at the most be as large, in difficult cases, as 2 m{\AA} for weak
lines, i.e.$\sim$ 20\%, and 4-5 m{\AA} for stronger lines, i.e.$\sim$
10\%. This translates to typically 0.08 dex in the resulting abundance
derived from a line of strength 10 m{\AA} and 0.04 dex for a line of
strength 50 m{\AA}. Apart from the effects of the 
continuum errors and blends, the error
in derived abundances due to errors in the measurement of the
equivalent width of a line is negligible ($<$ 0.01 dex). In general, the
lines and continua are, due to the high S/N and high resolution, easy to
fit and the errors given above should be regarded as upper limits.

\subsection{Oscillator strengths}

The oscillator strengths, derived from the observed solar spectrum,
can be affected by misidentification, by undetected blends and by
errors in continuum fitting and measurements of equivalent widths in
the solar spectrum. As in the stellar spectra, location of the continuum is
a  much larger source of error than the actual measurement of a
line. (Note, however, that the solar spectra have higher S/N, usually
$\sim 400$, and thus, identification of the continuum becomes easier
as well as identification of lines. The Sun is also more metal-poor
than the programme stars, which makes identification of the continuum
in the Sun easier.) Using the results in Sect. \ref{sec:5.1} we find
that errors in $\log gf$-values may be as large as 0.08 dex, but
a more characteristic number is 0.04 dex.

Since we perform a purely differential analysis errors due to misplaced
continua, neglected blends, etc. 
should partly cancel in the first approximation and not affect the
resulting differential abundances very much, as long as we study stars
similar to the Sun.

\subsection{Blends}

In the selection of lines we have carefully avoided all lines that
could be subject to blending with nearby lines as given in Moore et
al. (1966). For ions with several lines measured we have also looked
for lines which produces abnormally high abundances as compared with
the majority of the lines. This led us to exclude three
Fe\,{\sc ii} lines from our final analyses: 6383.71, 6383.45, and
6627.32 {\AA}.

\subsection{Fundamental parameters of the model atmospheres}

Edvardsson et al. (1993a) estimate the error in the effective  temperature
due to errors in $b - y$ to range from {--50 K to +100 K} and the
corresponding error in $\log g$ to be $\pm$0.2 dex. 

The effects of such errors in $T_{\rm eff}$ and $\log g$ are exemplified
in Table \ref{tabellpartest}. As expected, abundances derived from
ions are most sensitive to changes in surface gravity while abundances
derived from  atoms are most sensitive to changes in effective
temperature.  In general, errors in derived abundances are  smaller
than 0.10 dex for atoms when varying the effective temperature by $\pm
100$ K and less than 0.02 dex when the surface gravity is varied by
$\pm 0.20$ dex; they are smaller than 0.02 dex for ions when the
effective temperature is varied and less than 0.10 dex when the
surface gravity is varied.

\subsection{Photometric versus spectroscopic metallicities}
\label{fem}                                                                                
                                                            
\begin{figure}
\resizebox{\hsize}{!}{\includegraphics{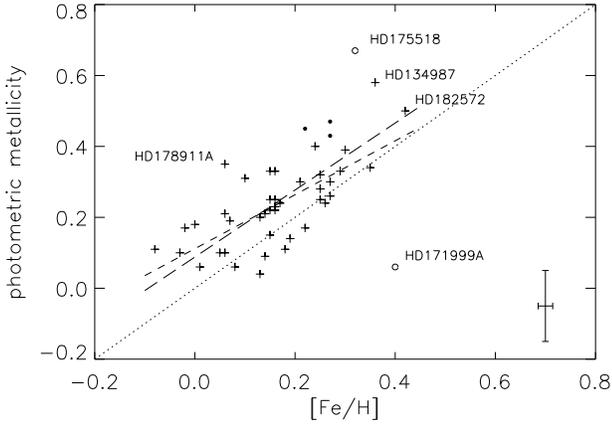}}
\caption[]{Metallicities determined from photometry vs iron abundances
 derived in our spectral  analysis. The one-to-one relation is
 indicated by a dotted line and a least-square fit to the data points,
 taking the error in [Fe/H] to be $\pm$0.02 and the error in the
 photometric metallicities to be $\pm$0.1, is also shown, dashed
 line. The $\chi^2$-proabability
 for this fit is 0.92. A fit made without taking the errors into account
is shown by a long-dashed line. The stars from Barbuy \& Grenon (1990),
 $\bullet$, and HD171999A and HD175518, denoted by $\circ$ symbols,
 were excluded from the fit (see Sect. \ref{fem}). }
\label{fephot}
\end{figure}

For dwarf stars that are significantly more metal-rich than the Sun
(e.g. [Me/H] $\approx$ 0.2 dex),  the
metallicity used in the calculation of the model atmosphere is
important, since it governs the line blanketing and thus the
temperature structure of the model atmosphere. We may expect that
final derived abundances are sensitive to this parameter. Following
Edvardsson et al. (1993a) we decreased (and in a few cases increased) the
metallicities to the values derived for [Fe/H] in a preliminary
abundance analysis and reiterated the abundance determination.  (We
have determined iron abundances from lines arising from both Fe\,{\sc i} and
Fe\,{\sc ii}. For Fe\,{\sc i} usually more than 30 lines were analysed and
for Fe\,{\sc ii} three to four lines. The formal error in the mean [Fe/H]
derived from Fe\,{\sc i} for a certain star is usually smaller than 0.02 dex.)
In the mean we find that we had to reduce the abundances by 0.07  dex
from the initial photometric values, with a spread of 0.01 dex, Table
\ref{parameters}.  Apparently, our photometric metallicities tend to
overestimate the metal content in these metal-rich stars as compared
to the iron abundance derived from spectral abundance analysis,
Fig.\ref{fephot}.

Usually, the differences between the metallicities estimated from
photometry, are close enough to those derived from spectroscopy that
we have not found it necessary to reiterate the  determination of
effective temperature and $\log g$. (The determination of metallicity
was, however, changed according to the spectroscopic result so that
the final model used in the analysis had [Me/H] consistent with the
resulting spectroscopic [Fe/H].)  However, for some of the stars the
photometry indicates rather extreme metallicities compared with the
spectroscopic iron abundances.  For HD171999A we have only measured 6
Fe\,{\sc i} equivalent widths (since this star was only observed with
one CCD setting, see Sect. \ref{sect:observations}) and thus the
spectroscopically determined iron abundance is not as good as for the
other stars. However, we note that the line-to-line scatter is small,
0.03 dex.  For HD175518 it is questionable if its photometric
metallicity is realistic. In any large catalogue there will always be
a few stars with $2-3 \sigma$ errors in the photometry. Since
spectroscopic iron abundances indicate a much lower metallicity this
is most probably an example of that.

We have  studied, for HD175518, the effects on derived abundances if
[Me/H] is lowered to 0.2 dex as indicated by the spectroscopy, thus
affecting the estimates of the rest of the fundamental parameters.
The lower metallicity implies a lower effective temperature; [Me/H] =
0.22 dex corresponds to a decrease of $T_{\rm eff}$ by $\sim$200 K.
From Table \ref{tabellpartest} we see that most elements will change
by $\sim$ 0.1 dex and thus the star will mainly move horizontally, by
$\sim$ 0.1 dex, in our relative abundance diagrams. HD175518 is an
extreme case in our sample and the abundances of the other stars and
general abundance trends for the whole sample should not be affected
by comparable amounts.

\subsection{Effects of hyperfine structure}

\begin{figure}
\resizebox{\hsize}{!}{\includegraphics{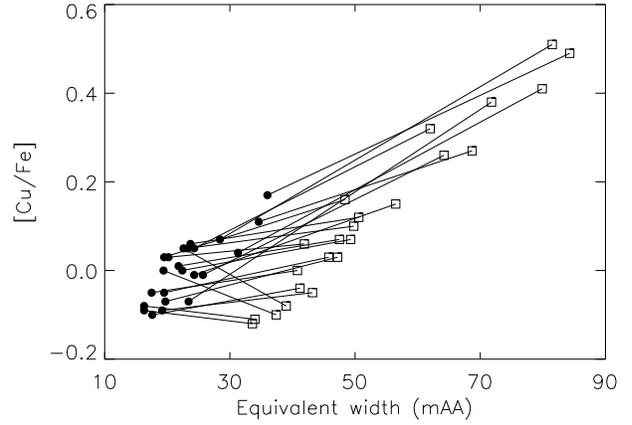}}
\caption{Copper abundances derived from the lines at 
5220 {\AA}, $\bullet$ symbols, and 7933 {\AA}, $\Box$ symbols.
 The widths of these 
lines are 16 and 36 m{\AA}, respectively, 
in the solar flux spectrum. }
\label{cufe}
\end{figure}

Some of the elements analysed are well known to be subject to
hyperfine structure. We have not taken this into account when deriving
the chemical abundances. Instead,  for those elements in particular we
have aimed at selecting weak enough lines, so that the neglect of
hyperfine structure in the abundance analysis should not affect the
calculated equivalent widths and thus not the  derived abundances. 

\paragraph{Copper.}

 Figure \ref{cufe} shows copper abundances derived from the two lines
used in our study as functions of equivalent width.  Our data clearly
show that the stronger copper line is
 subject to hyperfine structure and that these lines
should be analysed using synthetic spectroscopy taking the hyperfine
structure splitting into account. Results by Summers  (1994)  may
suggest departures from LTE  in the population of levels in the copper
atom. We do not, however, have enough data to make further empirical
investigations of such departures from LTE.   We omit copper
abundances from the following discussion.

\paragraph{Manganese and Cobalt.}

The manganese and cobalt lines used in this study are not
saturated. When plotting abundances derived from each line as a
function of equivalent width no distinct pattern was found; indicating
that the omission of hyperfine structure in the treatment of the lines
is not problematic.  In many stars lines with equivalent widths of
about 20 and 60 m{\AA}, respectively,  gave manganese abundances that
are in excellent mutual agreement. 

\paragraph{r- and s-process elements.} 

The lines used to derive abundances for the heavy s- and r-process
elements  are sufficiently weak in our programme stars to be safely used
as abundance criteria in spite of being subject to hyperfine
structure. Among the light s-process elements we note, however, that only Y\,{\sc
ii}, and maybe Y\,{\sc i}, have lines strong enough and secure enough
that derived abundances can be used with confidence. The $\log gf$-values
for zirconium are very uncertain because of the faintness of the lines
in the Sun. 

\begin{figure}
\resizebox{\hsize}{!}{\includegraphics{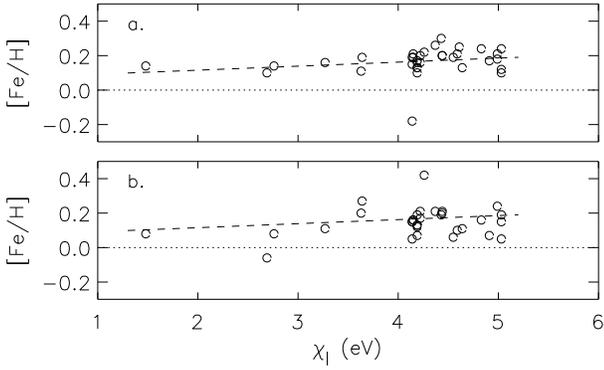}}
\caption[]{Two examples of how the slope coefficients in
Fig. \ref{teffkoeff} were obtained. Examples are for {\bf a.} HD91204
and {\bf b.} HD125968.  The dashed lines represent linear least-square
linear fits. }
\label{chifits}
\end{figure}

\begin{figure}
\resizebox{\hsize}{!}{\includegraphics{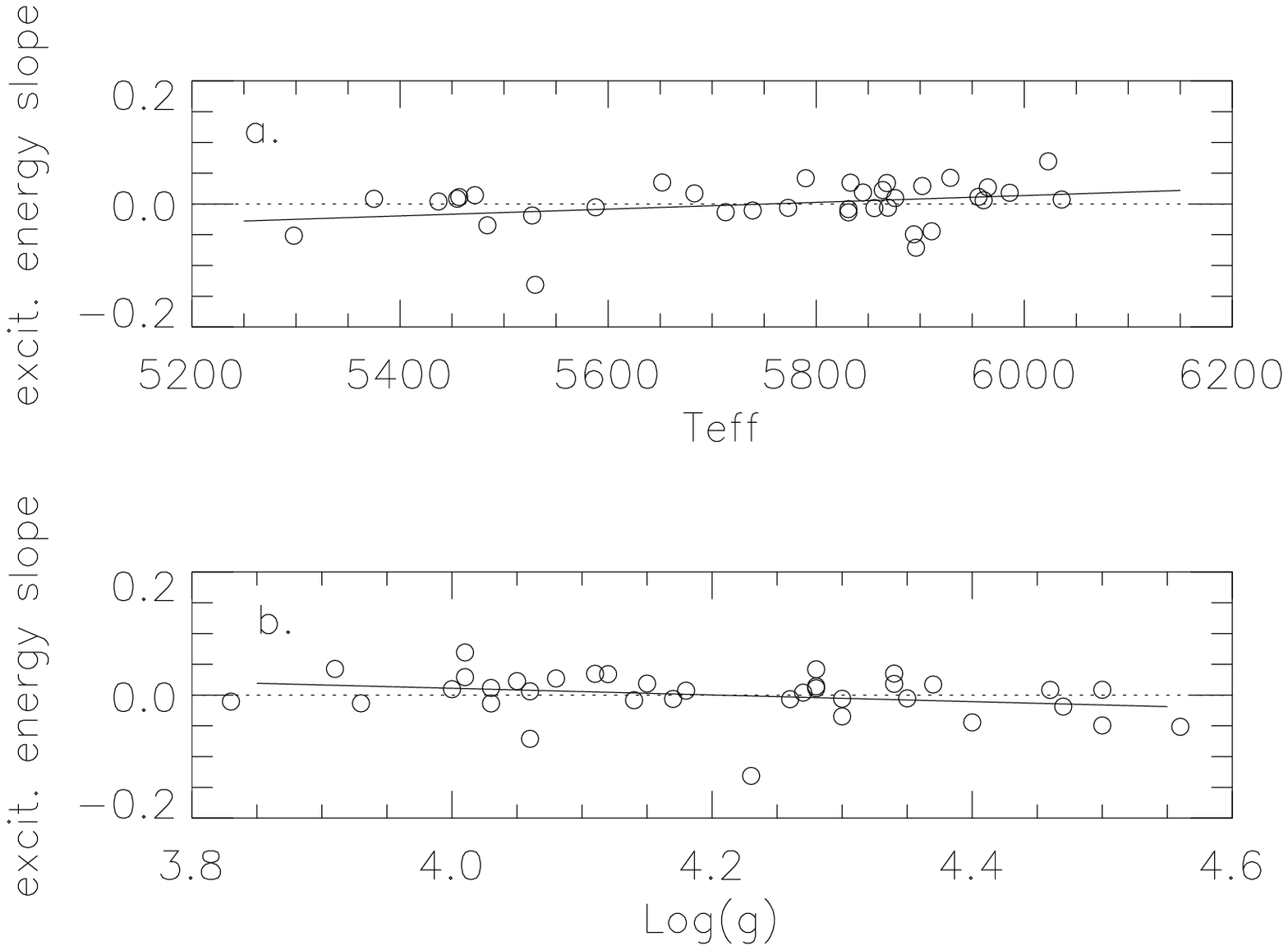}}
\resizebox{\hsize}{!}{\includegraphics{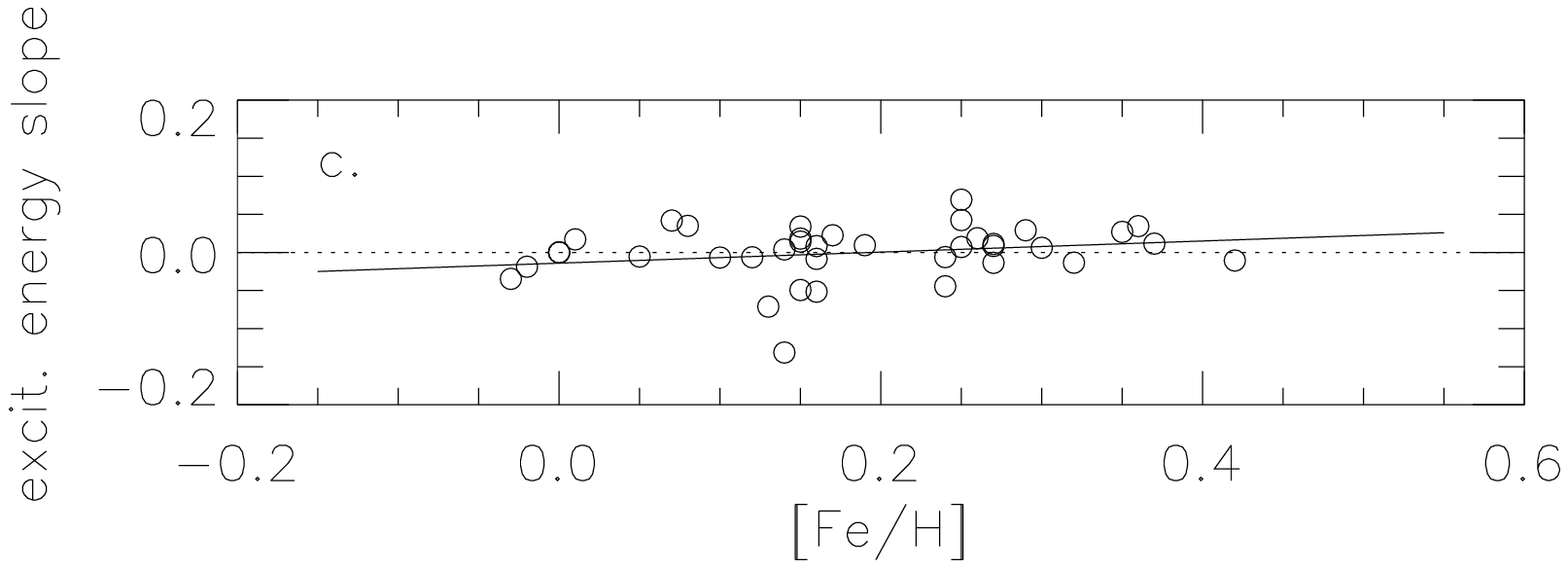}}
\caption[]{The slope coefficients from the
excitation energy balance diagram for each star plotted vs effective
temperature, surface gravity and spectroscopically derived iron
abundances. The star with the lowest $k$ is
HD180890. Linear least square fits yield:\\ {\bf a}. $k =
-0.31 + 5.5 \cdot 10^{-5}\cdot T_{\rm eff}$ \\ {\bf b}.  $k
= 0.23 - 0.054\cdot \log g$ \\ {\bf c}. $k = -0.014 + 0.073
\cdot$[Fe/H].} 
\label{teffkoeff}
\end{figure}

\begin{figure}
\resizebox{\hsize}{!}{\includegraphics{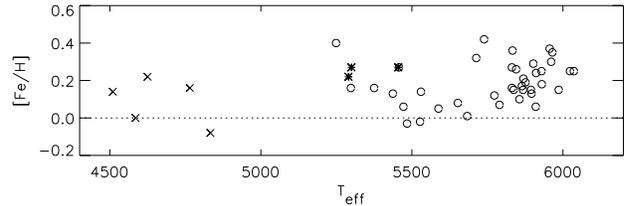}}
\caption[]{Iron abundances, derived in this study, as a 
function of effective temperatures.} 
\label{feteff}
\end{figure}

\subsection{Checks on Boltzmann and Saha equilibria}
\label{sec:exbal}

Deviations from the Boltzmann excitation equilibrium, which might
suggest an error in the  effective temperature, can be traced by
studying abundances derived from individual lines as a function of the
lower excitation potential for the lines.   32 Fe\,{\sc i} lines
measured in most of the stars, marked in Table \ref{linelist}, were
selected for this purpose. For all stars (excluding four stars with
too few lines observed)  the abundance of each iron line was plotted
as a function of the excitation energy of the lower level,
$\chi_{_{l}}$, and a least square linear fit was made to the points of
the form [Fe/H]= $a + k\cdot\chi _{_l}$, Fig. \ref{chifits}. The mean
value of the slope coefficients, $k$, is 0.00 (excluding the K dwarf
stars), suggesting that the systematic errors in effective
temperatures  are small. 

Next, these linear coefficients were plotted as functions of effective
temperature, surface gravity and spectroscopically derived iron
abundance, Fig. \ref{teffkoeff}. We have compared our results with
those obtained by Edvardsson et al. (1993a) (their Fig. 9a-f) and find that
the two studies span the same range of excitation energy slopes,
$k$. There are small but systematic deviations from the excitation
equilibrium,  varying with effective temperature. A change in $T_{\rm
eff}$ of +100 K results in a $\Delta k$ of +0.006 dex eV$^{-1}$. The
change of excitation energy slope with surface gravity estimates seems to
reflect the variation of surface
 gravity with effective temperature (surface gravity
increases as effective temperature decreases). 

In many studies surface gravities are determined by requiring
ionization equilibrium. This is typically made by changing the surface
gravity until the iron abundances derived from Fe\,{\sc i}  and
Fe\,{\sc ii} lines yield the same abundance. We have not, as discussed
earlier, used this method. As we will see this has led us to discover
what appears to be a case of significant overionization in K dwarf
stars and an opposite smaller effect for the hotter stars (see
Fig. \ref{overion}).

What would the effects be if we assumed ionization equilibrium, and
adjusted the surface gravities accordingly?  We can estimate changes
in the stellar abundances from the results of Table
\ref{tabellpartest} and Table \ref{abundances}. From this we find that
half of the stars should have their surface gravities increased by
$0.25 - 0.35$ dex to achieve ionization equilibrium for iron. This
means that the the iron abundance will change with $\approx
-0.03$. Abundances of other elements will change with similar amounts
but with differing signs, which means that for some elements [X/Fe]
will change by up to 0.1 dex and for others not at all. However, we
note that the oxygen abundances are very sensitive to the surface
gravity and may change by up to 0.2 dex.  As a comparison we estimate
the maximum error in the derived oxygen abundance caused by
incorrectly set continua to be less than 0.1 dex.

An adjustment downwards of the gravities by about 0.3 dex would increase the
conflict with the gravity estimates from the Ca\,{\sc i} 6162 {\AA} line
wings. We consider such a revision less probable.

\subsection{K dwarf stars -- Overionization} 
\label{kdwarfs}

Our results admit  a comparison for five elements (scandium, vanadium,
chromium, iron and yttrium) of abundances derived from ions to
abundances derived from the corresponding atom, as function of
effective temperature, within a rather wide range of effective
temperature.

\begin{figure}
\resizebox{\hsize}{!}{\includegraphics{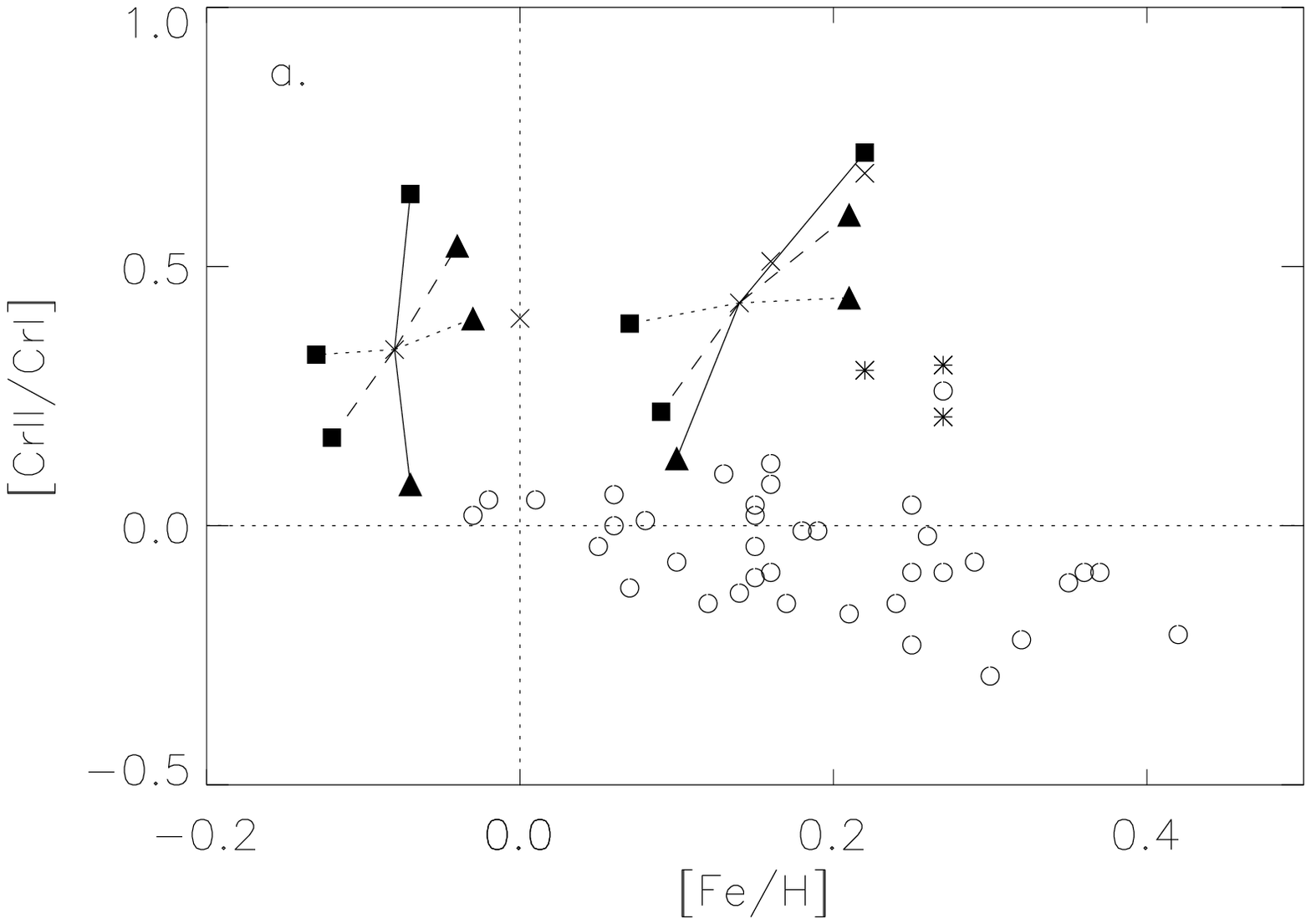}}
\resizebox{\hsize}{!}{\includegraphics{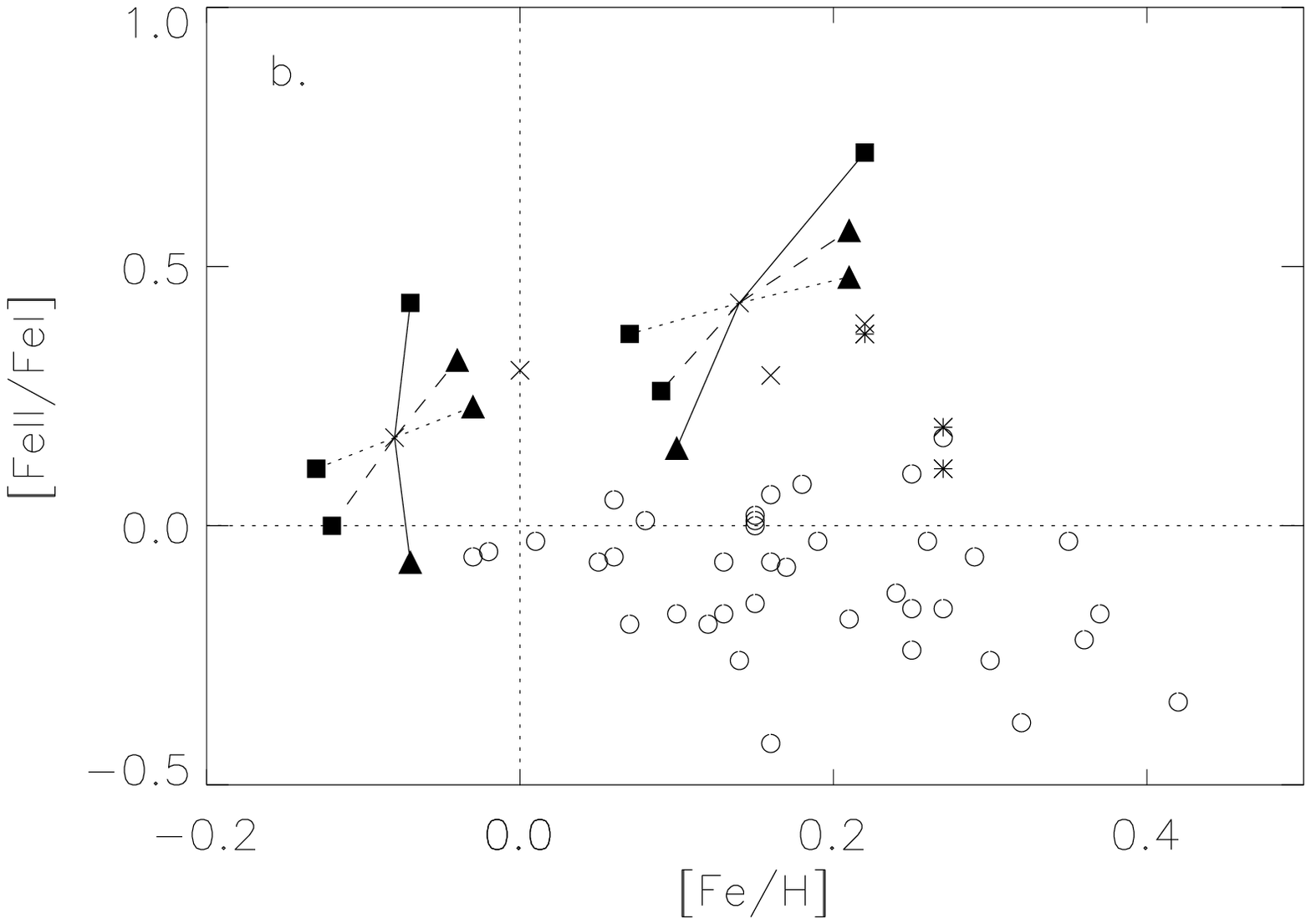}}
\caption[]{Abundance ratios of  chromium and iron  as functions of
[Fe/H].  [Cr\,{\sc ii}/Cr\,{\sc i}], denotes [Cr/H]$_{\rm LTE}$ as
determined from Cr\,{\sc ii} lines minus [Cr/H]$_{\rm LTE}$ as
determined from Cr\,{\sc i} lines, and similarly for iron. Results for
chromium are shown in  panel {\bf a.} and for iron  in panel {\bf
b.}. $\times$ symbols denote the K dwarf stars and $\ast$ symbols
denote the stars from Barbuy \& Grenon (1990). We exemplify, with
HD61606A (${\rm [Fe/H]} = -0.08$)and HD103932 (${\rm [Fe/H]} = 0.16$),
how the resulting loci of the K dwarf stars will be shifted in the
diagram if the parameters of the stellar model atmospheres are
changed;  changes of $T_{\rm eff} \pm 200$K (solid lines), $\log g \pm
0.4$ dex (dashed lines) and [Fe/H] $\pm 0.2$ dex (dotted lines) are
shown. Filled triangles denote increased values of the respective
parameters while filled squares denote decreased values. }
\label{overion}
\end{figure}

We  find an apparent overionization as compared to expectations from
 LTE calculations for the five K dwarf stars in our study,
 Fig. \ref{overion}.  Our results are at first sight unexpected,
 especially for the two K dwarf stars with iron abundances of $\sim$
 0.3 dex relative to the solar iron abundance, as derived from
 Fe\,{\sc i} lines. In stars more metal-rich than the Sun the
 photoionizing non-local UV-flux is more efficiently blocked than in
 more metal-poor stars. However, as discussed in Sect. \ref{nlte}, the
 stronger blocking may be more than compensated for by the increased
 temperature gradient which may enhance the non-locality of the
 radiation field.

\begin{table*}
\caption[]{The table shows results of different
derivations of the effective temperatures for the five
 K dwarf stars; in the second column
 as derived  from the calibration by
Olsen (1984), and which are used in this study; in the third 
spectral classes;  in the fourth and fifth  differences between
effective temperature derived in this work and from MK class
calibrations by Bell \& Gustafsson (1989) and Johnson (1966); in the sixth and
seventh $B-V$ and effective temperatures derived from $B-V$ using  the
calibration of Johnson (1966). The three last columns
contain effective temperatures derived by other authors,
Neff et al. (1995), Morell (1994) and Arribas \& Martinez Roger (1989). }
\begin{tabular}{lrrrrrrrrr}
\hline\noalign{\smallskip}
 & $uvby- \beta$   &Spectral & Bell \&  & Johnson& \multicolumn{2}{c}{Johnson} 
&  Neff & Morell & \multicolumn{1}{c}{Arribas \&} \\ 
  && class    & Gustafsson  & & &  & et al. &  & Martinez Roger  \\   
\noalign{\smallskip}
\hline\noalign{\smallskip}
    ID    &    $T_{\rm eff}$& & $\Delta T_{\rm eff}$ & $\Delta  T_{\rm eff}$
& $B-V$&$T_{\rm eff}$&$T_{\rm eff}$&$T_{\rm eff}$&\multicolumn{1}{c}{$T_{\rm eff}$}\\
\noalign{\smallskip}
\hline\noalign{\smallskip}
HD32147   & 4625 & K3V& +200  & +100  & 1.06 & 4619  & 4570 &&\multicolumn{1}{c}{4670$\pm$150}\\
HD61606A  & 4833 & K2V& +300  & +100  & 0.96 & 4863  &      &       &  \\
HD103932  & 4510 & K5V& --300 & --100 & 1.12 & 4473  &      &       &  \\
HD131977A & 4585 & K4V& --100 & 0     & 1.10 & 4522  & 4575 & 4570  &     \\
HD136834  & 4765 & K3V& +100  & 0     & 1.00 & 4765  &      &        &     \\
\noalign{\smallskip}
\hline
\end{tabular}
\label{UVBteff}
\end{table*}

We have carefully inspected the Cr\,{\sc ii} and Fe\,{\sc ii} lines
in the K dwarf spectra and excluded all lines which may be
subject to severe blends in these cool stars. For Fe\,{\sc ii} we used
the lines at 6456.39 and 6516.39 {\AA}   and for Cr\,{\sc ii} the
lines at 5305.86 and 5310.69 {\AA}.  In spite of using blend
free lines  the apparent overionization remains.  The internal
consistency between abundances derived from the two Fe\,{\sc ii} lines is
very good and this is also the case for Cr\,{\sc ii}.

For scandium, vanadium and
yttrium blends remain a possible source of error, but the
similarity with the trends for iron and chromium  suggests a
common cause of the apparent overionization for all these elements. 

The iron abundances derived from the atom show no obvious trend with
effective temperature, see Fig. \ref{feteff}.

A probable reason for these effects is overionization (see Sect. 
\ref{nlte}). Before discussing this, however, we shall 
explore the possibility that errors in
the temperature scale could also contribute significantly.

\subsubsection{Errors in effective temperatures for K dwarf stars}

 The calibration of the photometry in Olsen (1984) is, for $b-y >
0^{\rm m}.510$, based on a sample of 15 K and M dwarf stars using
stellar parameters from Cayrel de Strobel \& Bentolila (1983). The K
and M dwarf stars in Olsen's sample span a small range in $\delta m_1$
and $\delta c_1$. This is reflected in the change in the calibration
at $b-y = 0^{\rm m}.510$. For $b-y > 0^{\rm m}.510 $ the calibration
is degenerate in metallicity.  Olsen (1984) quotes an error of
$\sim$100K for the effective temperature as derived from $b-y$.

Additional photometry is scarce for our K dwarf stars. However, $UBV$
photometry exists and we have checked our effective temperatures using
the calibration of $B-V$ by Johnson (1966), Table
\ref{UVBteff}. These effective temperatures agree well with those
obtained from $uvby-\beta$ photometry.  We note, however, that the
increased blocking for metal-rich stars, as compared to the
calibration stars used by Johnson (1966), may cause the present effective
temperatures to be underestimated. We have also derived (crude)
effective temperatures from calibrations of the MK classification.
Comparing the effective temperatures derived from photometry with
calibrations of effective temperatures as functions of spectral
classification (Bell \& Gustafsson 1989 and Johnson 1966) we  estimate
$T_{\rm eff}$-values  that deviate as compared with our standard
values as indicated in Table \ref{UVBteff}, columns 4 and 5.

We have searched the literature for independent derivations of the
effective temperature for these stars. Those  found  agree well with
the photometric results, Table \ref{UVBteff}. As described earlier
we have also used the excitation energy balance to check our effective
temperatures. A change in effective temperature of +200K for HD32147
changed the slope of the least-square fit to the data points in the
diagram abundance-versus-excitation energy from +0.04 to +0.00.  A
change of --200K brought about a similar change but in the opposite
direction. Thus, the  excitation equilibrium indicates that +200K is
an acceptable change of the effective temperature for this star. The
line-to-line scatter in derived abundances  from Fe\,{\sc i} lines for this
star is among the largest, $\pm$0.13. The same changes in effective
temperatures give similar values for HD61606A. Since the other three K
dwarf stars are in the same effective temperature range and have
similar surface gravities as the two stars discussed here, and since
HD32147 is the most metal-rich and HD61606A is the most metal-poor of
the five K dwarf stars, changes in the fundamental parameters of the
remaining three stars will produce similar changes in abundances. 

To conclude,  a change in effective temperature of +200 to +400K may
be  allowed as judged from  the excitation equilibrium. As is obvious
from Fig. \ref{overion}, a change of this order of magnitude would
restore the LTE ionization balance for iron and chromium.  The large
line-to-line scatter in derived abundances for lines with high
excitation energies makes attempts to derive effective temperatures
from excitation equilibria very dependent on one or two points in the
lower end of the excitation energy range, spanned by the lines as
illustrated in Fig. \ref{chifits}. Effective temperatures derived in
other studies  and from $B-V$ colours  deviate by less than this from
our values. 

As already noted the surface gravities for the K dwarf stars seem
rather well determined, see Table \ref{parameters}. We also note that
our $\log g$ value for HD131977A agrees well with that given by
Morell (1994). We conclude that realistic errors in $\log g$ are
not enough to account for the departure from ionization equilibrium.

To conclude, we cannot from our analysis exclude that the
apparent pattern of overionization, at least partially, is  due to
a temperature scale that is several hundred K too low.  However, our
analysis, together with evidence from other studies, suggest that
deviations from LTE is a more plausible  cause for the effects.

\subsection{Non-LTE}
\label{nlte}

No detailed study has been devoted to the non-LTE effects on abundance
determinations for metal-rich dwarf stars, cooler than the Sun.  A general
result of the available studies for solar-type stars is,  however,
that  several different effects are at play and may counteract each
other, and this makes all extrapolation to the present study of
metal-rich dwarf stars  from studies of other types of stars or studies of
other elements questionable.

Among the significant effects are (cf. Bruls et al. 1992) resonance
line  scattering, photon suction, ultraviolet overionization,
(infra)red  over-recombination and optical line pumping. The
resonance-line scattering effects, in which  photon losses cause the
source functions of resonance lines to drop far below  the Planck
function at depths greater than those where the line optical depth is
unity, may lead to  severe overestimates of abundances -- e.g.,
Carlsson et al. (1994) find that in Li-rich cool stars the Li
abundance may well be overestimated by  a factor of 3 as a result of
this. 

Photon suction may, for metal-rich cases in particular, lead to
overpopulation of, e.g., the ground state and thus inhibit the effects
of overionization.  This is the result of a compensation of photon
losses in the upper photosphere in  resonance lines, as well as in
connected ladders of transitions, by a downward population
replenishment flow from the continuum reservoir. It is of great
significance for atoms with a majority of corresponding ions and  with
pronounced cascade ladders. For complex atoms is should be of greatest
significance for the high-lying levels that  thus can compensate
population depletion processes at lower excitation energy, e.g. caused
by overionization.

The ultraviolet overionization has been a major worry in analyses of
late type stellar spectra for two decades. It arises  because the mean
intensity $J_{\nu}$ drops below the Planck function $B_{\nu}$ in the
line-forming regions of the atmosphere on the blue side of the
spectrum peak. Overionization is known to occur for Fe\,{\sc i} in the
Sun from levels a few electron volts below the continuum (see Rutten
1988 and references therein) and may more or less effect other metals
as well (see, e.g.,  Baumueller \& Gehren 1996, Bruls 1993).
Overionization was suggested by Auman \& Woodrow (1975) to be
significant for a number of elements with lower ionization energies in
cool stars. Major problems in modeling it are, however, the
difficulties in predicting the  ultraviolet flux of late-type stars
with the crowding  of spectral lines and the possible  existence of an
"unknown opacity" (cf., e.g. Gustafsson 1995), as well as the
dependence of the results of the notoriously uncertain collision cross
sections, e.g. for collisions with H atoms (cf. Steenbock \& Holweger
1984).  Empirically, M\"ackle et al. (1975)  and Ruland et al. (1980),
found a tendency for the low-excitation lines ($\chi \sim 2$ eV) of
Fe\,{\sc i} and Ti\,{\sc i} in K giant star spectra  to give
systematically lower abundances than the  high-excitation lines ($\chi
\sim 4.5$ eV).  The abundance difference was typically found to be
0.15 dex.  Subsequently, Steenbock (1985) succeeded in reproducing
this result with statistical-equilibrium calculations.  He found the
effect to mainly reflect an overionization in upper layers (notably
$\tau_{\rm 5000} < -2$) of the atmospheres,  where the low excitation
lines are formed. The effect  is much smaller in the solar spectrum,
leading to systematic errors in a differential analysis where red
giants are compared with the Sun.
 
For metal-rich stars, the blocking by the crowd of spectral lines  in
the ultraviolet could be expected to -- at first sight -- strongly
reduce  the overionization effect, but this may be compensated for by
a steeper temperature gradient in their atmospheres as a result of
line blanketing effects, as in the case of Li\,{\sc i}, Carlsson et
al. (1994), or of Ca\,{\sc i}, Drake (1991). The latter study is
particularly instructive  for judging the results of the present
investigation. Drake finds that for G and K-type stars the
overionization effects on Ca\,{\sc i} abundances  increase with
decreasing effective temperature, with increasing acceleration  of
gravity and with increasing metallicity. At least the last two results
may seem contrary to intuition. They reflect the significance of
H\,{\sc i} absorption shortwards of the Balmer discontinuity, which
blocks more of the ionizing UV flux for the giants than for the
dwarfs, and the afore-mentioned effects of metal-line blanketing on
the temperature structure. For the K dwarf stars the  effects on Ca\,{\sc
i} abundances may, according to Drake's results, well result in an
underestimate by a factor of two or more if LTE is assumed.

Over-recombination is important for photoionization transitions from
levels  close to the continuum (i.e.. in the infrared),  since for
them the angle-averaged $J_{\nu}$ may drop below  the local Planck
function deep in the photosphere.  This may produce net
recombinations, and overpopulation of the upper levels.  

Optical (ultraviolet) pumping occurs in strong lines, e.g. the
resonance lines and is analogous to overionization in  that it is
driven by $J_{\nu} > B_{\nu}$. It  is important, not the least when it
occurs in  ultraviolet resonance lines and excites the atoms to states
which may be much more easily photoionization due to a much richer
radiation fields  available at longer wavelengths, as was early
suggested by  Aumann \& Woodrow (1975). This is most important for
trace  elements, and for metal-poor stars.

The complex interplay between these different mechanisms affects most
levels of the atom, at great atmospheric depths, for atoms where the
strong lines get efficiently optically thin in the photosphere,
i.e. for relatively rare elements like the alkalis, while for the
more abundant atoms like Fe and Mg the stronger, e.g. resonance,
transitions are in detailed balance through most of the
photosphere. For these, the relatively  simple overionization
phenomenon is probably dominating, except for transitions very close
to the continuum for which e.g. photon suction may be significant. 

In a recent study Gratton et al. (in prep.) have used detailed
statistical-equilibrium calculations to explore the departures from
LTE for solar-type dwarfs as well as for red giants of different
metallicities, and their effects on abundance determinations for O,
Na, Mg, and Fe. These authors find relatively small effects for stars
cooler than the Sun for O\,{\sc i} -- LTE abundances from the IR triplet
lines should be corrected downwards by less than 0.1 dex for stars
with $T_{\rm eff}  \la 6000~{\rm K}$. For Na\,{\sc i}  the subordinate
lines are weakened by overionization and cascade by about 0.1 dex for
the solar-type dwarfs.  The dominating effect for Mg\,{\sc i}  is
overionization, and the non-LTE abundance corrections are thus
generally   positive. Typically the corrections are ~0.1 dex in the
dwarf stars.  For Fe\,{\sc i}, where again overionization is
dominating the   abundance corrections $\simeq 0.1$ dex. Most of these
effects are found to be greater for $T_{\rm eff} > 6000~{\rm K}$. It should
be noted, however, that models for  metal-rich dwarfs with $T_{\rm
eff} < 5000~{\rm K}$ were not included in this study.  

Summing up the discussion of non-LTE we conclude that the effects on
abundances are expected to be mainly due to overionization for
most of the elements. For the alkali atoms, as well as for the rare
earths, more complex effects may also be significant. Typically, errors
of about 0.1 dex may be expected in the differential results but 
the complexity of the interplay between different effects, and in 
particular the results obtained by Drake (1991) for Ca\,{\sc i}, suggest that 
greater effects may be present, in particular for the 
metal-rich K dwarf stars.

\subsection{Collecting errors}

We have shown that errors in fundamental parameters give errors in
mean resulting abundances of less than 0.1 dex. For elemental
abundances derived from several lines this may be the dominating
error, while for abundances derived from one single line errors due to
blends and fitting of continua may be the main contributors to the
overall error. Deviations from LTE in the excitation and 
ionization balance may also be of importance, probably more so for
abundances based on few lines, in particular for the K dwarf stars.
We collect our best estimates of errors due to different sources in
Table \ref{allerrors}.

\begin{table}
\caption[]{The effects of  error sources  explored in this work on
estimates of abundances relative to the Sun.}
\medskip
{
\begin{tabular}{lrrrrrrrrr}
\hline\noalign{\smallskip}
 Source of error & Error in resulting relative abundance  \\
\noalign{\smallskip}
\hline\noalign{\smallskip}
Measurement of $W_{\lambda}$ & negligible\\
Continuum fitting & $<$ 0.09 dex, usually 0.05 dex\\
$T_{\rm eff}, \log g $& $<$ 0.1 dex\\
Non-LTE effects & $0.1 - 0.2$ dex?\\
Oscillator strengths & $<$0.1 dex\\
\noalign{\smallskip}
\hline
\end{tabular}}
\label{allerrors}
\end{table}

\subsection{Comparison of results for stars in common with other studies}
		
\begin{table}
\caption[]{Comparison of iron abundances between our work and the
abundances quoted in the catalogue by Cayrel de Strobel et
al. (1997). The second column gives our results and the third the
mean, and the spread,  of the iron abundances given by Cayrel de
Strobel et al. In the  fifth give the number of derivations used. A
straight mean has been taken to represent the mean abundance from the
catalogue. }
\medskip
{ \setlength{\tabcolsep}{1.0mm}
\begin{tabular}{lrclrr}
\hline\noalign{\smallskip}
    ID       & \multicolumn{1}{c}{[Fe/H]} &$<$[Fe/H]$>$& $\pm$s  & \# \\
\noalign{\smallskip}
\hline
\noalign{\smallskip}
HD30562  & 0.19$\pm$0.09  & 0.14&  0.0        & 2 \\
HD32147 &  0.22$\pm$0.13   &    0.02 & 0.0  & 2\\
HD67228  & 0.16$\pm$0.08  & 0.05 &  & 1 \\
HD131977A & 0.00$\pm$0.12  & 0.01&          & 1\\
HD144585 & 0.27$\pm$0.05&           0.23 &&1\\
HD182572 & 0.42$\pm$0.05  & 0.32 &0.14 & 7\\
HD186427 & 0.12$\pm$0.05  & 0.06&0.04 & 5\\
\noalign{\smallskip}
\hline
\end{tabular}}
\label{cayrelcatalogue}
\end{table}

The majority of our stars have not been
studied  before through spectroscopic abundance analysis. 

For those of our stars (HD30562, HD32147, HD67228, HD1319777,
HD182572, HD186427) that are in the catalogue by Cayrel de Strobel et
al. (1997) the agreement between iron abundances derived in this study
and those listed in the catalogue is good, cf. Table
\ref{cayrelcatalogue}.

HD32147 has been given much attention in the discussion of Super
Metal Rich (SMR) stars. SMR stars have been defined as stars with
${\rm [Fe/H]} < 0.2$ dex (for a discussion and references on SMR stars see
Taylor 1996). Low resolution work and photometric determinations of
[Fe/H] have been carried out for this stars, but this is, to our
knowledge, the first high dispersion analyses of the star. Our [Fe/H]
of 0.28 dex implies that this star is really an SMR star.

Our results for HD182572 are compared with the results of the detailed
analysis by McWilliam (1990) in Table \ref{hd182572}. We note that
considerable  discrepancies remain even after correcting for the
difference in effective temperatures.

\begin{table}
\caption[]{Comparison of abundances, [X/H],  derived for HD182572
(HR7373) in our study and by McWilliam (1990). McWilliam uses
($T_{\rm eff}$/log{\it g}/[Fe/H]/$\xi_t$) $=$  (5380/3.92/0.15/1.9) and
we (5739/3.83/0.42/1.9). The results by  McWilliam have been
scaled to the same solar abundances as we use, Table \ref{linelist},
(this is most important for iron). In the last column our values are 
scaled to the effective temperature used by McWilliam. }
\medskip
{ \setlength{\tabcolsep}{1.0mm}
\begin{tabular}{lrrrr}
\hline\noalign{\smallskip}
           & \multicolumn{1}{c} {Mc William}  & \multicolumn{2}{c}{This work} \\
           &                   & 5739~K & 5380~K\\
\noalign{\smallskip}
\hline
\noalign{\smallskip}
Fe\,{\sc i}  &  0.31  & 0.42 	   & 0.24 \\
 Si\,{\sc i} &  0.28  & 0.51 	   & 0.47 \\
 Ca\,{\sc i} &--0.11  & 0.42	   & 0.21 \\
 Sc\,{\sc ii}&  0.14  & 0.36 	   & 0.36 \\
 Ti\,{\sc i} &--0.02  & 0.50	   & 0.21\\
 V\,{\sc i}  &--0.02  & 0.44	   & 0.12\\
 Co\,{\sc i} &  0.18  & 0.58	   & 0.29\\
 Ni\,{\sc i} &  0.00  & 0.46	   & 0.24\\
 Eu\,{\sc ii}&  0.18  & 0.13	   & 0.13\\
\noalign{\smallskip}
\hline
\end{tabular}}
\label{hd182572}
\end{table}

\begin{table}
\caption[]{The first column for of each star contains the result of this
work, the second the difference between the two studies, Diff = this
work - Edvardsson et al. (1993a). Stellar parameters used by
Edvardsson et al. are for HD30563 ($T_{\rm eff}$/log{\it
g}/[Fe/H])=(5886/3.98/0.17), for HD67228 (5779/4.20/0.04) and for HD144585
(5831/4.03/0.23). The values of the parameters used in this study may
be found in Table \ref{parameters}.} 
\label{comparison}
\medskip
{
\begin{tabular}{lrrrrrrrrr}
\hline\noalign{\smallskip}
       & \multicolumn{2}{c}{\bf HD30562 } & \multicolumn{2}{c} {\bf HD67228}  
& \multicolumn{2}{c} {\bf HD144585} \\
 &    & Diff. &   & Diff.&   & Diff.\\
\noalign{\smallskip}
\hline\noalign{\smallskip}
 [O\,{\sc i}]  &      &          & 0.21  &          &       &   \\    
 Na\,{\sc i} & 0.21 & +0.01    & 0.23  & +0.06    &0.36   &$\pm$0.00 \\  
 Mg\,{\sc i} & 0.33 & --0.01   & 0.23  & +0.08    &       &   \\  
 Al\,{\sc i} & 0.25 & --0.07   & 0.22  & +0.03    &       &   \\  
 Si\,{\sc i} & 0.21 & --0.02   & 0.27  & +0.11    &0.24   &--0.03 \\  
 Ca\,{\sc i} & 0.17 & --0.01   & 0.15  & +0.12    &0.24   &--0.03\\
 Ti\,{\sc i} & 0.13 & --0.02   & 0.09  & +0.02    &0.29   & +0.02 \\  	  
 Fe\,{\sc i} & 0.19 & +0.05    & 0.16  & +0.12    &0.27   & +0.04 \\      
 Fe\,{\sc ii}& 0.16 & +0.10    & 0.22  & +0.06    &0.11   &--0.05  \\     
 Ni\,{\sc i} & 0.17 &$\pm$0.00 & 0.14  &--0.01    &0.32   & +0.06 \\      
 Y\,{\sc ii} & 0.09 &          & 0.12  &$\pm$0.00 &0.06   & +0.06 \\      
 Nd\,{\sc ii}&      &          & --0.10 &          &--0.14 & --0.06 \\
\noalign{\smallskip}
\hline
\end{tabular}}
\end{table}

\begin{table}
\caption[]{Comparison of derived abundances, [X/H], between
Friel et al. (1993) and this work for HD186427, 16 Cyg B. Friel et al.  use
the following stellar parameters (T$_{\rm eff}$/log{\it
g}/[Fe/H])=(5770/4.30/0.05)  and we use (5773/4.17/0.12).  } 
\medskip
{
\setlength{\tabcolsep}{1mm}
\begin{tabular}{lrrrrrrrrr}
\hline\noalign{\smallskip}
  & \multicolumn{1}{l}{Na\,{\sc i}} & \multicolumn{1}{l}{Al\,{\sc i}} 
& \multicolumn{1}{l}{Si\,{\sc i}} & \multicolumn{1}{l}{Ca\,{\sc i}}\\
\noalign{\smallskip}
\hline\noalign{\smallskip}
Friel et al. & \multicolumn{1}{l}{0.07}        & \multicolumn{1}{l}{0.12 }       
& 0.06$\pm$0.03 &0.07$\pm$0.05 \\
This work      &0.13$\pm$0.04&0.12$\pm$0.04& 0.10$\pm$0.04 &0.02$\pm$0.09 \\
\noalign{\smallskip}
\hline\noalign{\smallskip}
& \multicolumn{1}{l}{Ti\,{\sc i}} & \multicolumn{1}{l}{Fe\,{\sc i}} 
& \multicolumn{1}{l}{~Fe\,{\sc ii}} & \multicolumn{1}{l}{Ni\,{\sc i}}\\
\noalign{\smallskip}
\hline\noalign{\smallskip}
 & 0.10$\pm$0.04 & 0.05$\pm$0.03 & 0.02$\pm$0.04&0.05$\pm$0.03\\
 & 0.07$\pm$0.10 & 0.05$\pm$0.05 &--0.06$\pm$0.04&0.04$\pm$0.10\\
\noalign{\smallskip}
\hline
\label{hd186427}
\end{tabular}}
\end{table}

\begin{table}
\caption[]{Comparison between results of our analyses for HD30592 and HD67228, 
with the results of Tomkin et al. (1997). The second and fourth column 
give the results of this work, while the third and fifth give the difference
between this work and that of Tomkin et al. Stellar 
fundamental parameters of Tomkin et al.
were ($T_{\rm eff}/\log g/{\rm [Me/H]}/\xi_t$) = (5930/4.20/0.21/1.55) and 
(5835/4.10/0.09/1.75) for HD30592 and HD67228, respectively. Our
parameters are given in Table \ref{parameters}.}
\medskip
{
\begin{tabular}{lrrrrrrrrr}
\hline\noalign{\smallskip}
 & \multicolumn{2}{c}{\bf HD30562}  &   \multicolumn{2}{c}{\bf HD67228}\\
   &&        Diff. &  &     Diff.\\
\noalign{\smallskip}
\hline\noalign{\smallskip}
[O\,{\sc i}]  &             & 0.21 & 0.01\\
Na\,{\sc i} & 0.21& --0.02 &  0.23& +0.03\\
Mg\,{\sc i} & 0.33&  0.12 &  0.23&  0.09\\
Al\,{\sc i} & 0.25& --0.01 &  0.22&  0.03\\
Si\,{\sc i} & 0.21& --0.04 &  0.27&  0.12\\
Ca\,{\sc i} & 0.17& --0.06 &  0.15&  0.08\\
Sc\,{\sc ii}& 0.31&  0.06 &  0.33&  0.16\\
Ti\,{\sc i} & 0.13& --0.12 &  0.09& --0.02\\
V\,{\sc ii}  & 0.10& --0.16 &  0.12&  0.01\\
Cr\,{\sc i} & 0.18& --0.02&&\\
Cr\,{\sc ii}& 0.17& --0.05 &  0.18&  0.02\\ 
Fe\,{\sc i} & 0.19& --0.02 &  0.16&  0.07\\
Fe\,{\sc ii}& 0.16& --0.04 &  0.22&  0.10\\
Ni\,{\sc i} & 0.17& --0.08 &  0.14&  0.01\\
YI\,{\sc i} & 0.09& --0.12 &  0.12&  0.06\\
Zr\,{\sc i} & 0.65&  0.39 &  &\\
Eu\,{\sc ii}& 0.14& --0.17 &&\\
\noalign{\smallskip}
\hline
\label{tomkin}
\end{tabular}}
\end{table}

We have analysed three stars  previously  studied by Edvardsson et
al. (1993a). The results are  compared in Table \ref{comparison}.  For
HD30562 and HD144585 the results agree to within the errors quoted.
The results for HD67228 show larger discrepancies than the other stars
for magnesium,  silicon, calcium and iron.  A higher microturbulence
parameter, as used in Edvardsson et al. (1993a), would decrease our
results by 0.01 or 0.02 dex (ionized iron by  0.04 dex). Edvardsson et
al. (1993a) find that a change in $T_{\rm eff}$ by +100 K gives an
iron abundance 0.06 higher for their stars. Thus, the increase of 52 K
needed to transform the results of Edvardsson et al. (1993a) to our
temperature scale means an increase of their iron abundance by 0.03
dex. Also the silicon and calcium  abundances are affected in the same
way as iron while all the other abundances remain as before, within
the errors, in the two studies. The log$gf$-values agree well (0.06
difference, we have the higher value). The spectrum we have obtained
for this star is of high quality (S/N $\sim$ 200). We do not find the
discrepancy between the two studies alarming. HD67228 has also been
studied by Andersen et al. (1984) in a study on lithium isotope ratios
in F and G dwarf stars. They derive an [Fe/H] of 0.05 from spectral
lines using a model with ($T_{\rm eff}$/log{\it
g}/[Fe/H])=(5850/4.2/0.05). 

Two stars from our sample, HD30562 and HD67228, were  recently
analysed in detail by Tomkin et al. (1997) on the basis of different
spectra; however, obtained with the same instrument and analysed
independently with model atmospheres computed with the same computer
program.  The results are compared with those of our analyses in Table
\ref{tomkin}.  In view of the errors in these analyses we find the
agreement satisfactory. 

HD186427 (16 Cyg B) have been extensively studied, in particular in
connection with searches for solar twins. A recent spectroscopic study
has been performed by Friel et al. (1993). The results are  in good
agreement, Table \ref{hd186427}. However, we find lower iron
abundances derived from Fe\,{\sc ii} lines than Friel et al. (1993)
do. This difference is probably mainly due to the different surface
gravities used. 

To conclude, we find that, for those few stars in our programme in
common with other studies, abundances are rather well
reproduced. This gives confidence when we now apply
 our results to the exploration of
the chemical evolution of the Galaxy.

\section{Abundance results}\label{sec:res}

\begin{table*}
\caption[]{Elemental abundances our programme stars. For each star and
each ion  the derived elemental abundance, the line-to-line scatter
(the error in the mean = line-to-line scatter/$\sqrt n_{\rm lines}$)
and the number of lines used in the analysis are  given. The table is
continued on the following pages. This table is only published electronically.}
\label{abundances}
\end{table*}

\begin{figure*}
\resizebox{\hsize}{!}{\includegraphics{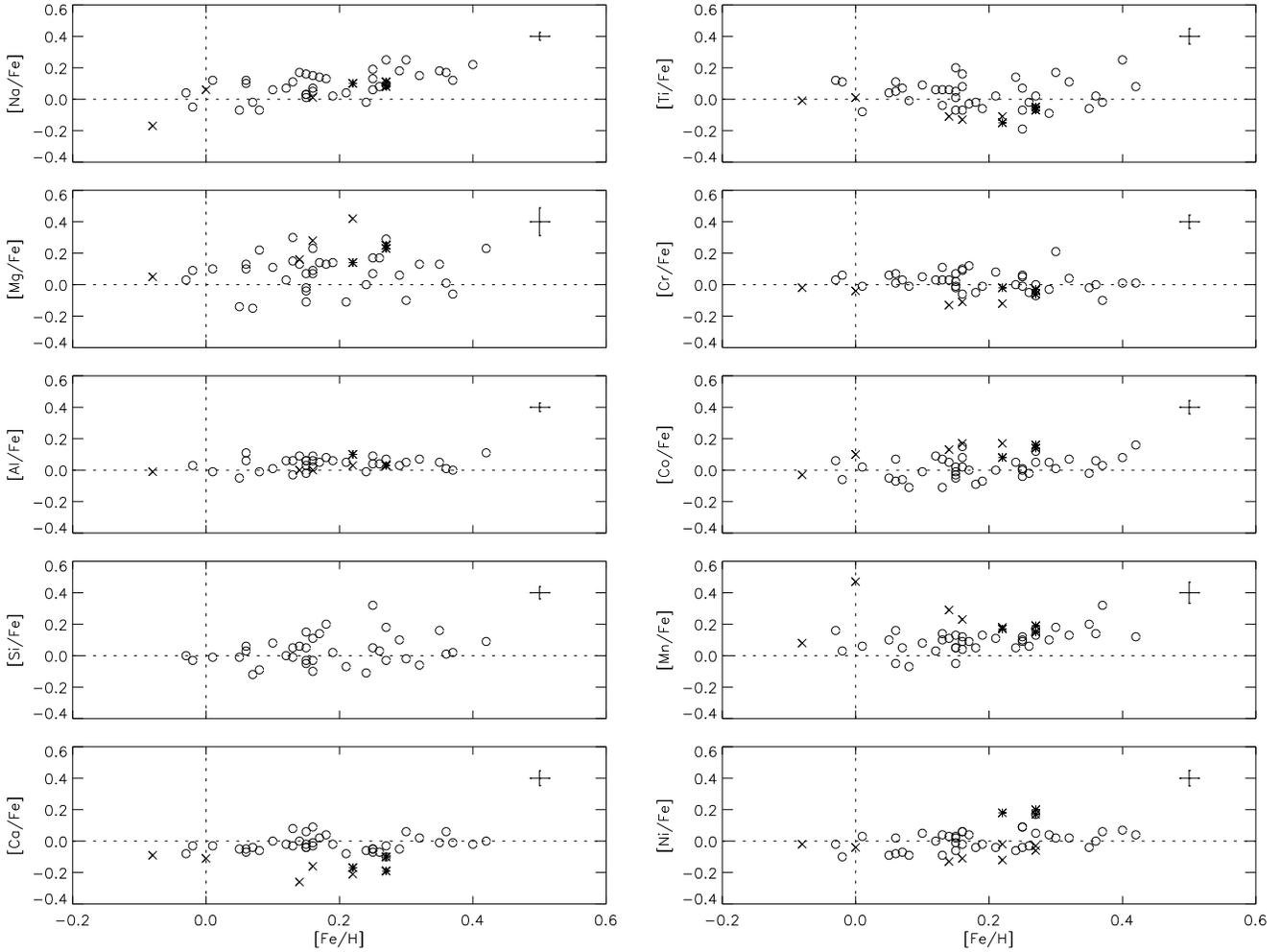}}
\caption[]{Our abundances relative to Fe. 
$\times$ symbols denote the five K
dwarf stars while $\ast$ symbols denote the stars from 
Barbuy \& Grenon (1990).}
\label{abundall}
\end{figure*}

\begin{figure*}
\resizebox{\hsize}{!}{\includegraphics{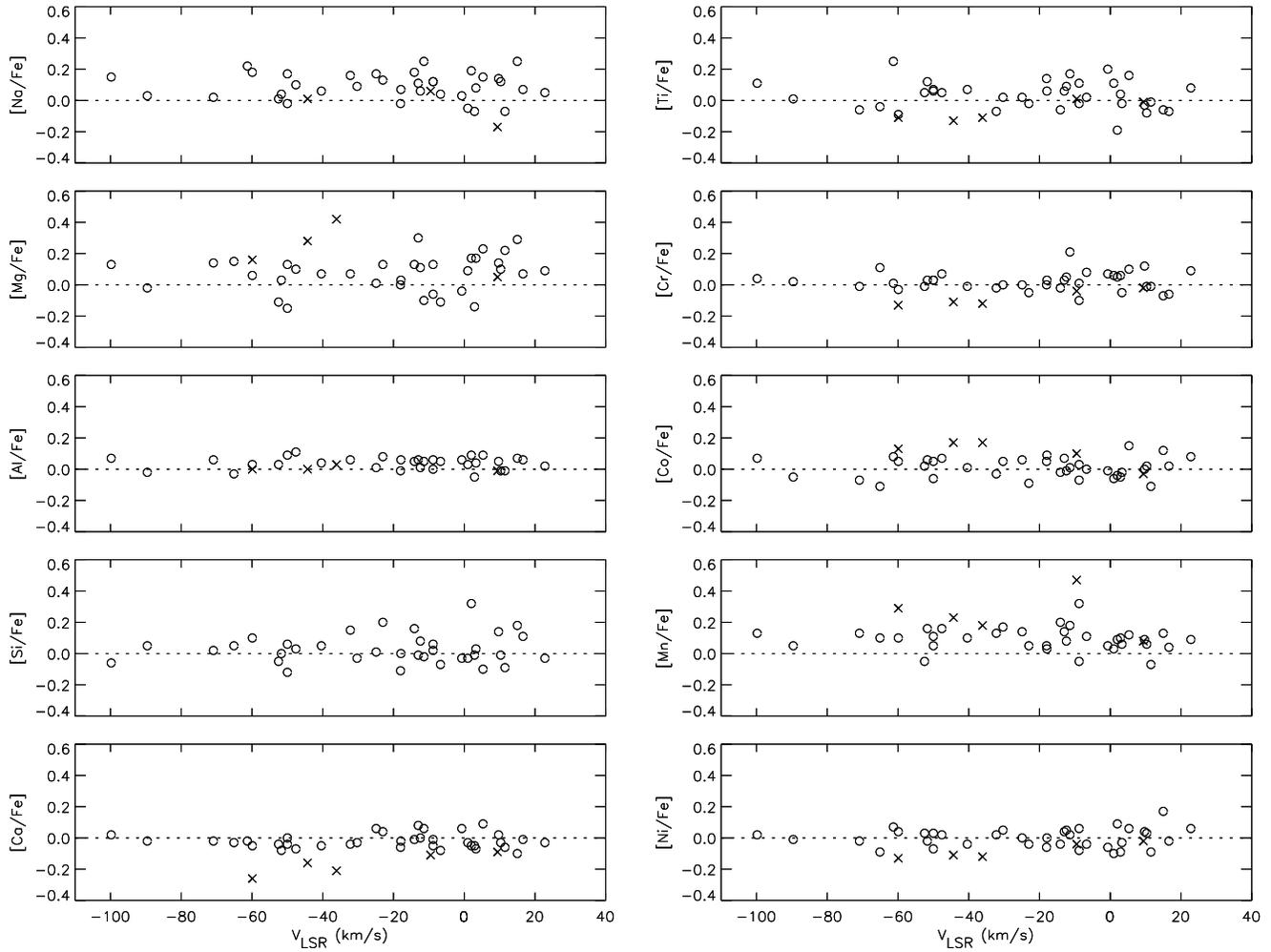}}
\caption[]{Our abundances relative to $V_{\rm LSR}$. $\times$ symbols
denote the five K
dwarf stars.}
\label{abundall_v}
\end{figure*}

\begin{figure*}
\resizebox{12cm}{!}{\includegraphics{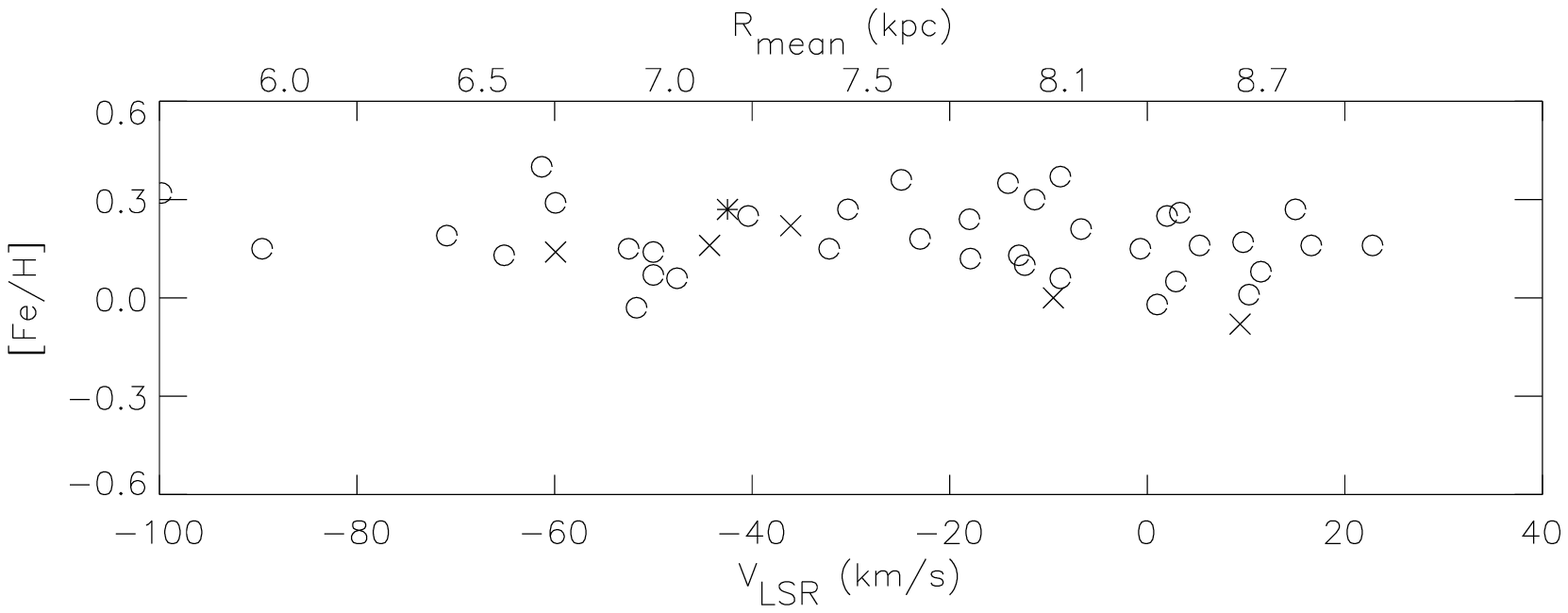}}
\hfill
\parbox[b]{55mm}{
\caption[]{Iron abundances relative hydrogen vs $V_{\rm LSR}$. 
$\times$ signs denote the five K dwarf stars and HD87007 is denoted by
a $\ast$ sign.}}
\label{fehvlsr}
\end{figure*}

Our abundance results are presented in Table \ref{abundances} and
relative abundances are plotted in Figs. \ref{abundall},
\ref{abundall_v} and \ref{fehvlsr}.  Below, we shall compare our
results with model calculations for the galactic chemical
evolution. Sometimes, we shall also quote results of other studies, in
particular that of Edvardsson et al. (1993a) (which is grossly
compatible in terms of calibration and methods with the present one),
but also with others, in particular for Population II dwarf stars.

Comparison of stellar abundance results from different studies may not
always be straightforward. There are many inconsistencies that may
confuse the interpretation of such comparisons. The usage of different
lines for the abundance analysis, different stellar parameters,
different model atmospheres, and, if the study is differential to the
Sun, different solar model atmospheres; all of these may lead to
offsets and may cause the compiled data to show trends which are
unreal.  Nevertheless, below we shall compare our abundances with
results from other studies to put our results into a broader picture
of  the galactic chemical evolution.


\subsection{Iron}

In observational studies of galactic chemical evolution iron is often
used as the reference element. The reason for this is twofold; iron is
believed, but this is debated, to be a fair chronometer for the
nucleosynthesis in the Galaxy, and the spectra of dwarf stars show 
many iron lines, easy to measure. The evolutionary picture for iron
is complicated by the fact that iron is produced in both core collapse
and type Ia supernovae. From this point of view oxygen, which is  only
produced in core-collapse supernovae, may be preferable as reference
element. However, as we will discuss  in Sect. \ref{sec:syre},
 oxygen abundances are not trivial to derive. We will therefore
conform with common practise and use iron as reference element.

Our resulting iron abundances are well determined with a line-to-line
scatter in  [Fe/H] of typically 0.09 dex and a formal  error in the
mean iron abundance of typically less than 0.02 dex for each star.  We
find that [Fe/H] does not vary with $V_{\rm LSR}$, i.e. $R_{\rm m}$,
Fig. \ref{fehvlsr}. This is an important observation when we consider
other elemental abundances relative to Fe later,
cf. Sect. \ref{sec:na}, and a  first indication that the mixing of gas
over the $\Delta R_{\rm m}$ spanned has been quite efficient. The five
K dwarf stars show a similar behaviour as the rest of the sample.

\subsection{Oxygen}
\label{sec:syre}

Oxygen is the third most abundant element in stars and therefore plays
a significant role for stellar opacities and energy generation. Therefore,
 determination of stellar  ages depends strongly on the
assumed initial oxygen abundance in the star, see
e.g. VandenBerg (1992). Oxygen abundances also affect  the
determination of time-scales in the galactic chemical evolution and
star-formation rates.  Thus, it is important to know the amount
of oxygen throughout the history of the Galaxy.

We have studied three oxygen criteria; the forbidden line at
 6300{\AA}, the  6158{\AA}  line and the  7774{\AA} triplet. These
 criteria are commonly used; however, discrepancies between the
 abundances derived from the different criteria  for the Sun, as well
 as for other late-type stars, have prevailed in spite of much work
 (cf. e.g. Eriksson \& Toft (1979) and Kiselman (1993) and references
 therein). Non-LTE and granulation are two proposed sources of the
 discrepancy between abundances derived from the [O\,{\sc i}] line and
 from the triplet lines. 

 The formation of the [O\,{\sc i}] line is  expected not to be subject
to departures from LTE. The lower level of the transition is the
ground level of the atom and the majority of the oxygen  should
be found in the ground state of the  atom under solar photospheric
conditions. There are, however, suspicions that the analysis of the
line might be subject to systematic errors due to the adoption of
 plane parallel  stellar
atmosphere models since granulation effects are not taken into account
in these models, Kiselman \& Nordlund (1995). 

\begin{figure*}
\resizebox{12cm}{!}{\includegraphics{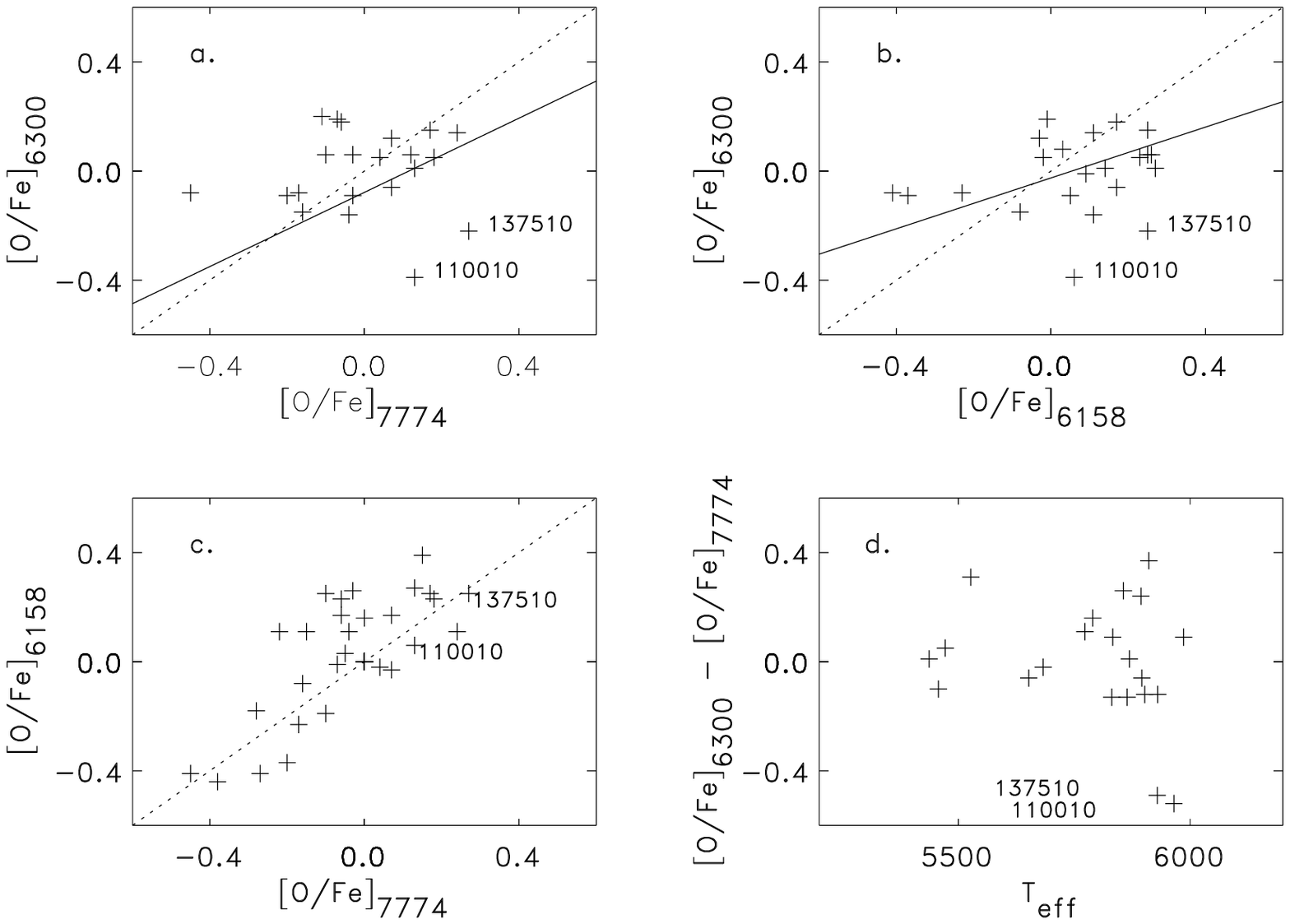}}
\hfill
\parbox[b]{55mm}{
\caption[]{Oxygen abundances derived from different abundance criteria
are compared, panel {\bf a.}, {\bf b.} and {\bf c}. On each axis is
the wavelength of the criterion indicated. The 7774 {\AA} oxygen
abundance represents the mean of the results for the three triplet
lines. The one-to-one relations are indicated by dotted lines and the
relations found by Edvardsson et al. (1993a) by solid lines. In panel
{\bf d.} we show the difference between abundances derived from 6300
{\AA} and the triplet lines as a function of effective temperature.}}
\label{oxy.corell}
\end{figure*}

Edvardsson et al. (1993a) found a correlation between oxygen abundances
derived from the [O\,{\sc i}] line and the abundances
derived from the triplet lines as well as a correlation with the
abundance derived from the 6158{\AA} line. These relations are shown,
together with our data, in Fig. \ref{oxy.corell}.  We do not find a
clear correlation between  abundances derived from [O\,{\sc i}] and
the abundances derived from the 7774{\AA} triplet and 6158{\AA} lines
for our stars. This circumstance, and the fact that the
[O\,{\sc i}] line is not expected to be affected by departures from
LTE, lead us to use only the abundance derived from
the [O\,{\sc i}] line in our analysis. 

When comparing our results with those of Nissen \& Edvardsson (1992)
we find in general a higher oxygen abundance. This discrepancy can be
understood: Nissen \& Edvardsson (1992) have used an  oscillator
strength of --9.75 (Lambert 1978) while ours is --9.84. This means
that our abundances should be scaled down by 0.09 dex to be put on the
same scale as the Nissen \& Edvardsson (1992) abundances. 

Three of our stars, HD37986, HD77338 and HD87007, have previously been
studied by Barbuy \& Grenon (1990). These authors derive oxygen
abundances and metallicities for a group of 11 dwarf stars. The stars
were selected on the basis of their kinematics and  claimed to
represent the ``local bulge population''. The stars fell clearly above
the oxygen trend expected from simple models of galactic chemical
evolution (their Fig. 1). The results were interpreted as possible
evidence for a rapid, and probably early, enrichment of the gas in the
galactic Bulge.  For most of the 11 stars in their study, Barbuy \&
Grenon (1990) derived the same abundance from the forbidden line as
from the triplet lines. This is not the case for the majority of the
stars in our study, see Fig. \ref{oxy.corell}. As shown in Table
\ref{barbuystars}, for the three stars in common with Barbuy \& Grenon
(1990),  we find that our iron abundances are lower by 0.2 dex as
compared to their results, while the oxygen abundances derived from
the triplet lines stay the same relative to iron. The largest
discrepancy in derived oxygen abundance is found for the forbidden
line in HD87007. Unfortunately, the error in our determination of this
abundance is rather large. The spectrum is one of our poorer, with a
S/N of only $\sim$ 80. The estimated error in [O/H] from noise might
then be as large as 0.5 dex. Thus, it is possible that the high
abundance derived by us from the forbidden line for HD87007 is due to
errors. For the two other stars the forbidden oxygen line was,
unfortunately,  heavily obscured by telluric lines and could not be
used. 

\begin{figure*}
\resizebox{12cm}{!}{\includegraphics{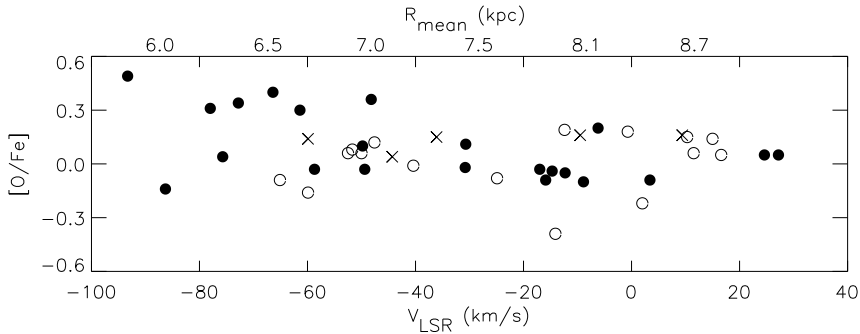}}
\hfill
\parbox[b]{55mm}{
\caption[]{Oxygen abundances, $\circ$ and $\times$ this work and
$\bullet$  Nissen \& Edvardsson (1992), as function of $V_{\rm LSR}$. 
At the top of
the figure is the mean galactocentric distance, $R_{\rm m} = (R_{\rm
perigal.} + R_{\rm apogal.})/2$ kpc, indicated for different
velocities. The relation between $V_{\rm LSR}$ and $R_{\rm m}$ is
taken from Fig. 1 in Edvardsson et al. (1993b). The star with the lowest
oxygen to iron abundance ratio is HD110010.}}
\label{ofev}
\end{figure*}

We only have access to velocity data  for one of the  stars, HD87007
(Barbuy, private communication). It has a $V_{\rm LSR}$ velocity of
$-42.5$ km/s which means that it would not satisfy the velocity
criteria of our high-velocity sample designed to represent the inner
part of the disk. The high $U_{\rm LSR}$ velocity (radial velocity in
the galactic plane of symmetry) of the star, however, gives a total
spatial velocity which fulfils the requirement of membership in our
high velocity sample. Castro et al. (1997) have studied 9 other 
 high velocity stars from the work by Grenon (1989). Their
results support our results. It is, however, difficult to make a 
clear comparison since velocity data for these stars have not been
published.

\begin{figure}
\resizebox{\hsize}{!}{\includegraphics{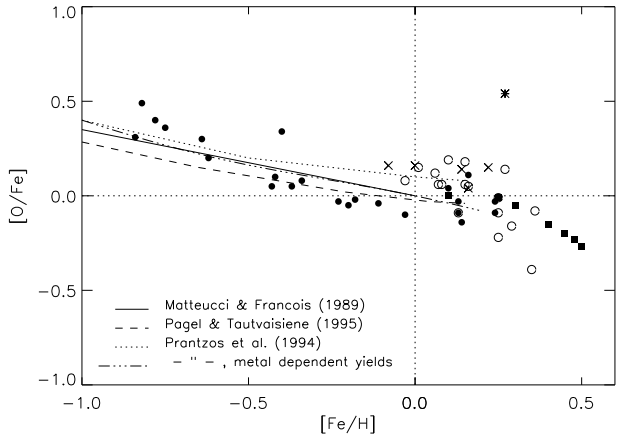}}
\caption[]{ Oxygen abundances from this work, $\circ$ symbols, and from
Nissen \& Edvardsson (1992), $\bullet$ symbols. $\times$ symbols 
denote our five K dwarf stars, $\ast$ symbol HD87007 and the 
filled boxes data from Castro 
et al. (1997). The star with lowest [O/Fe] is HD110010. }
\label{oxygen}
\end{figure}

In Fig. \ref{ofev} we plot [O/Fe] vs $V_{\rm LSR}$ for our stars as
well as the, generally, more metal-poor disk stars in the sample of
Nissen \& Edvardsson (1992). For the combined sample one may trace a
tendency for [O/Fe] to decrease with increasing $V_{\rm LSR}$, but
this is not shown by our sample alone.  From  Fig. \ref{oxygen} we
conclude that the oxygen abundance in general keeps declining relative
to the iron abundance also for ${\rm [Fe/H]}>$ 0.1 dex. This is not
inconsistent with what Nissen \& Edvardsson (1992) found in this
metallicity range.

\begin{table}
\caption[]{Stellar parameters and oxygen abundances for the stars in
common between this work and that by Barbuy \& Grenon (1990). For each
star we give our
iron abundandance as derived from spectral lines, [Fe/H], 
the metallicity as quoted by Barbuy \& Grenon, [M/H] and the
oxygen abundances derived from the different criteria as
indicated. For each criterion we give our resluts on the first line and
the results by Barbuy \& Grenon on the second.  [O/Fe]$_{\rm 6158}$
was not measured by them.}  {
\begin{tabular}{llrrrrrrrr}
\noalign{\smallskip}
\hline
\noalign{\smallskip}
ID &&{\bf HD37986}& {\bf HD77338} &{\bf HD87007}\\
\noalign{\smallskip}
\hline
\noalign{\smallskip}
& [Fe/H]& 0.27  & 0.22 & 0.27\\
&[M/H]&   0.47  &   0.45 &   0.43\\
\noalign{\smallskip}
\hline
\noalign{\smallskip}
&[O/Fe]$_{\rm 6300} $&  &   & 0.54\\
&& 0.15  & 0.20 & 0.00\\
&[O/Fe]$_{\rm 7770}$& 0.15  & 0.29 & 0.27\\
&& 0.23  & 0.20 & 0.20\\
&[O/Fe]$_{\rm 6158}$& 0.39& --0.04 & 0.27\\
\noalign{\smallskip}
\hline
\end{tabular}}
\label{barbuystars}
\end{table}

Oxygen is  produced in massive stars exploding as supernovae of types
II, Ib and Ic, Woosley \& Weaver (1995) and Thielemann et al. (1996).
Therefore, oxygen is expected to rapidly build up at early times in
the Galaxy or in any region which has experienced substantial star
formation ``lately''.  [O/Fe] starts to decline once the iron
producing supernovae start contributing more significantly to the
enrichment of the interstellar gas.  This decline starts at
[Fe/H]$\sim-1$ dex in the galactic disk, or at even lower metallicities.

In  Fig. \ref{oxygen} we compare our data with three different
theoretical models of the galactic chemical evolution. The model by
Matteucci \& Fran\c cois (1989) clearly shows the envisaged decline in
[O/Fe] after [Fe/H] = --1.0 dex, i.e. when supernovae type Ia start to
contribute to the enrichment of the gas. This can be compared with a
more recent model, taking the effects of metallicity dependent
supernova yields into account, by Prantzos \& Aubert (1995). The
difference between the two models by Prantzos \& Aubert  (1995) is not
large for lower metallicities but from around solar metallicity there is an
increasing discrepancy between their two models. If this suggested
trend continues to higher [Fe/H] the model using metallicity dependent
yields would be favoured by our data.  The  model by Pagel \&
Tautvai\v sien\. e (1995) is a simple analytic model, which assumes
supernovae type Ia to give their yields after a fixed time delay. In
spite of its simplicity it fits the data remarkably well and may
suggest that the basic understanding of the processes involved is
correct. 

 Tsujimoto et al. (1995) studied the abundance gradients
in the galactic disk by means of a  viscous disk model of galactic
chemical evolution. As can be seen in their Fig. 8 differences in
oxygen abundances as a function of radial distance from the galactic
centre are predicted to be so small that it would hardly be possible
to resolve this in a study as ours where such a small part of the
radius is spanned. This is compatible with our results in Fig. \ref{ofev}.
 We note that their model predicts (their Figs. 7
and 8) [O/Fe] vs [Fe/H] to flatten out at solar metallicities. If
this suggested trend is continued when the models are further evolved,
it would be difficult to reconcile them with our data, and with the 
results by Castro et al. (1997).

\subsection{Sodium}
\label{sec:na}

\begin{figure*}
\resizebox{\hsize}{!}{\includegraphics{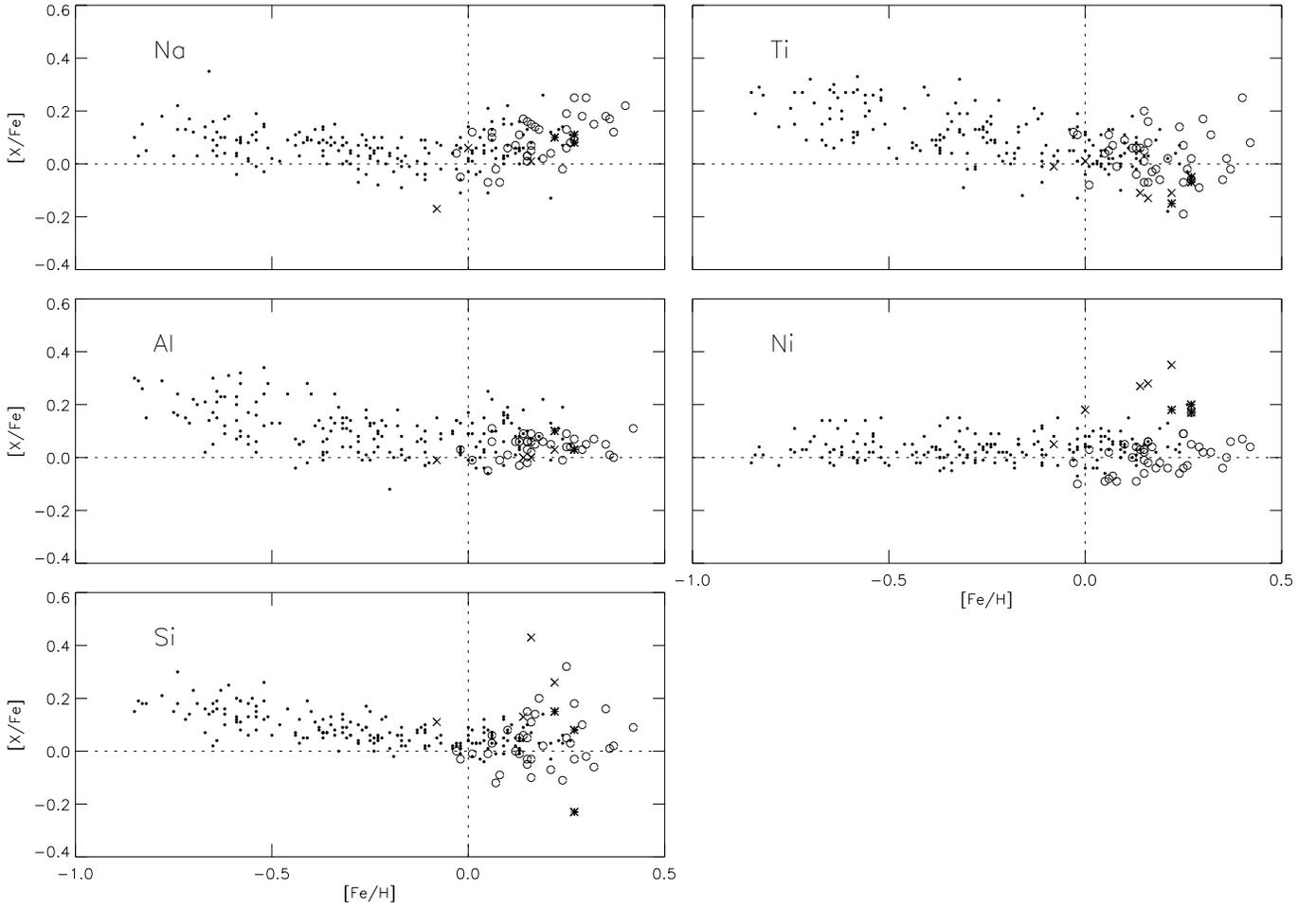}}
\caption{Abundances from this study compared with results from
Edvardsson et al. (1993a). On the vertical axes we give [X/Fe] where X
is the element indicated in the upper left corner of each
panel. $\circ$ symbols denote our results except the five K dwarf stars which
are denoted by $\times$ symbols and the three stars from Barbuy \& Grenon
(1990) which are denoted by $\ast$ symbols. Results from Edvardsson et al. are
denoted by $\bullet$ symbols.} 
\label{abund_bdp}
\end{figure*}

\begin{figure*}
\resizebox{12cm}{!}{\includegraphics{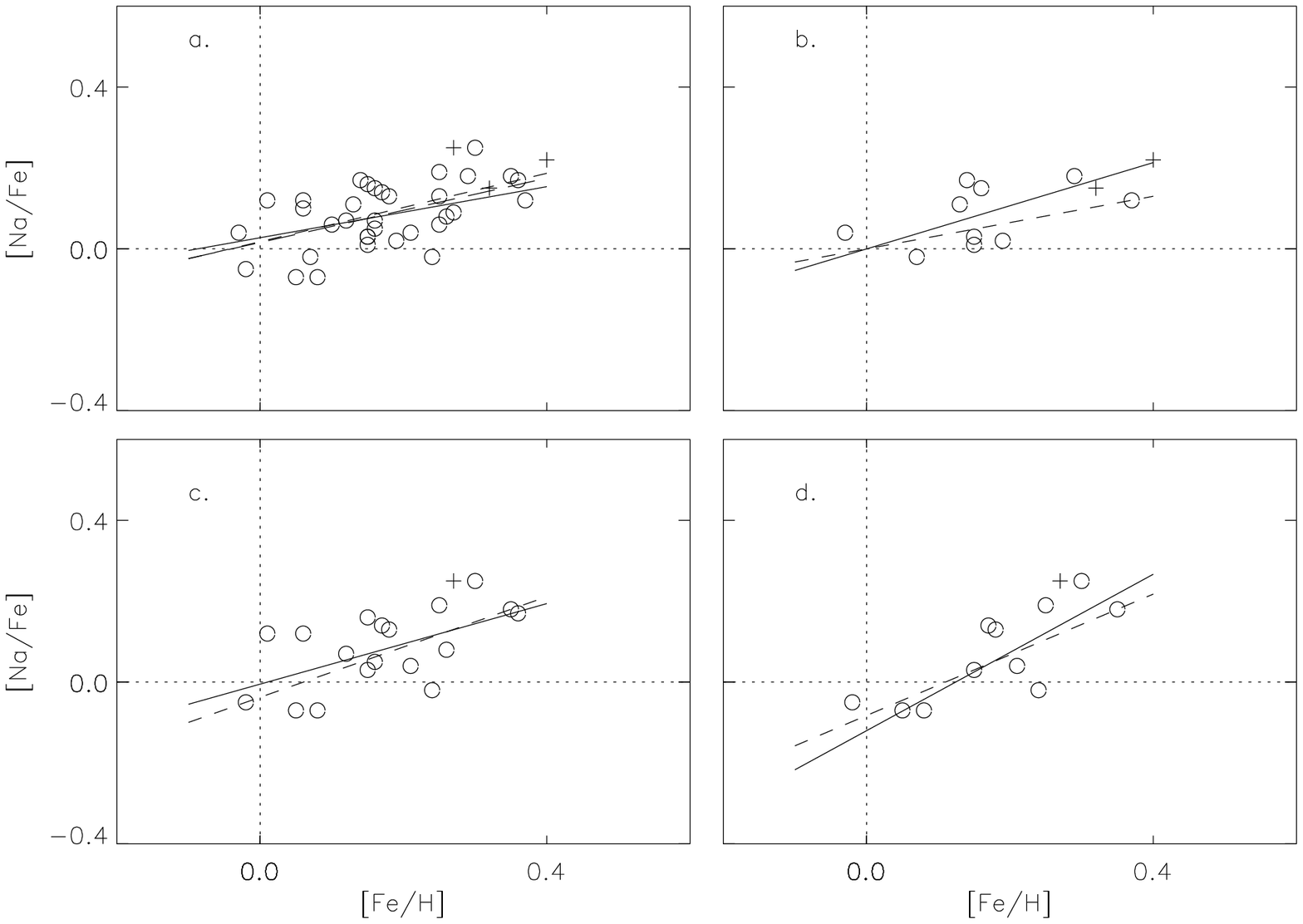}}
\hfill
\parbox[b]{55mm}{
\caption{In this figure we explore the different velocity samples as
indicators of differences in galactic chemical evolution in adjacent
parts of the galactic disk. The five K dwarf stars and the stars from
Barbuy \& Grenon (1990) are excluded from this discussion. In panel
{\bf a.} we show [Na/Fe] for all stars. Error-bars are smaller than the
symbols. In panel {\bf b.} are the stars with $V_{\rm LSR} < - 50$ km/s and/or
$Q_{\rm LSR} > 60$ km/s shown.  In panel {\bf c.} stars with $Q_{\rm LSR} <
45 $ km/s  and in panel {\bf d.} stars with $Q_{\rm LSR} < 30 $ km/s. Solid
lines show the result of a weighted linear least square fit to the
data and the dashed lines show least distance fits to the same
data. Stars denotes by $+$ symbols have sodium abundance determined from
one line only, and are not used in the error weighted fits.}}
\label{nafesamples}
\end{figure*}

One of the aims with our study was to determine whether or not the
upturn of [Na/Fe] vs. [Fe/H] found by  Edvardsson et al. (1993a) is real and
if so, if it is an effect of a mixture of stars from different
populations, cf Figs. \ref{abundall}, \ref{abundall_v}, \ref{abund_bdp}
and \ref{nafesamples}. We confirm
that the upturn is real. However, the upturn is less steep in our
study than what one could trace from the scattered diagram of
Edvardsson et al. (1993a). 

In Fig. \ref{nafesamples} we  have divided our sample into stars
representing the disk interior to the solar orbit and the solar orbit
and made linear least-square fits to the data. In panel a. we show all
47 stars and a least-square fit with the K dwarf stars and the stars
from Barbuy \& Grenon (1990) excluded. In panel b. we show the stars
that represent the disk interior to the solar orbit.  There is no
appreciable difference found between this sample and the whole
sample. In panel c. and d. we have defined the solar orbit sample in
two different ways. In c. it contains all stars with $V_{\rm LSR} <
45$ km/s and in d. with $V_{\rm LSR} < 30$ km/s. The sample in panel
c. also has a behaviour indistinguible from that of panel a. However,
the $V_{\rm LSR} < 30$ km/s seem to show a somewhat steeper trend.

An interesting question is now what difference in [Na/Fe] one would
have reason to expect for stars formed at different $R_{\rm m}$. Using
the data of Edvardsson et al. (1993a) we estimate that the minimum
value of  $d{\rm [Na/Fe]/d R}_{\rm m}$ is about $0.05$ dex/kpc. Extrapolating
this to the metal-rich stars and to $R_{\rm m} \sim 6$ kpc one finds at
the most a difference of 0.1 dex  between our two samples. Implicit in
this assumption is then that the  population of metal-rich stars at 6
kpc, which is very sparsely represented in the sample of Edvardsson et
al. (1993a), is not qualitatively different from that in the solar
neighbourhood. The results obtained in the present study supports this
and indicate that the difference in [Na/Fe] is, in fact, at the most 0.05
dex. 
 
In this connection, one should also note that orbital diffusion may
well mask the possible differences between stars formed at 6 kpc and 8
kpc.  E.g., stars of solar age in an orbit with $R_{\rm m}=8.5$ kpc
may have migrated from an orbit with $R_{\rm m}=6-7$ kpc (cf. Wielen
et al. 1996). The mixture of stars with different original orbits, in
combination with a radial galactic gradient of [Fe/H], was proposed by
Wielen et al. (1996) to explain the unexpectedly high scatter in [Fe/H] of
0.2 dex for  solar type stars, with similar age and similar present
$R_{\rm m}$, found by Edvardsson et al. (1993a). From the work of
Wielen et al. (1996) we estimate that two samples of stars with $<R_{\rm m}>= 
8.5$ and 6.5 kpc, respectively, would then be mixed by orbital  diffusion
so much that the population effects in abundances only would  show up
to about half the expected size as compared with the situation  if
orbital diffusion is not present. Although the reason for the great
inhomogeneities in the gravitational potential, needed to account for the
orbital diffusion of this magnitude, is not known, we conclude that the
effects looked for by dividing the total  sample of stars according to
the velocity criteria  used here, might be diminished considerably by
this phenomenon. 
 
\begin{figure*}
\resizebox{\hsize}{!}{\includegraphics{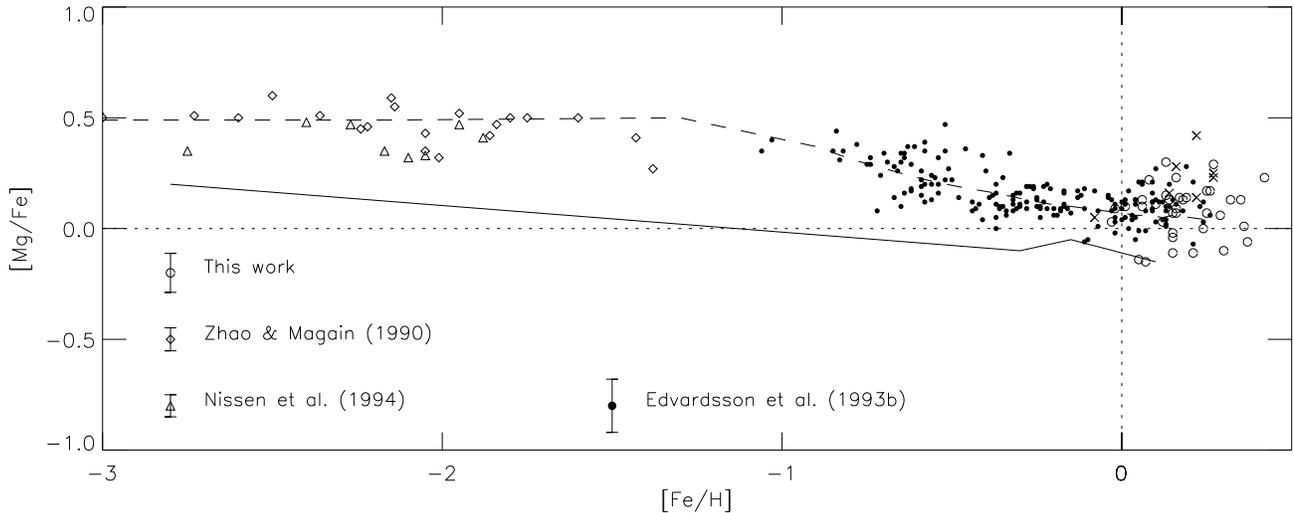}}
\caption[]{Magnesium vs iron abundances from several studies, as given
in the figure, as well as model calculations of the galactic chemical
evolution from Pagel \& Tautvai\v sien\. e (1995) (dashed line) and
Timmes et al. (1995) (solid line). $\times$ signs denote the K dwarf
stars in our sample. Indicated error bars refer to the error in the
mean.}
\label{mgfefel}
\end{figure*}

\subsection{Aluminium}

Aluminium is produced in heavy single stars, with $ M > 8 M_{\odot}$.
 The scatter, as well as the internal line-to-line scatter, in our
 data is considerably smaller  than in Edvardsson et al. (1993a). The upturn
 in [Al/Fe] vs [Fe/H] indicated in their data is not obviously present
 in ours, see Fig. \ref{abund_bdp}. The two studies use different lines for
 the abundance analysis. The small scatter in [Al/Fe] for [Fe/H] $>
 0.0$ dex is also evident in Morell (1994).  The lines used by
 Morell (1994) and us, 6696.03{\AA} and 6698.66{\AA}, are situated
 in a part of the stellar spectrum which is clean. Thus, we expect no
 problems with continuum fitting and blends and errors arising 
from measurements
 of the line strengths should be negligible. The trend in
 Fig. \ref{abund_bdp} and its similarity with, e.g., that of Ca
 (Fig. \ref{cafecoolfig}) suggests a similar origin in core-collapse supernovae.

\subsection{Magnesium}

Due to the large line-to-line scatter in our magnesium data it is not
possible to determine here if the large scatter found by
Edvardsson et al. (1993a) is real or not. We note, however, that we get
roughly the same amount of scatter as they do, Fig. \ref{mgfefel}.
There is no evidence in our data for a correlation between 
kinematics and  [Mg/Fe] ratios of the stars, Fig. \ref{abundall_v}.

Magnesium is, like oxygen, usually  assumed to be  formed only in
core-collapse supernovae through hydrostatic carbon burning. Timmes et
al. (1995) are not so successful in describing the over-all evolution
of magnesium abundances in the Galaxy. This may suggest that the
production of Mg is not fully understood at present. The simple-minded
model with the delayed yield formalism is more successful in this
respect, Pagel \& Tautvai\v sien\. e (1995).

\subsection{Silicon}

We only use two lines to determine silicon abundances while Edvardsson
 et al. (1993a) used eight lines.  We find a much larger scatter in
 our data than they do, Fig. \ref{abund_bdp}.   It has not been
 possible, from our data, to determine the origin of neither the
 line-to-line scatter nor the star-to-star scatter. When inspecting
 the lines one by one no line seems to stand out in terms of derived
 abundances. Nor do we find  any signs of the scatter to be an effect
 of different stellar populations mixing, see e.g.
 Fig. \ref{abundall_v}.

\subsection{Calcium}

\begin{figure*}
\resizebox{\hsize}{!}{\includegraphics{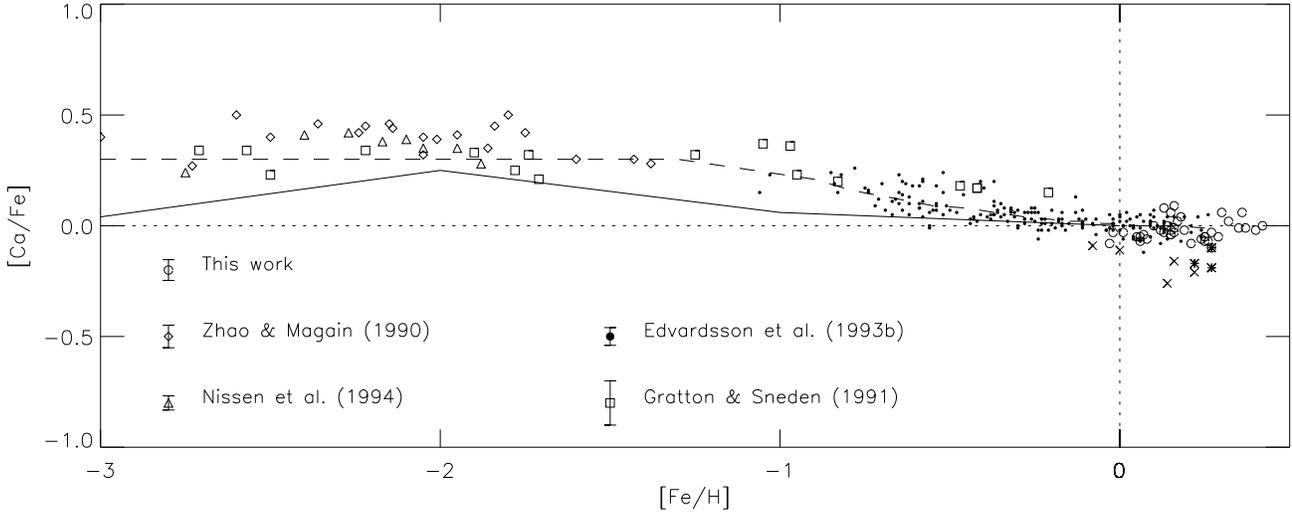}}
\caption[]{Calcium abundances from several studies, as given in the
figure, as well as model calculations of the galactic chemical
evolution from Pagel \& Tautvai\v sien\. e (1995) (dashed line) and
Timmes et al. (1995) (solid line). Typical error bars are indicated
for each study.  $\times$ symbols denote K dwarf stars.}
\label{calcium2}
\end{figure*}

Comparison of our calcium abundances with those of Edvardsson et
al. (1993a), in Fig. \ref{calcium2}, confirms their finding that the
[Ca/Fe] flattens out towards higher metallicities. Also, there is no
obvious difference between stars with different galacto-centric mean
distances, Fig. \ref{abundall_v}.

\begin{figure}
\resizebox{\hsize}{!}{\includegraphics{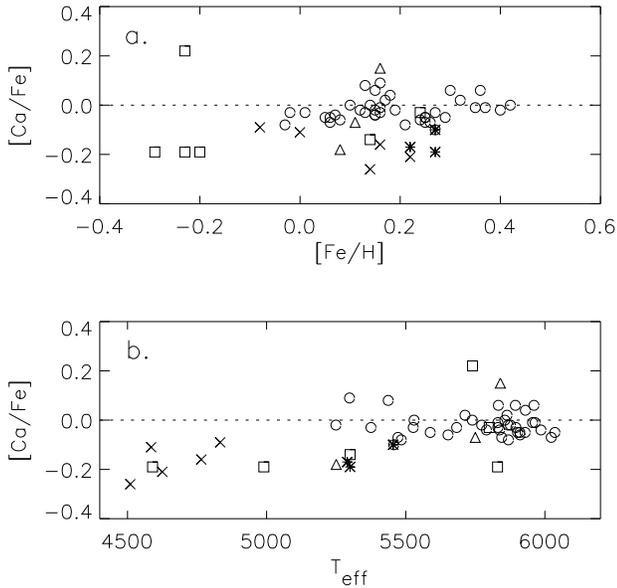}}
\caption[]{Calcium abundances for our stars, except the K dwarf stars,
$\circ$ symbols. The K dwarf stars are denoted by $\times$ symbols. Abundances for
solar-type stars from Abia et al. (1988), $\Box$ symbols, and Gratton \& Sneden (1987),
$\bigtriangleup$ symbols, and the stars from Barbuy \& Grenon (1990),
$\ast$ symbols.}
\label{cafecoolfig}
\end{figure}

As is evident from  Figs. \ref{abundall} and \ref{cafecoolfig} the K
  dwarf stars exhibit a behaviour which is very different from  the
  rest of the sample.  They seem to have a mean calcium relative to
  iron abundance  $\sim 0.2$ dex lower than the mean abundance for the
  rest of the stars. When plotting  [Ca/Fe] as a function of effective
  temperature we see that the stars with low Ca abundances  have the
  lowest effective temperatures, cf. Fig.  \ref{cafecoolfig}. 

In a high resolution, high signal-to-noise abundance study of dwarfs
and giants in the disk by Abia et al. (1988), there are six dwarf stars in
the same metal and effective temperature range as our K dwarf stars.
In  Fig. 4a in Abia et al. (1988) the same features as in
Fig. \ref{cafecoolfig} can be seen, i.e. cool, metal-rich stars show
up as underabundant in calcium. In another study, Gratton \& Sneden (1987),
of light elements in field disk and halo stars we also find support
for such a behaviour of calcium in cool dwarf stars. 

The key to these low [Ca/Fe] measures may lie in
overionization. Drake (1991) performed non-LTE calculations for
calcium for a range of stellar parameters.  He showed that the
difference between an abundance  derived under non-LTE and  LTE
conditions varies strongly with effective temperature and surface gravity,
and less strongly with metallicity. 
Drake (1991) finds that the non-LTE effect on abundances of Ca in G
and K dwarf stars increase considerably with decreasing effective
temperature. From his Figs. 4, 7 and 8, we estimate the correction
factor for weak lines in a dwarf star of at least solar metallicity
with an effective temperature of 4500K to be on the order of  0.3
dex. Such an adjustment would indeed put the K dwarf stars right on
the line, [Ca/Fe] = 0.0 dex. This suggests that the calcium abundance may
vary in lockstep with the iron abundance also for metal-rich  K dwarf
stars.

\subsection{Titanium}

[Ti/Fe] was shown by Edvardsson et al. (1993a) to be a slowly
decreasing function of [Fe/H]. The decline may  continue also for
higher iron abundances.  We use $10-12$ lines to derive titanium
abundances for our stars, while Edvardsson et al. (1993a) used four.
In spite of our, presumably, smaller random errors in  the abundance
determination for each star, as is shown in Fig. \ref{abund_bdp}  we
still find the same and  comparatively large scatter in the abundances
found by Edvardsson et al. (1993a). 

Inspection of derived stellar abundances as a function of  excitation
energy for the lower level in the transition  for each line indicated
no presence of  non-LTE effects or blends.  However, also for titanium
we found no evidence that the large star-to-star scatter should be a
result of a mixing of stars with different mean-perigalactic
distances, i.e. different $V_{\rm LSR}$ velocities,  see
Fig. \ref{abundall_v}.

\subsection{Scandium and Vanadium}

\begin{figure}
\resizebox{\hsize}{!}{\includegraphics{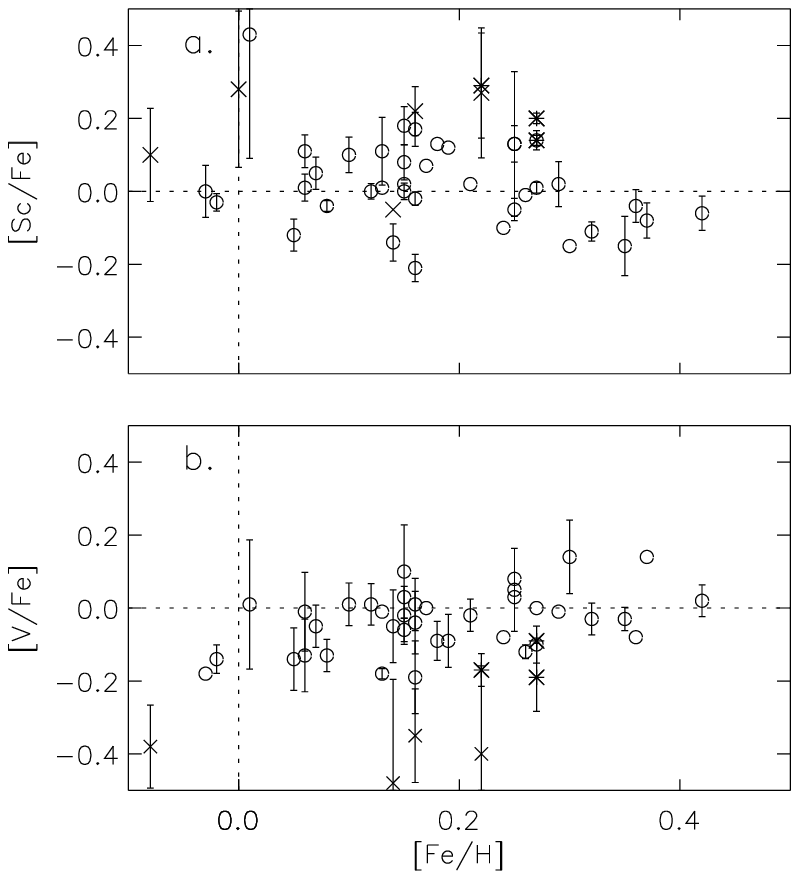}}
\caption[]{Scandium abundances from Sc\,{\sc ii} and vanadium abundances
from V\,{\sc I}. Error bars indicate the error in the mean. Stars with no
error  bar means that the abundance was derived from a single
line. $\times$ symbols denote K dwarf stars and $\ast$ symbols the stars from 
Barbuy \& Grenon (1990). }
\label{sc2fefel}
\end{figure}

Abundances derived from Sc\,{\sc i} lines are unreliable and we therefore
only present abundances determined  from Sc\,{\sc ii} lines,
Fig. \ref{sc2fefel}. Scandium exhibits some scatter but seems to vary in
lockstep with iron.

For vanadium the atom is represented by two lines and the ion by
one. We present the data derived from lines of the
atom. Also vanadium appears to vary in lockstep with iron over the
metallicity range studied.

\subsection{Chromium}

\begin{figure*}
\resizebox{\hsize}{!}{\includegraphics{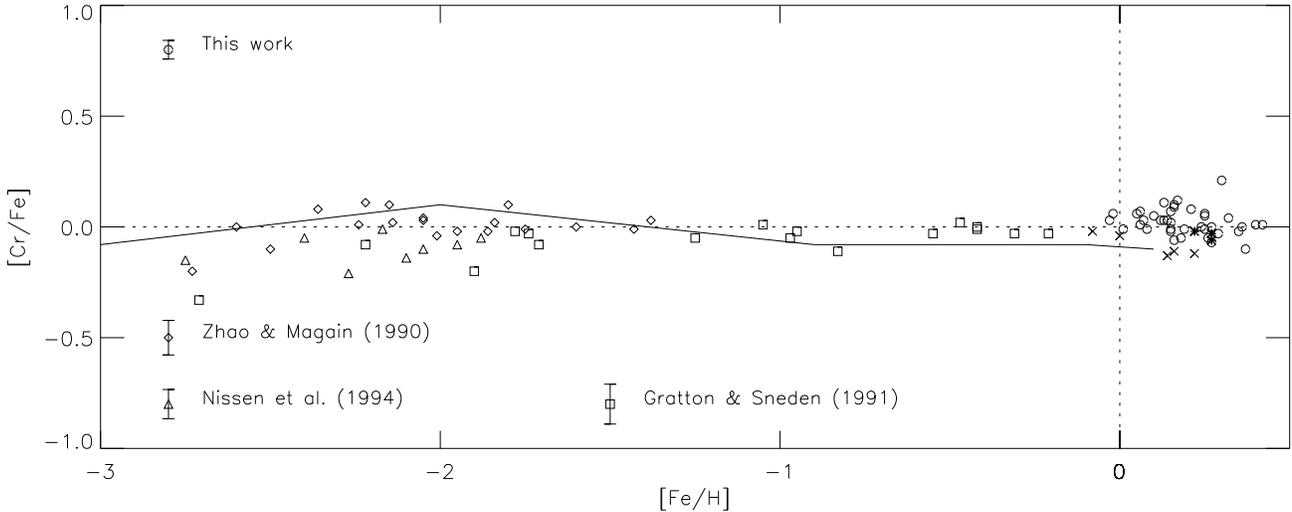}}
\caption[]{Chromium data from several sources, as given in the figure,
showing the galactic chemical evolution of chromium. The 
solid line is from Timmes et al. (1995). The chromium  rich star in
our sample is HD87646. The error bars indicate the error in the mean.
 $\times$ symbols denote our K dwarf stars.  Chromium was not
studied by Edvardsson et al. (1993a).}
\label{chrom}
\end{figure*}

For the most metal-rich stars chromium,
 as well as other iron peak elements, varies in lockstep with iron,
 Fig. \ref{abundall}.  The overall evolution of [Cr/Fe] seems to be
 well described by Timmes et al. (1995) using the supernova yields by
 Woosley \& Weaver (1995), Fig. \ref{chrom}. The flatness of the
 relation between [Cr/Fe] and [Fe/H] can be understood as a
 consequence of that massive stars with solar initial metallicities
 produce enough chromium to balance the iron production by SNIa,
 cf. Timmes et al. (1995).

\subsection{Manganese and Cobalt}

Few studies of stellar abundances have been made for these elements.
Manganese abundances were measured by  Gratton (1989) for 25
metal-poor giants and dwarfs. Cobalt abundances were obtained by
Gratton \& Sneden (1991) for 17 metal-poor (mostly) giant and dwarf
stars and by Ryan et al. (1991) for  19 dwarf and giant stars. 

Five lines of the manganese atom were used for determination of
abundances, three weak and two stronger lines. [Mn/Fe] scales with
[Fe/H], but with a tendency to increase for ${\rm [Fe/H]} > -1.0$
dex. This increase seems to continue beyond ${\rm [Fe/H]} > 0.0$ dex.
We use seven lines arising from the atom to determine cobalt
abundances. From our data cobalt seems to vary in lock-step with iron
for $0.0~{\rm dex} < {\rm [Fe/H]} < 0.3$ dex. 

\begin{figure*}
\resizebox{12cm}{!}{\includegraphics{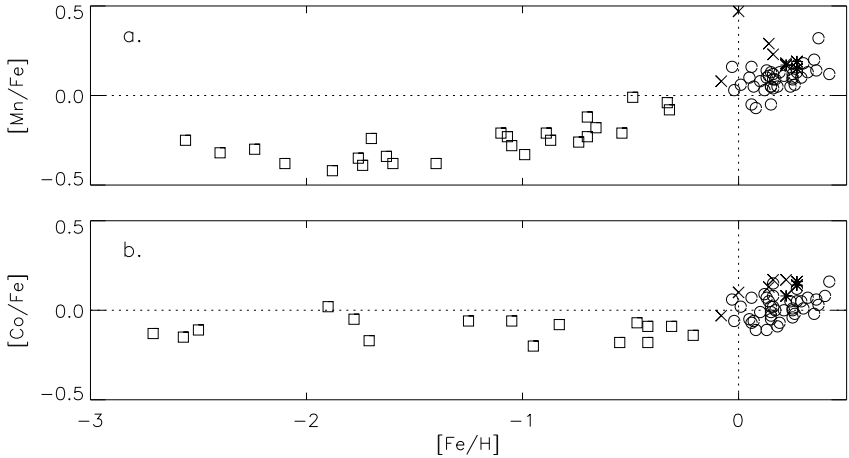}}
\hfill
\parbox[b]{55mm}{
\caption[]{Manganese, panel {\bf a.}, and cobalt, panel {\bf b.},
 abundances from this work, $\circ$, $\times$, and $\ast$ symbols, and
 from Gratton (1989) (manganese) and Gratton \& Sneden (1991)
 (cobalt), $\Box$ symbols.}}
\label{comn}
\end{figure*}

Manganese and cobalt belong to the group of iron-peak elements. These
elements are thought to be formed during explosive silicon burning in
supernova explosions and nuclear statistical equilibrium (Woosley \&
Weaver, 1995). 

Timmes et al. (1995) find that the rise in [Mn/Fe] vs [Fe/H] from
[Fe/H] $\sim - 1$ dex is due to the over-production of manganese in
supernovae type Ia and in heavy stars with solar metallicity, as
compared with the iron production, while for cobalt the production of
iron in supernovae type Ia is balanced by production of cobalt in
supernovae resulting from massive stars with initially solar
metallicity.

\subsection{Nickel}

We have used  12 lines from the nickel atom to obtain abundances;
 Edvardsson et al. (1993a) used 20 lines. Like  Edvardsson et al.
 (1993a) we find that nickel varies in lock-step with iron, and this
 continues also for higher metallicities. For the stars in common
 between  the studies the nickel abundances derived are in excellent
 agreement, although our study does not show the slight offset found
 by Edvardsson et al. (1993a).

Here, we also note an interesting behaviour of the K dwarf  stars,
namely that they show larger nickel abundances than the rest of the
sample. The large number of lines together with the fact that the
lower excitation energies for the lines span a range of values
($1.9-5.3$ eV) and that the formal error for each star is small makes
departures from excitation equilibrium an unlikely explanation for
this effect. 

 Overionization, for most stars but less for the cool ones, or  blends
are  possible but neither very probable explanations. The phenomenon
needs further systematic study.

\subsection{r- and s-process elements}

Most of the heavy elements (A$>$70) are  formed through the r- and
s-processes. For some of them one of the processes contributes much
more than the other.  The s-process contributes most, for the solar
system composition, to Y (73\%) and Zr (79\%) while Eu is to 90\%
formed in the r-process, according to Anders \& Grevesse (1989). Eu is
one of the few r-process elements with clean lines observable in the
visual part of stellar spectra. Therefore, it is well suited for
studies of the sites for the r-process. The relative abundances of
s-elements produced in thermally pulsing asymptotic giant branch stars
are set by the degree of the exposure to neutrons. Heavier neutron
flux enhances the abundances of the heavier elements (Ba, Nd, Hf)
relative to the lighter ones (Y, Zr, Mo).   Molybdenum is formed by a
mixture of processes (p- , r- and s-processes), see Anders \& Grevesse
(1989).

From this knowledge one would expect the r-process elements to have
high abundances in old stars and show a declining trend when compared
to iron. This is indeed seen for europium, see figures and discussions
in Mathews et al. (1992) and Woolf et al. (1995). For the s-process
elements, on the other hand, one would expect old stars to have low
s-element abundances while the more recently formed stars would show
an increase in their s-element abundance, due to the long time scales
for the evolution of the s-process sites. Such a tendency was also
traced by  Edvardsson et al. (1993a).

\paragraph{Results}

\begin{figure}
\resizebox{\hsize}{!}{\includegraphics{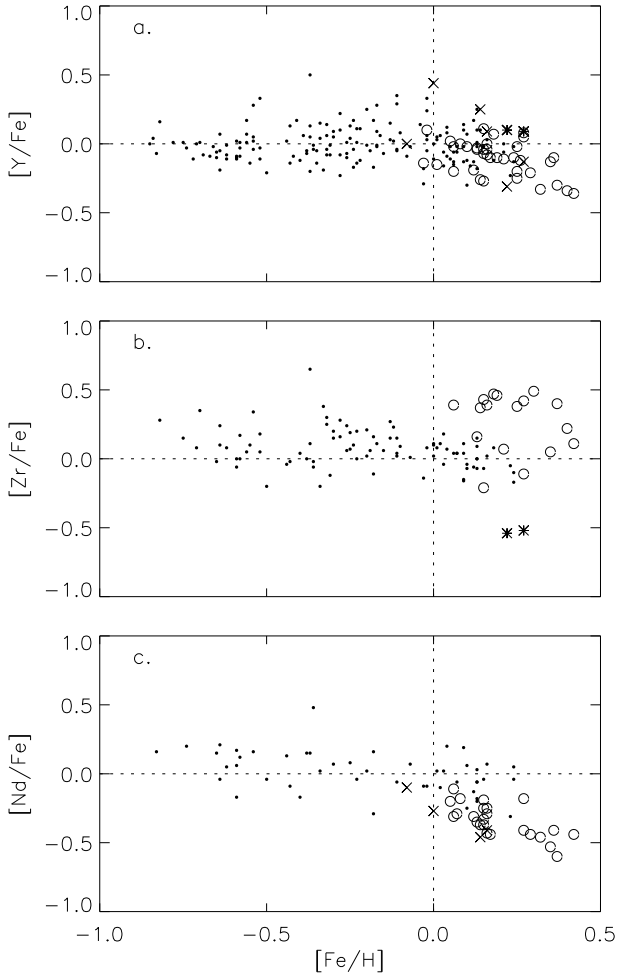}}
\caption[Zirconium  abundances]{Comparison of data from this study and
from Edvardsson et al. (1993a) for three elements, yttrium, zirconium
and neodymium. $\circ$ symbols denote our abundances, $\times$ symbols
denote the K dwarf stars, $\ast$ symbols the stars from Barbuy \&
Grenon (1990) and $\bullet$ symbols denote abundances from Edvardsson
et al. (1993a). }
\label{yzn}
\end{figure}

\begin{figure}
\resizebox{\hsize}{!}{\includegraphics{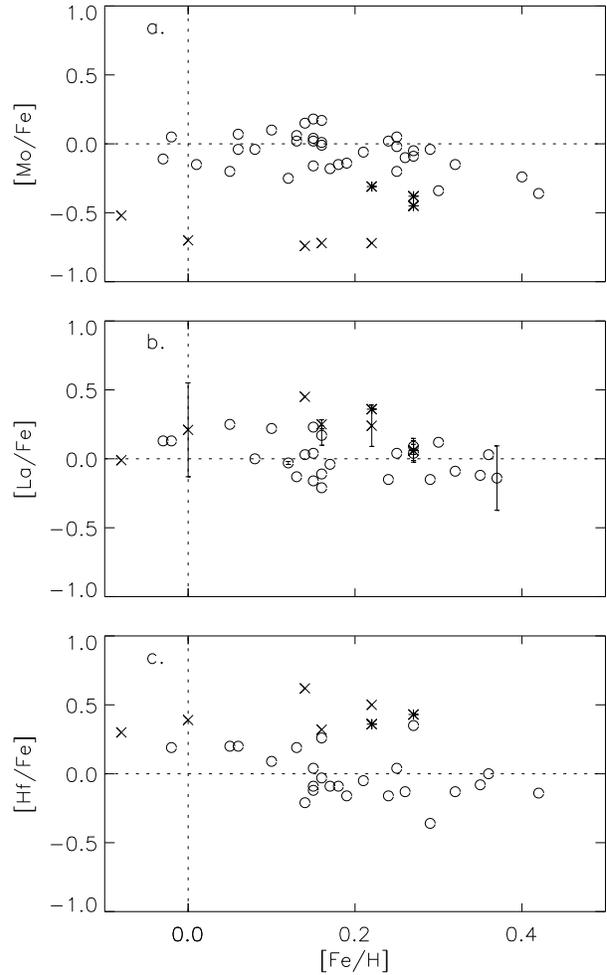}}
\caption[]{Molybdium, lanthanum and hafnium abundances relative to
iron. $\times$ symbols denote K dwarf stars and $\ast$ symbols the
stars from Barbuy \& Grenon (1990). Note the different ranges on the
vertical axes.}
\label{mofe}
\end{figure}

Our stellar spectra contain few lines of these heavy elements that are
accessible in the visual and may be securely
used as abundance criteria.

We have derived abundances for a number of s- and r-process elements,
using a small number of Y\,{\sc  ii} lines of suitable strength, one
line each for  Zr\,{\sc i} and Mo\,{\sc i}, two for La\,{\sc ii} and
one each for Nd\,{\sc ii}, {Eu\,{\sc ii} and Hf\,{\sc ii}. We find
that the metal-rich stars have roughly solar abundances of these
elements relative to iron, however, with some possible departing
trends.

Up to [Fe/H]$\sim$0.2 dex we confirm the result by Edvardsson et al. (1993a)
that Y varies in lock-step with iron. For the more metal-rich stars,
however, there may be a decline in [Y/Fe] (see Fig. \ref{yzn}).
The results may, however,  be due to  possible effects of overionization
in yttrium. 

As is clear from Fig. \ref{yzn} for zirconium there is probably a
systematic trend with effective temperature, resulting in a large
scatter (or even a division of stars into groups) and unreliable
abundances. Blends may may also contribute to this scatter.  The
results for our stars do generally indicate lower neodymium abundances
than those of Edvardsson et al. (1993a) and again a decrease of
[Nd/Fe] with increasing [Fe/H].

\begin{figure}
\resizebox{\hsize}{!}{\includegraphics{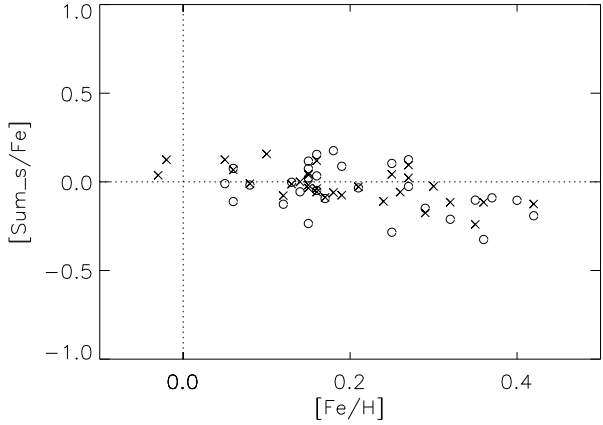}}
\caption[]{The sums of light and heavy s-process elements are
shown. The sums has been weighted as follows  $light = 0.5\cdot [Y/Fe]
+ 0.3\cdot [Zr/Fe] + 0.1\cdot [Nd/Fe], \\ heavy = 0.25 \cdot [Mo/Fe] +
0.5 \cdot [La/Fe] + 0.25\cdot [Hf/Fe]$. Only stars with at least two
of the elements in the sum measured are shown. $\circ$ symbols denote the sum
of the light s-process elements and $\times$ symbols denote the heavy elements.  }
\label{sum_sr}
\end{figure}

For molybdenum, lanthanum and hafnium we have no studies to compare
with and can therefore not say very much about the general evolution.
The molybdenum abundances are derived from one Mo\,{\sc i} line. Mo and
Hf, and probably also La, however, show the familiar pattern, ascribed
to  overionization in the K dwarf stars.

In conclusion, the abundances of Mo, La, and Hf seem to roughly vary
in lock-step with Fe, however, with some indications that the
abundance ratios decrease with increasing [Fe/H].

In order to improve the statistics we have derived two quantities,
$<s-light/Fe>  = 0.5\cdot [Y/Fe] + 0.3\cdot [Zr/Fe] + 0.1\cdot
[Nd/Fe]$, and $<s-heavy/Fe>= 0.25 \cdot [Mo/Fe] + 0.5 \cdot [La/Fe] +
0.25\cdot [Hf/Fe]$. The weights in these expressions reflect the
number of spectral lines measured of each element. The results are
plotted vs [Fe/H] in  Fig. \ref{sum_sr}, where the K dwarfs have
been excluded.  A downward slope of roughly the same magnitude for the
"light" and the "heavy" elements is seen. This trend is in fact also
present in most of the corresponding diagrams for the individual
elements, although the scatter in larger, see Figs. \ref{yzn} and
\ref{mofe}. A slope of this magnitude can also be traced for the
s-elements in Edvardsson et al. (1993a), Fig. \ref{yzn}. 
 
A tendency of this type may, if true, indicate that s-elements
enrichment occurs less frequently in metal-rich AGB stars. One may
speculate that this might be because mass loss could finish their
evolution earlier than for more metal-poor stars.

\subsection{Europium.}

In some studies the iron abundance derived from {Fe\,{\sc ii}} lines
are preferred as reference element for europium rather than abundances
derived from {Fe\,{\sc i}} lines. From our data this does not seem to
be an obvious choice. Particularly, this is not so for the K dwarf
stars, since europium with its low ionization energy, 5.7 eV, will
remain highly ionized for all our stars irrespective of the effective
temperature while almost all iron will be in the neutral state in the
cool stars. Thus abundances derived from Fe\,{\sc ii} for these stars
would be vulnerable to departures from LTE in the ionization
equilibrium. In our study we derive iron abundances from, in general,
more than 30 lines  from {Fe\,{\sc i}} and from four or three lines
from Fe\,{\sc ii}. Thus, also statistically  we could expect the
atomic abundance to be better determined. 

In Fig. \ref{eufe} data from Woolf et al. (1995) and our data are
plotted with iron abundances derived from {Fe\,{\sc i}}  and
{ Fe\,{\sc ii}} as reference, respectively.
Europium shows a declining trend with metallicity from [Fe/H] = $-1.0$ to
0.0 dex and this trend now seems to continue unchanged for [Fe/H] $> 0.0$ dex.

Europium is well-correlated with oxygen as well as with the
$\alpha$-elements, Fig. \ref{euo}.  This supports the idea that
europium, oxygen and the $\alpha$-elements are all formed in the same
type of events, supernovae type II. 

\begin{figure}
\resizebox{\hsize}{!}{\includegraphics{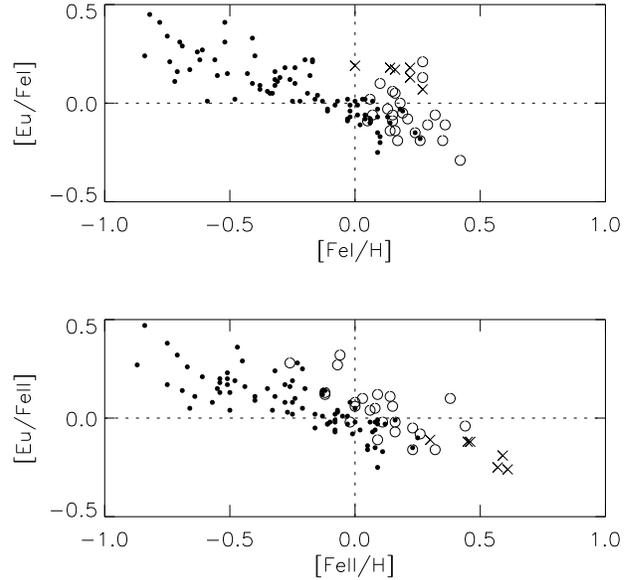}}
\caption[Europium abundances]{Europium abundances. $\circ$ symbols
this work, $\times$ symbols K dwarf stars this work, $\bullet$ symbols
Woolf et al.(1995). Iron abundances are derived from Fe\,{\sc i} and
Fe\,{\sc ii}, in the two panels respectively.  }
\label{eufe}
\end{figure}

\begin{figure}
\resizebox{\hsize}{!}{\includegraphics{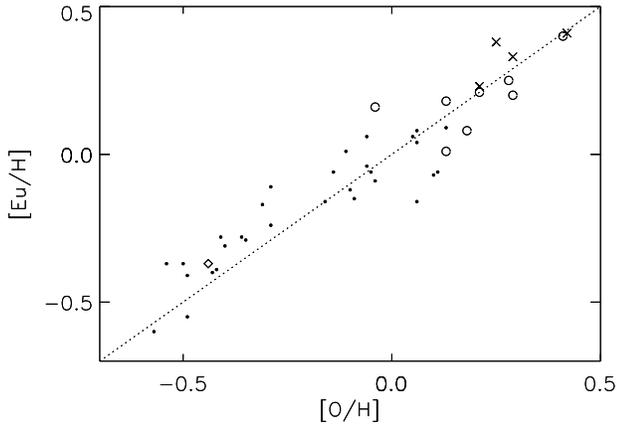}}
\caption[Europium abundances]{Europium abundances and oxygen
abundances compared. $\circ$ this work (excluding the K dwarf stars),
$\times$ K dwarf stars, $\bullet$ Woolf et al.(1995). The line with slope
$+1$ is indicated by a dotted line. Oxygen abundances are in our work
derived from the [O\,{\sc i}] line at 6300 {\AA} while the oxygen
abundances used together with the europium data from Woolf et al.(1995)
are from Edvardsson et al. (1993a) and are derived (mostly) from the triplet
lines at 7774 {\AA}, scaled to the [O\,{\sc i}] abundances using
results from Nissen \& Edvardsson (1992). }
\label{euo}
\end{figure}

It should finally be mentioned that the abundance trends (or lack of
trends) discussed here are consistent with the results obtained for 9
solar-type dwarf stars in the interval $0.05~{\rm dex}  < {\rm [Fe/H]}<0.25$ dex by
Tomkin et al. (1997). One exception from this, however, is Eu for
which the latter results sooner suggest a slight increase in the
[Eu/Fe] with increasing [Fe/H] on the basis of measurements of the
same Eu lines as used here.

\section{Conclusions -- Summary}

We have determined elemental abundances for 47 G and K dwarf stars
with iron abundances  ranging from solar to [Fe/H]=0.4 dex. The stars
selected either move on relatively eccentric orbits spending most of
their time inside the solar circle, even having been born there, or
move on solar like orbits. Among the stars on  solar like orbits there
are also a number of Extreme Population I stars, with photometric
logarithmic  metallicities ${\rm [Me/H]} > 0.4$ dex. These samples were
selected in an attempt to study differences in the chemical evolution
in different parts of the galactic disk. Our main results and
conclusions are summarized below.

The results fall roughly into two categories; galactic chemical
evolution as studied from G dwarf stars and issues related to 
spectroscopy of dwarf stars and in particular to overionization in K dwarf stars.

\paragraph{Galactic Chemical Evolution} Our data do, in general, 
fit well into the overall picture of galactic chemical evolution as it
is currently understood from empirical studies. They give further
 constraints on theoretical  models of galactic
chemical evolution for the region above solar metallicity, where
metallicity dependent supernova yields further complicate the
picture. In particular we find:

\begin{enumerate}
\item that  Mg, Al, Si, Ti, Ni, Ca, and Cr follow iron also at these high
metallicities,

\item that [Fe/H] does not vary with $V_{\rm LSR}$, nor any other element.
This indicates that the chemical evolution has lately been very
homogeneous in the regions under study. This is in contrast with
findings for $\alpha$-elements in the more metal-poor part of the disk
population, Edvardsson et al. (1993a).

\item that oxygen continues to decline with increasing [Fe/H]. This has
also been confirmed by Castro et al. (1997). 

\item that the upturn in [Na/Fe] vs [Fe/H], as observed by
Edvardsson et al. (1993a), is reproduced and reinforced. The origin of the
upturn is, however, still somewhat unclear, but metallicity dependent
supernova yields certainly play a role. In contrast to sodium we find
no such upturn for aluminium. This is somewhat surprising since sodium
and aluminium are thought to be produced in the same environment in
the pre-supernova star.
\end{enumerate}

Our lack of success in tracing variations in  relative abundances,
 e.g. of [O/Fe], [Na/Fe], [Si/Fe], [Ti/Fe] etc to vary with the
 stellar velocities,  i.e. with the present orbital mean distance from
 the galactic centre ($R_{\rm m}$). Thus, the probable mixture of the
 metal-rich stellar sample, e.g., of one old and one more recent
 population is not revealed as differences in relative
 abundances. That may be because of several reasons. One is that the
 relative number of stars from each population may not change very
 strongly with galactic radius. Another may be that the yields of
 heavy elements from the sites of nucleosyntheis are such that, e.g.,
 the interstellar Na/Fe ratio may stay rather constant through the
 evolution of the galactic disk.  A third possibility is that galactic
 orbital diffusion is strong enough to  bring these differences below
 the limit of detection.      Thus, the observed scatter in such
 abundance ratios can not be explained as the result of the mixing of
 stars with different formation sites at different distances from the
 galactic centre. 

For oxygen and europium our data suggest a continuation in the
declining trend found by earlier studies at lower metallicities. This
suggests that enrichment from supernovae Ia is still important, also at
these relatively high  metallicities. 

The scatter in titanium abundances found in Edvardsson et al. (1993a)
is reproduced in our data in spite of the fact that we use more lines
to determine abundances from. It seems probable, but remains to be
proven, that this scatter is ``cosmic''. 

Stars with very high metallicities, so called Super Metal Rich, SMR,
stars, have attracted some attention over the years. These stars have
been claimed to either represent a late stage of the local galactic
disk evolution or to be members of the Galactic bulge, which have now
been scattered out to the solar neighborhood. Barbuy \& Grenon (1990)
obtained photometric metallicities and oxygen abundances for eleven
such stars. They found these stars to have high photometric
metallicities and to be over-abundant in oxygen. Castro et al. (1997)
have obtained abundances for  9 similar stars. They find that oxygen,
calcium and titanium relative to iron declines with increasing [Fe/H],
while sodium, silicon and nickel relative to iron increases. Several of
these stars have fairly low temperatures and we interpret their results for
calcium and nickel as further evidence for the phenomenon possibly arising
from overionization. No velocity data have so far been published for
these stars and therefore it is difficult to compare with our results.
We have, however, obtained spectra and derived abundances for three of
the stars in  Barbuy \& Grenon (1990).  These stars remain puzzling -
their proposed very high [Fe/H] have not been verified, but at least
some of them seem to have, as these authors suggest, a high [O/Fe]
ratio. This should be further explored.

\paragraph{Overionization} For the K dwarf stars 
we do detect what appears to be a  case of overionization for, in
particular, iron and chromium, such that abundances derived
from Fe\,{\sc ii} and Cr\,{\sc ii} are much higher that those derived
from  Fe\,{\sc i} and Cr\,{\sc i}.  Fe\,{\sc i} abundances do not vary with the
effective temperature. Also for calcium and nickel the K dwarf
stars depart markedly from the trend outlined by the G dwarf stars. 
Signs of similar departures for rare earth atoms are also traced.

\begin{acknowledgements}

Erik Hein Olsen is thanked for providing his catalogue prior to
publication, invaluable for the planning of the project and subsequent
analysis of the data. Poul-Erik Nissen is thanked for valuable comments
and suggestions and for taking time to teach SF how to most efficiently
reduce echelle spectra. Pierre Magain is thanked for compiling and
sharing the individual line-to-line scatter for the studies from his
and Zhao's work. Bengt Edvardsson is thanked for providing programs for
the photometric calibration and comments during the work.  

Karin Eriksson  carried out a substantial fraction of
all the measurements of equivalent widths and Matthias
Palmer and Mikael Nilsson the determinations of
surface gravities from the strong 6162 {\AA} Ca\,{\sc i} line.
We are much indebted to all three for these contributions.

The Swedish Royal Academy of Sciences is gratefully thanked for giving
SF a grant to travel to the USA and thus being able to partake in the
observational work at McDonald Observatory.  This work was carried out
while BG held the Beatrice Tinsley visiting professorship at the
University of Texas, which is also gratefully acknowledged. The
Observing Committee at McDonald Observatory generously granted
observing time.

\end{acknowledgements}



\begin{thebibliography}{}

\bibitem[{Abia et al.}(1988)]{} Abia C., Rebolo R., Beckman J.E., Crivallari L., 1988, A\&A 206, 100

\bibitem[{Allen}(1973)]{} Allen C.W., 1973, Astrophysical quantities,
The Athlone Press

\bibitem[{Andersen et al.}(1984)]{} Andersen J., Gustafsson B., Lambert D.L., 1984, A\&A 136, 65

\bibitem[]{} Anders E., Grevesse N., 1989, Geochimica et Cosmochimica Acta 53, 197

\bibitem[{Arribas \& Martinez Roger}(1989)]{} Arribas S., Martinez Roger C., 1989, A\&A 215, 305

\bibitem[{Asplund et al.}(1995)]{} Asplund M., Gustafsson B, Kiselman D.,  Eriksson K., 1997, A\&A 318, 521

\bibitem[]{aumann75} Aumann J.R., Woodrow J.E.J., 1975, ApJ 197, 163

\bibitem[{Barbuy \& Grenon}(1990)]{} 
 Barbuy B., Grenon M., 1990, in Bulges of Galaxies, ESO/CTIO workshop, ed. Jarvis, B.J., Terndrup, D.M., 83

\bibitem[]{} Baumueller D., Gehren T., 1996, A\&A 307, 961

\bibitem[{Bell \& Gustafsson}(1989)]{} Bell R.A., Gustafsson B., 1989, MNRAS 236, 653

\bibitem[{Bi\'emont et al.}(1991)]{} Bi\'emont E., Baudoux M., Kurucz R.L., Ansbacher W., Pinnington E.H., 1991, A\&A 249, 539

\bibitem[{Blackwell \& Willis}(1977)]{} Blackwell D.E., Willis R.B., 1977, MNRAS 180, 169

\bibitem[{Blackwell et al.}(1995)]{} Blackwell D.E., Lynas-Gray A.E., Smith G., 1995, A\&A 296, 217

\bibitem[]{}Bruls J.H.M.J., 1993, A\&A 269, 509

\bibitem[]{}Bruls J.H.M.J., Rutten R.J., Shchukina N.G., 1992, A\&A 265, 237

\bibitem[]{}Carlsson M., Rutten R.J., Bruls J.H.M.J., Shchukina N.G., 1994, A\&A 288, 860

\bibitem[{Cayrel de Strobel \& Bentolila}(1983)]{cayrel83} Cayrel de Strobel G., Bentolia C., 1983, A\&A 119, 1

\bibitem[{Cayrel de Strobel et al.}(1997)]{} Cayrel de Strobel G., Soubiran C., Friel E.D., Ralite N., Francois P., 1997, A\&AS, 124, 299

\bibitem[]{} Castro S., Rich R.M, Grenon M., Barbuy B., McCarthy J.K., 1997, AJ, 114, 376

\bibitem[{Delbouille et al.}(1973)]{} Delbouille L., Neven L., Roland G., 1973, Photometric atlas of the solar spectrum from $\lambda 3000$ to $\lambda 10000$

\bibitem[{Drake}(1991)]{} Drake J.J., 1991, MNRAS 251, 369

\bibitem[{Edvardsson}(1988)]{} Edvardsson B., 1988, A\&A 190, 148

\bibitem[{Edvardsson et al. }(1993a)]{edvardsson93} 
Edvardsson B., Andersen J., Gustafsson B., Lambert D.L., Nissen P.E., Tomkin J. 1993a, A\&A 275, 101

\bibitem[{Edvardsson et al.}(1993b)]{edvardsson93b}
Edvardsson B., Gustafsson B., Nissen P.E., Anderssen J., Lambert D.L., Tomkin, J.,1993b, in Panchromatic View of Galaxies, eds. Hensler, G., Theis, Ch. \& Gallagher, J., 401

\bibitem[]{} Eggen O.J., 1960, MNRAS 120, 430

\bibitem[{Eriksson \& Toft}(1979)]{} Eriksson K., Toft S.C., 1979, A\&A 71, 178

\bibitem[{Friel et al.}(1993)]{} Friel E., Cayrel de Strobel G., Chmielewski Y., Spite M., L\`ebre A., Bentolila C., 1993, A\&A 274, 825

\bibitem[]{1989} Gratton R.G., 1989, A\&A 208, 171

\bibitem[{Gratton \& Sneden}(1987)]{} Gratton R.G., Sneden C., 1987, A\&A 178, 179

\bibitem[{Gratton \& Sneden}(1991)]{} Gratton R.G., Sneden C., 1991, A\&A 241, 501

\bibitem[{Grenon }(1989)]{} Grenon M, 1989, Astrophysics and Space Science 156, 29

\bibitem[{Griffin }(1968)]{} Griffin R.F., 1968, Photometric atlas of the spectrum of Arcturus, $\lambda \lambda 3600-8825$ {\AA}, Cambridge Philosophical Society

\bibitem[]{} Gustafsson B., 1995, in Astrophysical Applications of Powerful New Databases, ASP  Conf Ser. 78, eds. Adelman S.J. \& Wiese W.L., 347

\bibitem[{Gustafsson et al.}(1974)]{} Gustafsson B, Kjaergaard P., Andersen S., 1974, A\&A 34, 99

\bibitem[{Gustafsson et al.}(1975)]{} Gustafsson B., Bell R.A., Eriksson K., Nordlund {\AA}, 1975, A\&A 42, 407


\bibitem[{Hannaford et al.}(1992)]{} Hannaford P., Lowe R.M., Grevesse N., Noels A., 1992, A\&A  259, 301

\bibitem[{Hoffleit \& Jascheck}(1982)]{} Hoffleit D., Jascheck C, 1982, The Bright Star Catalogue, Yale University Observatory

\bibitem[{Holweger}(1971)]{} Holweger H., 1971, A\&A 10, 128

\bibitem[{Holweger et al.}(1991)]{} Holweger H., Bard A., Kock A., Kock M., 1991, A\&A 249, 545

\bibitem[{Johnson}(1966)]{} Johnson H.L., 1966, ARA\&A 4, 193

\bibitem[{Kiselman}(1993)]{3} Kiselman D. 1993, Thesis (Uppsala University) and A\&A 275, 269


\bibitem[{Kiselman \& Nordlund}(1995)]{} Kiselman D., Nordlund {\AA}.,  1995, A\&A 302, 578



\bibitem[{Kurucz}(1989)]{kurucz89} Kurucz R.L., 1989, Magnetic tapes with atomic line data for Ca through Ni, private communication

\bibitem[{Kurucz et al.}(1984)]{} Kurucz R., Furenlid I., Brault J., Testerman L., 1984, Solar Flux Atlas from 296 to 1300 nm, National Solar Observatory, Sunspot, New Mexico   

\bibitem[]{} Lambert D.D., 1978, MNRAS, 182, 249

\bibitem[{M\"ackle et al.}(1975)]{} M\"ackle R., Holweger H., Griffin R., Griffin R., 1975, A\&A 38, 239

\bibitem[{Mathews et al.}(1992)]{} Mathews G.J., Bazan G., Cowan J.J., 1992, ApJ 391, 719

\bibitem[{Matteucci \& Fran\c cois}(1989)]{} Matteucci F., Fran\c cois P., 1989, MNRAS 239, 885


\bibitem[{McWilliam}(1990)]{} McWilliam A., 1990, ApJS 74, 1075

\bibitem[{Moore et al.}(1966)]{} Moore C.E., Minnaert M.J.G., Houtgast J., 1966, The solar spectrum 2935 {\AA} to 8770 {\AA}, National Bureau of Standards Monograph 61

\bibitem[{Morell}(1994)]{} Morell O., 1994, PhD Thesis, Acta Universitas Upsaliensis ISBN 91-554-3322-7

\bibitem[{Neff et al.}(1995)]{} Neff J.E., O'Neal D., Saar S.H., 1995, ApJ 452, 879

\bibitem[{Neuforge}(1992)]{}  Neuforge C., 1992, in Origin and Evolution of the Elements, \\ eds. Prantzos, N., Vangioni-Flam, E., Cass\'e, M., 63

\bibitem[{Nissen \& Edvardsson}(1992)]{} Nissen P.E., Edvardsson B., 1992, A\&A 261, 255

\bibitem[{Nissen et al.}(1994)]{} Nissen P.E., Gustafsson G., Edvardsson B., Gilmore G., 1994, A\&A 285, 440


\bibitem[{Olsen}(1994)]{} Olsen E.H., 1983, A\&AS 54, 55

\bibitem[{Olsen}(1984)]{} Olsen E.H., 1984, A\&AS 57, 443

\bibitem[{Olsen}(1994)]{} Olsen E.H., 1993, A\&AS 102, 89

\bibitem[{Olsen}(1994)]{} Olsen E.H., 1994, A\&AS 106, 257

\bibitem[{Pagel \& Tautvai\v sien\. e}(1995)]{} Pagel B.E.J., Tautvai\v sien\.  e G., 1995, MNRAS 276, 505



\bibitem[]{} Prantzos N., Aubert O., 1995, A\&A, 302, 69

\bibitem[{Ruland et al.}(1980)]{} Ruland F., Holweger H., Griffin R., Griffin R., Biehl D., 1980 A\&A, 92, 70

\bibitem[{Ruland et al.}(1981)]{} Ruland F., Griffin R., Griffin R., Biehl D., Holweger H., 1981, A\&AS 42, 391

\bibitem[]{} Rutten R.J. 1988, in Physics of Formation of Fe\,{\sc II} Lines Outside LTE, ed. R. Viotti, A. Vittone, M. Friedjung, IAU Coll. 94, Reidel Publ. Company., p. 185

\bibitem[{Schuster \& Nissen}(1988)]{} Schuster W.J., Nissen P.E., 1988, A\&AS 73, 225

\bibitem[{Smith}(1981)]{} Smith G., 1981, A\&A 103, 351
\bibitem[]{} Steenbock W., 1985, in Cool stars with Excess of Heavy Elements, ed.  M. Jaschek \& P.C. Keenan D. Reidel Publ. Company, p. 231

\bibitem[]{}Steenbock W., Holweger H., 1984, A\&A 130, 319

\bibitem[{Summers }(1994)]{} Summers K., 1994, Undergraduate thesis at the Astronomical \\ Observatory, Uppsala University

\bibitem[]{} Taylor B.J., 1996, ApJS 102, 105

\bibitem[{Thielemann et al.}(1996)]{} Thielemann K-F., Nomoto K., Hashimoto M., 1996, ApJ 460, 408

\bibitem[{Timmes et al.}(1995)]{} Timmes F.X., Woosley S.E., Weaver T.A., 1995, ApJS 98, 617

\bibitem[{Tomkin et al.}(1995)]{} Tomkin J., Woolf V.M., Lambert D.L., Lemke M., 1995 AJ, 109, 2204

\bibitem[]{} Tomkin J., Edvardsson B., Lambert D.L., Gustafsson B., 1997, A\&A, in press

\bibitem[{Tsujimoto et al.}(1995)]{} Tsujimoto T., Yoshii Y., Nomoto K., Shigeyama T., 1995, A\&A 302, 704

\bibitem[{Tull et al.}(1995)]{} Tull T.G., MacQueen P.J., Sneden C., Lambert D.L., 1995, PASP 107, 251

\bibitem[{VandenBerg}(1992)]{} VandenBerg D.A., 1992, ApJ 391, 685

\bibitem[{van Altena et al.}(1991)]{} van Altena W.F, Truen-Liang Lee J., Hoffleit D., 1991, Yale Trigonometric Parallaxes Preliminary, Yale, University Observatory 

\bibitem[]{} Wielen R., Fuchs B., Dettbarn C., 1996, A\&A 314, 438
 
\bibitem[{Woolf et al.}(1995)]{} Woolf V.M., Tomkin J., Lambert D.L., 1995, ApJ 453, 660

\bibitem[{Woosley \& Weaver }(1995)]{} Woosley S.E., Weaver T.A., 1995, ApJS 101, 181 

\bibitem[]{} Wyse R., Gilmore G., 1995, AJ 110, 2771

\bibitem[{Zhao \& Magain }(1990)]{zhao90} Zhao G.,
Magain P., 1990, A\&A 238, 242









\end{thebibliography}
\end{document}